\documentclass{elsarticle}
\usepackage{graphicx}
\usepackage{bm}
\usepackage{amssymb,amsmath}
\usepackage{epstopdf}
\usepackage{epstopdf}

\usepackage{amssymb,latexsym,amsmath,amsthm,verbatim}
\usepackage{graphicx,epsfig,epstopdf,amssymb,color}

\newtheorem{theorem}{Theorem}[section]

\newtheorem{remark}[theorem]{Remark}

\newcommand{\uu}{{\bf u}}

\newcommand{\xx}{{\mathbf x}}
\newcommand{\zz}{{\mathbf z}}
\newcommand{\yy}{{\mathbf y}}
\newcommand{\sss}{{\mathbf s}}
\newcommand{\bb}{{\mathbf b}}
\newcommand{\aaa}{{\mathbf a}}

\newcommand{\VV}{{\bf V}}

\newcommand{\cE}{{\cal E}}

\newcommand{\real}{\mathbb{R}}

\newcommand{\inttwo}{{\int_{\real^2}}}

\title{A multi-moment vortex method for 2D viscous fluids}
\author[UCLA]{David Uminsky\corref{cor1}}
\ead{duminsky@math.ucla.edu}
\author[BU]{C. Eugene Wayne}
\ead{cew@math.bu.edu}
\author[UCLA]{Alethea Barbaro}
\ead{alethea@math.ucla.edu}

\cortext[cor1]{Corresponding author}
\address[UCLA]{UCLA Dept. of Mathematics, Box 951555, Los Angeles, CA 90095-1555}
\address[BU]{Dept. of Mathematics and Statistics, Boston University, 111 Cummington St., Boston, MA 02215}

\begin{document}
\begin{abstract}
 In this paper we introduce simplified, combinatorially exact formulas that arise in the vortex interaction model found in \cite{SIADS}. These combinatorial formulas allow for the efficient implementation and development of a new multi-moment vortex method (MMVM) using a Hermite expansion to simulate 2D vorticity. The method naturally allows the particles to deform and become highly anisotropic as they evolve without the added cost of computing the non-local Biot-Savart integral. We present three examples using MMVM. We first focus our attention on the implementation of a single particle, large number of Hermite moments case, in the context of quadrupole perturbations of the Lamb-Oseen vortex. At smaller perturbation values, we show the method captures the shear diffusion mechanism and the rapid relaxation (on $Re^{1/3}$ time scale) to an axisymmetric state. We then present two more examples of the full multi-moment vortex method and discuss the results in the context of classic vortex methods. We perform spatial convergence studies of the single-particle method and show that the method exhibits exponential convergence. Lastly, we  numerically investigate the spatial accuracy improvement from the inclusion of higher Hermite moments in the full MMVM.
\end{abstract}
\begin{keyword}
vortex methods, particle methods, vortex dynamics, Hermite functions, Navier-Stokes equations
\end{keyword}

\maketitle

\section{Introduction}\label{sec:Intro}


A new viscous vortex interaction model was introduced in \cite{SIADS,Uminsky:2009} whose main purpose is to study the dynamics of the viscous $n$-vortex problem. As a first application, the model was shown in \cite{nagem:2007}, to improve the approximation of the far field acoustic pressure field generated by
a pair of rotating vortices, when compared to classical point vortex methods.
One of the strengths of the model is the reduction of the 2D vorticity equation to a system of ODE's with purely quadratic nonlinearity which represent the evolution of the Hermite moments of each vortex. The main drawback of the model
as presented in \cite{SIADS} is that the coefficients in the ODE's are computed in terms of derivatives and limits of an explicit kernel function which are computationally intractable for higher resolution (more Hermite moments) computations.

We introduce in  \ref{sec:Combinatorial} simplified, combinatorial exact formulas for the coefficients of these terms, greatly reducing the computational costs of generating and implementing the ODEs. These simplified formulas allow the viscous vortex interaction model of \cite{SIADS} to be implemented as
a {\it new} multi-moment vortex method to simulate the two-dimensional vorticity equation. The novelty of this multi-moment vortex method (MMVM) is that the vortex particles are allowed to dynamically deform under the flow, {\it without} the usual difficulties of computing the Biot-Savart kernel for anisotropic basis elements, thus reducing the
convection error associated to classic vortex methods.

It is important to note that the multi-moment vortex method we propose here shares many features with traditional vortex methods for incompressible { \cite{Anderson:1985,Beale:1982, Chorin:1973} and compressible \cite{Eldredge:2002} flows which are comprehensively reviewed in \cite{Cottet:2000}. One of the advantages that this method shares with traditional vortex methods is that it is gridless. In particular, we can avoid the common computational issues of using a large grid approximation for an unbounded domain. Using vortex methods, one also avoids the necessity of imposing either artificial, or periodic boundary conditions which may introduce unwanted or unphysical effects. MMVM also naturally incorporates the effects of viscosity in a way most closely related to the core spreading method, utilized in the
work of Leonard, Rossi, Barba, Huang \cite{Leonard:1980,Rossi:1996,barba:2006,Huang:2009}. Core spreading, using a purely Lagrangian vortex method, was originally shown to converge to the wrong solution in \cite{Greengard:1985}.
An adaptive method, proposed by Rossi \cite{Rossi:1996}, has shown to fix this convergence problem and since then, innovative interpolation and initialization methods \cite{Barba:2010}, have been developed to more accurately address this convergence problem.

To incorporate anisotropic deformations, our method uses a Hermite moment expansion for each vortex element. For two-dimensional {\em inviscid} fluids,
Melander et al. \cite{Melander:1984,Melander:1986} also developed a model based on the evolution of moments of vortices. Instead of using Hermite moments, Melander et al. define a local coordinate system at each vortex and compute equations for the evolution of the centroid along with the time evolution of the local {\em geometric} moments. While in theory equations may be computed to any order for these expansions relatively
low order approximations are used in \cite{Melander:1984,Melander:1986} and convergence
issues are not addressed.
Using a Hermite moment expansion is natural, as they form a suitable basis and conditions for global in time convergence have been proven for MMVM \cite{SIADS}. In addition, it is assumed in \cite{Melander:1984,Melander:1986} that the maximum diameter of any localized vortex is much smaller than the minimum distance between any two vortices. This would specifically not allow the use of many overlapping vortices to simulate the fluid dynamics, i.e. a vortex method approach. Nonetheless, Melander et al. are able to show, using well separated vortices with moments up to second order, significant improvement over the point vortex method for several computed examples involving well-separated, co-rotating vortices.

To our knowledge, the most closely related work to MMVM is Rossi's deformable vortex method using elliptically deforming Gaussian vortex particles. Rossi implemented the method first for the advection diffusion equation \cite{Rossi:2005,Rossi:2006B} and also for the Navier-Stokes equations \cite{Rossi:2006A, Platte:2009}. For both the advection diffusion equation and the Navier-Stokes equations, Rossi's deformable vortex method was able to achieve higher order convergence, as compared to non-deforming vortices, but only if one also modifies the Lagrangian velocity field. Unfortunately, the method was hampered by the difficulty of computing the Biot-Savart law for anisotropic (elliptical) basis functions to reconstruct the velocity field. While several ideas
were proposed to alleviate this problem  \cite{Rossi:2006A, Platte:2009}, we note that one of the most promising advantages of our approach is the avoidance of this difficult computation altogether, as the Biot-Savart integral is explicitly known to all orders in our Hermite expansion.

For general flows the primary source of error from vortex methods arises from convection error which is caused by advecting vortices {\em without deformation}. Thus our method should greatly reduce this error since it allows for many different deformations, limited only by the order of the Hermite approximation. In fact, constructing basis functions that are allowed to deform have been conjectured as precisely what is need to improve the spatial accuracy \cite{Barba:2004} of vortex methods as seen already by Rossi's improvement from second to fourth order with elliptical deformed Gaussian basis functions \cite{Rossi:2006A}. Prior to Rossi's work elliptically deforming basis elements were used by Moeleker and Leonard \cite{Moeleker:2001} in a filtered advection-diffusion equation. In \cite{Moeleker:2001} the authors use a Hermite expansion but only to compute an approximation to the average velocity field, and not the vorticity field itself. This modification to the Lagrangian velocity field did not result in an increase the order of accuracy.

In this paper we will first analyze the single-particle MMVM ($n=1$) with a large number of Hermite moments, $m$. Unlike usual particle methods, our single vortex element must be large and model an entire vortex structure. { As a first application of the single-particle MMVM, we consider the problem of shear diffusion around a rotationally symmetric vortex. The shear diffusion manifests itself in this context through the mixing of vorticity and the axisymmetrization of the vorticity profile on a $Re^{1/3}$ time scale \cite{Bernoff:1994}. MMVM is a natural choice
to study this problem for several reasons. As we will discuss in Section \ref{sec:Review} our single-particle MMVM is equivalent to implementing a Hermite spectral method. Thus, while offering the usual good approximation properties of classical spectral methods it avoids the artificial periodic boundary conditions associated to classical spectral methods (Fourier, Chebyshev), see \cite{Canuto:2006}}. Aside from these general advantages, the single-particle MMVM offers some particular advantages for the study of two-dimensional fluid motion. Theoretical results have shown that the Hermite functions we use for our expansion give a natural basis with respect to which one can study the long-time asymptotics of solutions of the two-dimensional vorticity equation \cite{GallayWayne:2002}. Furthermore, the vorticity distribution (in contrast to the fluid velocity) has the property that if it is initially localized it will remain so for later times. This means that if our Hermite expansion of the initial vorticity distribution converges, it will do so for all later times \cite{SIADS}.
}

We next consider two examples using the full MMVM: A model for early time vortex merger behavior using  $n=2$ vortex elements, and second, we consider tripole relaxation using a course grid approximation $n=36$.  In both cases we observe increasingly accurate solutions as we increase the number of moments $m$.  MMVM has several important parameters to choose for implementation. Just as in regular vortex methods, one must choose the number of particles $n$, and the overlap ratio which is computed by dividing the particle core size by the particle spacing. In addition, we must also specify the number of Hermite moments $m$ for each particle to include in the simulation.  As we will see both parameters $n$ and $m$ can be increased to achieve greater accuracy and,
at least in the case of a  single particle ($n=1$) the spatial convergence is faster than polynomial in $m$.

We will work primarily with the 2D vorticity equation which can be derived from the Navier-Stokes equations for two dimensional incompressible fluids,
\begin{equation}\label{eq:fluid1}
\frac{\partial {\bf u}}{\partial t} + ({\bf u} \cdot \mathbf{\nabla}){\bf u} =
- \mathbf{\nabla} \frac{p}{\rho} + \nu \Delta {\bf u},
\end{equation}
\begin{equation}\label{eq:fluid2}
\mathbf{\nabla} \cdot {\bf u} = 0,
\end{equation}
where, ${\bf u}$ is the fluid velocity, $\rho$ is the fluid density and $\nu$ is the kinematic
viscosity.

If we take the curl of \eqref{eq:fluid1} and assume that the vorticity field {$\omega = \nabla \times \uu$} is sufficiently localized then the equations for viscous vorticity on the entire plane
are
\begin{eqnarray}\label{eq:2DIVP}\nonumber
&& \frac{\partial \omega}{\partial t} = \nu \Delta \omega - \uu \cdot \mathbf{\nabla} \omega ,\\
&& \omega = \omega(\xx,t) ,\ \xx \in \real^2 ,\ t > 0 \\ \nonumber
&& \omega(\xx,0) = \omega_0(\xx) \ ,
\end{eqnarray}
where we can recover the velocity of the fluid
via the Biot-Savart law
\begin{equation}\label{eq:fluids7}
{\bf u}(\xx) = \frac{1}{2\pi} \inttwo \frac{(\xx-\yy)^{\perp}}{|\xx-\yy|^2} \omega(\yy) d\yy\ ,
\end{equation}
and for a two-vector $\zz = (z_1,z_2)$, $\zz^{\perp} = (-z_2,z_1)$.




The rest of this paper is outlined as follow: In Section \ref{sec:Review}, we will briefly review the model from \cite{SIADS} which naturally defines the MMVM. In this section we begin with a review of the single-particle MMVM and then present the full MMVM.

In Section \ref{sec:threeEx}, we present three different examples of our method. We first carry out the single-particle MMVM implementation  in the context of perturbations of a monopole. We take our monopole to be the Lamb-Oseen vortex and perturb the vortex by a  quadrupole  perturbation. We show, using an enstropy calculation, that the rapid shear diffusion mechanism causes axisymmetric relaxation to occur on the  time scale of $Re^{1/3}$, much faster than  the purely diffusive time scale of $Re$.

Our next two examples in Section \ref{sec:threeEx} use the full MMVM. The first is an $n=2$ model of vortex merger. By using two vortex elements to represent the merging vortices we show that increasing the number of Hermite moments improves the short time fluid dynamic behavior of each vortex. The third example is a coarse grid approximation ($6\times6$) of a large quadrapole perturbation to the Lamb-Oseen vortex and we show that just including the second order Hermite moments significantly improves the early time simulation as compared to classic vortex methods. This example in particular illustrates the promise of implementing MMVM on a much larger scale.

In Section \ref{sec:Numerical}, we perform two convergence studies on the single-particle MMVM  and show that the convergence is exponential. We also quantify the improvement in spatial accuracy in the course grid calculation and end with a brief discussion of our computational implementation of MMVM.

We conclude in Section \ref{sec:conclusion} with a discussion of our results and future work in the further development of MMVM. Lastly, the simplified system of equations for MMVM is derived in \ref{sec:Combinatorial}.


\section{Review of the viscous vortex model }\label{sec:Review}

\subsection{Single-Particle MMVM}\label{sec:three}


In this section, we review the viscous vortex model,  presented in \cite{SIADS}, which expands the vorticity in terms of Hermite functions. We then formally state the single-particle MMVM, derived from the model in \cite{SIADS}.  We  begin with a brief review of the Hermite functions.

Define
\begin{equation}\label{eq:phi00}
\phi_{00}(\xx,t;\lambda) = \frac{1}{ \pi \lambda^2 } e^{-|\xx|^2/\lambda^2}
\end{equation}
where $\lambda^2 := \lambda_0^2 + 4 \nu t$ and $\lambda_0$ represents the initial core size of our localized vortex structure. { Note that for any value of $\lambda_0$, \eqref{eq:phi00} defines
an exact solution of the two dimensional vorticity equation known as the Lamb-Oseen vortex.
As a consequence, we can choose any value of $\lambda_0$ in the definition of our Hermite spectral method; $\lambda_0$ is most often chosen to represent a typical length scale in the
initial vorticity distribution.}

We now define the Hermite function of order $(k_1,k_2)$ by
\begin{equation}\label{eq:Hermitedef}
\phi_{k_1,k_2}(\xx,t;\lambda) = D_{x_1}^{k_1} D_{x_2}^{k_2} \phi_{00}(\xx,t;\lambda)
\end{equation}
and the corresponding moment expansion of a function by
\begin{equation}\label{eq:moment_expansion_def}
\omega(\xx,t) = \sum_{k_1,k_2=0}^\infty M[k_1,k_2;t] \phi_{k_1,k_2}(\xx,t;\lambda),
\end{equation}

We define the Hermite polynomials via their generating function:
\begin{equation}\label{eq:Hermitepoly}
H_{n_1,n_2}(\zz;\lambda) = \left.\left( (D^{n_1}_{\tau_1} D^{n_2}_{\tau_2}
e^{\left( \frac{2 {\mathbf \tau} \cdot \zz - |{\mathbf \tau}|^2}{\lambda^2}\right)} \right)\right\vert_{{\mathbf \tau}=0}.
\end{equation}
Note that the ``standard'' Hermite polynomials correspond to the case {$\lambda \equiv 1$}.
Then using the standard orthogonality relationship for the Hermite polynomials,
\begin{equation}
\inttwo H_{n_1,n_2}(\zz;1) H_{m_1,m_2}(\zz;1) e^{-z^2} {\rm d}\zz
= \pi 2^{n_1+n_2} (n_1!) (n_2!) \delta_{n1,m1} \delta_{n_2,m_2}~~,
\end{equation}
we see that the coefficients in the expansion \eqref{eq:moment_expansion_def} are
defined by the projection operators:
\begin{equation}\label{eq:coefficient}
 M[k_1,k_2;t] = P_{k_1,k_2} [\omega(t)] = \frac{ (-1)^{k_1+k_2} \lambda^{2(k_1+k_2)} }{2^{k_1+k_2} (k_1!) (k_2!) }
 \inttwo H_{k_1,k_2}(\zz;\lambda) \omega(\zz,t) {\rm d}\zz~~.
\end{equation}

If the function $\omega(\xx,t)$ in \eqref{eq:moment_expansion_def} is the
vorticity field of some fluid, the linearity of the Biot-Savart law implies that we
can expand the associated velocity field as:
\begin{equation}\label{eq:velocity_field_expansion}
\uu(\xx,t) = \sum_{k_1, k_2=0}^{\infty} M[k_1,k_2;t] \VV_{k_1,k_2}(\xx,t;\lambda)
\end{equation}
where
\begin{equation}\label{eq:velocity_moments}
\VV_{k_1,k_2}(\xx,t;\lambda) = D_{x_1}^{k_1} D_{x_2}^{k_2} \VV_{00}(\xx,t;\lambda)
\end{equation}
and $\VV_{00}(\xx,t;\lambda)$ is the velocity field associated with the Gaussian
vorticity distribution $\phi_{00}$. Explicitly we have:
\begin{equation}\label{eq:Vzerodef}
\VV_{00}(\xx,t;\lambda) = \frac{1}{2\pi}\frac{(-x_2,x_1)}{|\xx|^2}
 \Bigl(1 - e^{-|\xx|^2/\lambda^2}\Bigr)~~.
\end{equation}
\begin{remark}
For simplicity, in this review we do not allow the centroid of vorticity to evolve in the single-particle MMVM. Of course, this is not the case in the full multi-moment vortex method. This restriction essentially adds the condition that the center of our expansion is chosen in such a way that both first moments are zero.
\end{remark}

\subsubsection{Convergence}\label{subsec:onevortexconvergence}

One of the main theoretical results of \cite{SIADS} is the derivation of a criterion on the initial data which ensures convergence of the expansion (\ref{eq:moment_expansion_def}) for all $t>0$ under the flow of 2D vorticity \eqref{eq:2DIVP}. We state the proposition here:
\begin{theorem}\label{th:converence_criterion} Define the weighted enstrophy function
$$
\cE(t) = \int_{\real^2} \phi_{00}^{-1}(\xx,t;\lambda) (\omega(\xx,t))^2 {\rm d}\xx \ .
$$
If the initial vorticity distribution $\omega_0$ is such that
$\cE(0) < \infty$ for some $\lambda_0$, and $\omega_0$ is bounded
(in the $L^{\infty}$ norm) then the expansion \eqref{eq:moment_expansion_def} is convergent for the initial value problem \eqref{eq:2DIVP} for all times $t > 0$.
\end{theorem}
The proof relies on a differential inequality argument and the fact that the Hermite functions are eigenfunctions for a related self-adjoint linear operator, see \cite{SIADS} for details of the proof. With this criterion for convergence, we now turn our attention to the governing equations for the moments which must be solved to implement the MMVM.

\subsubsection{Differential equations for the moments}

Assuming that the function $\omega(\zz,t)$ is a solution of \eqref{eq:2DIVP}, we note here that by using  ansatz \eqref{eq:moment_expansion_def} for $\omega(\zz,t)$ it is clear that the single-particle MMVM is equivalent to a Hermite spectral method for solving \eqref{eq:2DIVP}, as the entire effort is to compute the evolution of the coefficients of the Hermite functions. To begin we first introduce the following summation notation:
\begin{equation}
\sum_{k_1,k_2}^m f(k_1,k_2) := \sum_{i=0}^m\sum_{\substack{k_1+ k_2=i \\ k_1\ge0, k_2\ge 0}} f(k_1,k_2).
\end{equation}
Thus we now define
\begin{equation}\label{eq:vorticity_trunc}
\omega^m(\xx,t) = \sum_{k_1, k_2}^m M[k_1,k_2;t] \phi_{k_1,k_2}(\xx,t;\lambda),
\end{equation}
and
\begin{equation}\label{eq:velocity_trunc}
\uu^m(\xx,t) = \sum_{k_1, k_2}^m M[k_1,k_2;t] \VV_{k_1,k_2}(\xx,t;\lambda).
\end{equation}
Here, $\omega^m$ and $\uu^m$ represent our Hermite approximation to solutions of \eqref{eq:2DIVP}.

{
We can now derive equations for the evolution of the coefficients in the moment expansion by
a standard Galerkin approximation.} For nonlinear PDEs, this derivation can be challenging since passing the nonlinearity through the projection operator \eqref{eq:coefficient} often cannot be explicitly computed. This is even more true in the case of vorticity where the nonlinearity includes a product of vorticity and velocity. However, the expressions for the coefficients in these expansions are surprisingly simple and explicit. Let {$P^m[\cdot]$} represent the projection on the subspace spanned by the Hermite functions of order $m$ or less. (Without detailing the argument here, we consider this as a subspace of the weighted $L^2$ space in which the norm is defined by the weighted enstrophy function in Theorem \ref{th:converence_criterion}. See \cite{SIADS} for a detailed discussion.) The standard Galerkin approximation to \eqref{eq:2DIVP} is then given by:
\begin{eqnarray}\nonumber \label{eq:Galerkin}
&& \partial_t \omega^m = \sum_{k_1,k_2}^m \frac{{\rm d}M[k_1,k_2;t] }{{\rm d}t}
\phi_{k_1,k_2}(\xx,t;\lambda) + \sum_{k_1,k_2}^m
M[k_1,k_2;t] \partial_t \phi_{k_1,k_2}(\xx,t;\lambda) \\ && \quad
= \sum_{k_1,k_2}^m M[k_1,k_2;t] \left( \nu \Delta \phi_{k_1,k_2}(\xx,t;\lambda) \right) \\ \nonumber
&&
\qquad
\qquad
{- P^m \left[\left( \sum_{\ell_1,\ell_2}^m M[\ell_1,\ell_2;t] \VV_{\ell_1,\ell_2}(\xx,t;\lambda) \right)
\cdot \mathbf{\nabla} \left(\sum_{k_1,k_2}^m M[k_1,k_2;t] \phi_{k_1,k_2}(\xx,t;\lambda) \right)\right]~~.}
\end{eqnarray}

Our first simplification comes from noticing that the final term in the first line cancels the
middle line of equation \eqref{eq:Galerkin} (from the definition of Hermite functions) and hence if we apply the projection operators
$P_{k_1,k_2}$,
defined in \eqref{eq:coefficient}, for $k_1+k_2 \le M$ we are left with the
system of ordinary differential equations for the moments
\begin{eqnarray}\label{eq:mom1}
\frac{{\rm d}M}{{\rm d}t}[k_1,k_2;t] &=& - P_{k_1,k_2} \left[ \left( \sum_{\ell_1,\ell_2}^m M[\ell_1,\ell_2;t]
\VV_{\ell_1,\ell_2}(\xx,t;\lambda) \right)\right. \\ \nonumber
&& \qquad \qquad
\cdot \mathbf{\nabla}\left. \left(\sum_{m_1,m_2}^m M[m_1,m_2;t] \phi_{m_1,m_2}(\xx,t;\lambda) \right) \right] ~~,
\end{eqnarray}
which, evaluated, takes the form,
\begin{eqnarray} \label{eq:moment_evolution}
 && \frac{d M}{dt}[k_1,k_2,t] = - \frac{(-1)^{(k_1+k_2)} \lambda^{2 (k_1+ k_2)}}{2^{k_1+k_2} (k_1 !) (k_2 !) } \sum_{\ell_1,\ell_2}^m \sum_{m_1,m_2}^m
 M[\ell_1,\ell_2,t] M[m_1,m_2,t] \\ \nonumber && \qquad \qquad \qquad \times \int_{\real^2} H_{k_1,k_2}(\xx)
 ( D_{x_1}^{m_1} D_{x_2}^{m_2} \VV_{00}(\xx;\lambda) ) \cdot \mathbf{\nabla}_{\xx} ( D_{x_1}^{\ell_1} D_{x_2}^{\ell_2} \phi_{00}(\xx;\lambda) ) {\rm d}\xx.
\end{eqnarray}

Explicitly computing equation \eqref{eq:mom1} is crucial to numerically implementing our single-particle MMVM (or equivalently our Hermite spectral method), since the projection operator is an integral over the entire plane. The difficulty in obtaining an
analytic expression for the RHS of \eqref{eq:mom1} is that products of $\VV_{\ell_1,\ell_2}$ and $\phi_{k_1,k_2}$ {cannot} be integrated in a straightforward manner.
It was shown in \cite{SIADS} that this difficulty can be surmounted and equation \eqref{eq:mom1} can be reduced to taking limits of derivatives of a relatively simple kernel function, $K$, thus taking the form:
\begin{eqnarray} \label{eq:moment_evolution_redux}
 && \frac{{\rm d} M}{{\rm d}t}[k_1,k_2,t] = \\
 \nonumber && \quad = - \frac{ (-1)^{(k_1+k_2)} \lambda^{2 (k_1+ k_2)}}{2^{k_1+k_2} (k_1 !) (k_2 !) } \sum_{\ell_1,\ell_2}^m \sum_{m_1,m_2}^m \Gamma[k_1,k_2,\ell_1,\ell_2,m_1,m_2;\lambda]
 M[\ell_1,\ell_2,t] M[m_1,m_2,t] \end{eqnarray}
where
\begin{eqnarray}\label{eq_coefficient_def}
&& \Gamma[k_1,k_2,\ell_1,\ell_2,m_1,m_2;\lambda] = \\
 && \qquad = D_{\tau_1}^{k_1} D_{\tau_2}^{k_2} D_{b_1}^{m_1} D_{b_2}^{m_2} D_{a_1}^{\ell_1} D_{a_2}^{\ell_2} K(a_1,a_2,b_1,b_2,\tau_1,\tau_2;\lambda)|_{ {\mathbf \tau=0,a=0,b=0}}
\end{eqnarray}
and
\begin{eqnarray}\label{eq:oneblobkernel}
&& K(a_1,a_2,b_1,b_2,t_1,t_2;\lambda) = \frac{1}{\pi \lambda^2} e^{-\frac{2}{ \lambda^2}\left(t_1a_1+t_2a_2 \right)} \times \\ \nonumber
&& \left( \frac{
    -{a_2}\,{t_1} +
    {b_2}\,{t_1}
     +{a_1}\,{t_2}- {b_1} \,
     {t_2} }{ \, \left( (b_1 + (t_1-a_1))^2 + (b_2 + (t_2-a_2) )^2 \right)
    } \right) \left( 1 - e^{{-\frac{1}{2 \lambda^2}\left( (b_1 + (t_1-a_1))^2 + (b_2 + (t_2-a_2) )^2 \right)}}\right).
\end{eqnarray}

Implementing the single-particle MMVM is now reduced to solving equations \eqref{eq:moment_evolution_redux}-\eqref{eq:oneblobkernel} subject to appropriate initial conditions. Unfortunately, evaluating equation \eqref{eq_coefficient_def} directly
is computationally intractable for large $m$. One of the main results of this paper is an
explicit combinatorial formula for { $\Gamma[k_1,k_2,\ell_1,\ell_2,m_1,m_2;\lambda]$}, derived in  \ref{sec:Combinatorial}. Using the result from \ref{sec:Combinatorial}, we can instead implement our single-particle MMVM  by solving the system of quadratic ODEs given by equations \eqref{eq:FinalSingleI}-\eqref{eq:moment_evolutionFinish}. 
We will implement this single-particle MMVM  in Section \ref{sec:Sheardiffusion}  to study the shear diffusion mechanism but next, we review the general MMVM model.

\subsection{The general MMVM}\label{sec:Deform}

To extend the single-particle MMVM we  separate the solution of the vorticity equation into $n$ vortex particles and derive separate evolution equations for each particle.
Returning to the initial value problem \eqref{eq:2DIVP}, we begin by decomposing the vorticity
distribution, writing (for $t \ge 0$),
\begin{equation}\label{eq:vorticity_decomposition}
\omega(\xx,t) = \sum_{j=1}^n \omega^j(\xx-\xx^j(t),t),
\end{equation}
with velocity field
\begin{equation}\label{eq:velocity_decomposition}
\uu(\xx,t) = \sum_{j=1}^n \uu^j(\xx-\xx^j(t),t),
\end{equation}
where $\uu^j(\yy,t)$ is the velocity field associated to $\omega^j(\yy,t)$, centered at $\xx^j(t)$, determined by the Biot-Savart Law.

\begin{remark}
This decomposition is not unique.  We can choose any number of pieces, $n$, into which we decompose the
vorticity. However we will require two conditions on the choice of decomposition:
the first is that the total vorticity of each vortex is non-zero, i.e.
\begin{equation}\label{eq:mass}
m^j = \int_{\real^2} \omega^j_0(\xx) {\rm d}\xx \ne 0 ,\ j=1, \dots ,N .
\end{equation}
The second condition requires that for $t\ge 0$
\begin{equation}\label{eq:xjdef}
\int_{\real^2}(\xx-\xx^j(t)) \omega^j(\xx-\xx^j(t),t) {\rm d}\xx = \mathbf{0} ~~{\rm for~all} ~~t>0, \ j = 1, \dots N\ .
\end{equation}
As we will see, imposing condition (\ref{eq:xjdef}) defines the motion of each vortex and is a departure from the
traditional Lagrangian perspective of assigning the condition:
\begin{equation}\label{eq:lagrangian}
\frac{{\rm d}\xx_i}{{\rm d} t} = \uu(x,t).
\end{equation}
Physically, condition (\ref{eq:xjdef}) imposes that the motion of
the centers of the vortices are determined by momentum
conservation, {\em not} by the passive advection of the local
velocity field.

There are two reasons why we favor \eqref{eq:xjdef} instead of the
more traditional choice \eqref{eq:lagrangian}. First, in
\cite{GallayWayne:2002}, it was proven that for the long-time
asymptotic behavior of solutions of the two-dimensional vorticity
equation, one has more rapid convergence to the Lamb-Oseen vortex solution
if the initial data satisfies \eqref{eq:xjdef}. Secondly, support
for this point of view is provided by the work of Lingevitch and
Bernoff in \cite{Bernoff:1995} who argue that condition \eqref{eq:xjdef}
is the ``natural'' definition for the center of a vortex immersed
in a background flow.

\end{remark}


 If we take the partial derivative of \eqref{eq:vorticity_decomposition}
and use the vorticity equation \eqref{eq:2DIVP}, we find:
\begin{eqnarray}\label{eq:allomega}
\partial_t \omega(\xx,t)&=& \sum_{j=1}^n \partial_t \omega^j(\xx-\xx^j(t),t) -
\sum_{j=1}^n \dot{\xx}^j(t) \cdot \mathbf{\nabla} \omega^j(\xx-\xx^j(t),t) \\ \nonumber
&=& \sum_{j=1}^n \nu \Delta \omega^j(\xx-\xx^j(t),t) - \sum_{j=1}^n \left(\sum_{\ell=1}^n
\uu^{\ell}(\xx-\xx^{\ell}(t),t) \right) \cdot \mathbf{\nabla} \omega^j(\xx-\xx^j(t),t)\ .
\end{eqnarray}

{Given this equation, it is natural to define $\omega^j$ as the solution of the equation}:
\begin{eqnarray}\label{eq:omegaj}\nonumber
\frac{\partial \omega^j}{\partial t}(\xx-\xx^j(t),t) &=& \nu \Delta \omega^j(\xx-\xx^j(t),t)
- \left(\sum_{\ell=1}^n
\uu^{\ell}(\xx-\xx^{\ell}(t),t) \right) \cdot \mathbf{\nabla} \omega^j(\xx-\xx^j(t),t) \\
&& \qquad + \dot{\xx}^j(t) \cdot \mathbf{\nabla} \omega^j(\xx-\xx^j(t),t)
\ ,\ j=1, \dots , N.
\end{eqnarray}
It remains to determine $\dot{\xx}^j(t)$.  If we differentiate condition \eqref{eq:xjdef} we find that the equations of motions for the centers are defined as:
\begin{equation}\label{eq:xjdot}
\dot{\xx}^j_n(t) := \frac{{\rm d} \xx^j_n}{{\rm d}t}(t) = \frac{1}{m^j} \sum_{\ell=1;\ell \ne j}^n \inttwo \left(
\uu^{\ell}_n (\zz + \xx^j(t) -\xx^{\ell}(t),t) \omega^j(\zz,t) \right) {\rm d}\zz\ ,~~n=1,2,
\end{equation}

Equation (\ref{eq:omegaj}) gives a set of $n$ partial differential equations which govern the evolution and deformation of the vorticity of each localized vortex structure and equation (\ref{eq:xjdot}) gives a set of $2n$ ODEs governing the motion of the centers of each localized vortex structure. Taken together, equations (\ref{eq:omegaj}) and (\ref{eq:xjdot}) are the model proposed in \cite{SIADS} and solutions of the model solve the 2D vorticity equation exactly.

\subsubsection{The expansion for several vortex centers}

Let us now briefly comment on the extension of the Hermite moment expansion of the previous section to the case in which there are two or more centers of vorticity by combining this
expansion with the multi-vortex representation of defined in the previous section.
In this case, we consider the equations \eqref{eq:omegaj} for the evolution
of each vortex and then expand each of the functions $\omega^j$ in Hermite
moments as in the previous section.  Thus, we define
\begin{equation}\label{eq:moment_expansion_def_tw0}
\omega^j(\zz,t) = \sum_{k_1,k_2}^{\infty} M^j[k_1,k_2;t] \phi_{k_1,k_2}(\zz,t;\lambda)
\end{equation}
for $j=1,~2,...,~n$.
We make a similar expansion for the velocity field in terms of the functions
$\VV_{\ell_1,\ell_2}$, and insert the expansions into \eqref{eq:omegaj}. Letting
$\zz = \xx-\xx^j(t)$ and recalling that $\partial_t \phi_{k_1,k_1} = \nu
\Delta \phi_{k_1,k_2}$, we obtain:
\begin{eqnarray}\label{eq:mom_multi}
&& \frac{{\rm d}M^j [k_1,k_2;t] }{{\rm d}t} = \\ \nonumber &&
\qquad - P_{k_1,k_2} \left[ \left( \sum_{j^{\prime}=1}^n
\sum_{\ell_1,\ell_2}^{\infty} M^{j^{\prime}} [\ell_1,\ell_2;t]
\VV_{\ell_1,\ell_2}(\zz+\sss_{j,j^{\prime}},t;\lambda) \right)
 \right. \\ \nonumber &&
\qquad \qquad \qquad \qquad \left. \cdot \mathbf{\nabla} \left(\sum_{m_1,m_2}^{\infty} M^j[m_1,m_2;t] \phi_{m_1,m_2}(\zz,t;\lambda) \right) \right] \nonumber\\
&& + P_{k_1,k_2} \left[\dot{\xx}^j(t)\cdot \mathbf{\nabla} \left(\sum_{m_1,m_2}^{\infty} M^j[m_1,m_2;t] \phi_{m_1,m_2}(\zz,t;\lambda) \right)  \right] \nonumber.
\end{eqnarray}
where $\sss_{j,j^{\prime}} = \xx^{j^{\prime}}(t) - \xx^j(t)$.

Each of the above projections is fully simplified in  \ref{sec:Combinatorial}.  Thus, implementing MMVM is now a matter of simulating equations (\ref{eq:Final_CrossCoef})-(\ref{eq:xjdot_Final}).  The simplified combinatorial formulae derived in \ref{sec:Combinatorial} allow for the efficient implementation of MMVM using basis functions constructed out of Hermite moments.  Remarkably, the convergence for the multi-vortex model is also established in \cite{SIADS}.

\section{Implementation and results}\label{sec:threeEx}

In this section we implement the MMVM in three examples, the first being an implementation of the single-particle MMVM in the context of studying shear-diffusion.  The second example is implementation of the full MMVM using $n=2$ particles to model the early stages of vortex merger. The last example builds toward a larger scale simulation of the full MMVM by using a coarse grid ($6 \times 6$) approximation to a tripole relaxation example.  In the last two examples we specifically consider the effect of increasing the number of moments $m$ used in the MMVM.

\subsection{The single-particle MMVM}

\subsubsection{Shear diffusion}\label{sec:Sheardiffusion}
It has been observed that small perturbations of an axisymmetric vortex monopole rapidly relax back to an axisymmetric state.
It is has been argued \cite{Bernoff:1994,lundgren:1982} that the shear diffusion mechanism is the cause of the rapid homogenization of small perturbations on the fast time scale $Re^{1/3}$. Shear diffusion or the rapid relaxation back to an axisymmetric state was first associated to the passive scalar problem on the time scale $Pe^{1/3}$, where $Pe$ is the analogous P\'{e}clet number, by Rhines and Young in \cite{Rhines:1982}. Unlike the passive advection diffusion equation, the vorticity is coupled to the streamfunction, and yet Lundgren argued in \cite{lundgren:1982} that the same shear diffusion mechanism also homogenizes vorticity on a time scale of $Re^{1/3}$. Bernoff and Lingevitch further refined this assertion in \cite{Bernoff:1994}, by showing that any non-axisymmetric, non-translational (i.e. zero and first Fourier modes) perturbation will decay on the $Re^{1/3}$ time scale. Moreover, the higher the Fourier mode of the perturbation, the more rapidly the
 the perturbation decays.


This shear diffusion mechanism is based on a study of the linearized problem and
 thus is not guaranteed to hold for larger perturbation values. Here, we study small perturbations of the Lamb-Oseen vortex with purely Fourier mode $m=2$  perturbation and for low to medium Reynolds number ($Re=500$--$4000$). Using our single-particle MMVM we are able to capture a rapid relaxation to an axisymmetric state that is consistent with the shear-diffusion time scale of $Re^{1/3}$.

To begin, our initial perturbation of the Lamb-Oseen vortex (monopole) is of the form:
\begin{equation}\label{eq:perturb1}
\omega'(\xx,0) = \frac{\delta}{4\pi}|\xx|^2\exp(-\frac{|\xx|^2}{4})\cos(2\theta).
\end{equation}
Perturbations of this exact form (but large values of $\delta$) were used in the study of the formation of tripoles by Rossi et al. in \cite{rossi:1997} and by Barba and Leonard in \cite{barba:2006}.  It is not hard to see that if we also choose the total circulation to be $M[0,0]=1$, then the initial conditions with the perturbation \eqref{eq:perturb1} can be re-written entirely in our Hermite basis as
\begin{equation}\label{eq:perturb2}
\omega(\xx,0) = \phi_{00}(x,0) + 4\delta(\phi_{20} -\phi_{02}).
\end{equation}
We point out here that we have implicitly chosen the intrinsic core size $\lambda_0=2$ for both the vortex and the expansion which we will now fix for the remainder of this section. In addition we define the Reynolds number to be $Re~=~M[0,0]/\nu~=~1/\nu$. From this we see that the initial data can be exactly represented in our expansion as:
\begin{eqnarray}\label{eq:IC_sd}
M[0,0](0)=1,~~M[2,0](0) = 4\delta,~~M[0,2](0) = -4\delta.
\end{eqnarray}

\subsubsection{Results}

We now use an order $24$ expansion to evolve initial conditions \eqref{eq:IC_sd} under our single-particle MMVM
to simulate the shear diffusion mechanism and axisymmetrization of the vorticity. Small perturbations of the quadrupole moment of the form \eqref{eq:IC_sd} create elliptical deformations of the Lamb-Oseen vortex as seen in Figure \ref{fig:IC} below:
\begin{figure}[!hbp]
\begin{center}
\includegraphics[width=.6\textwidth]{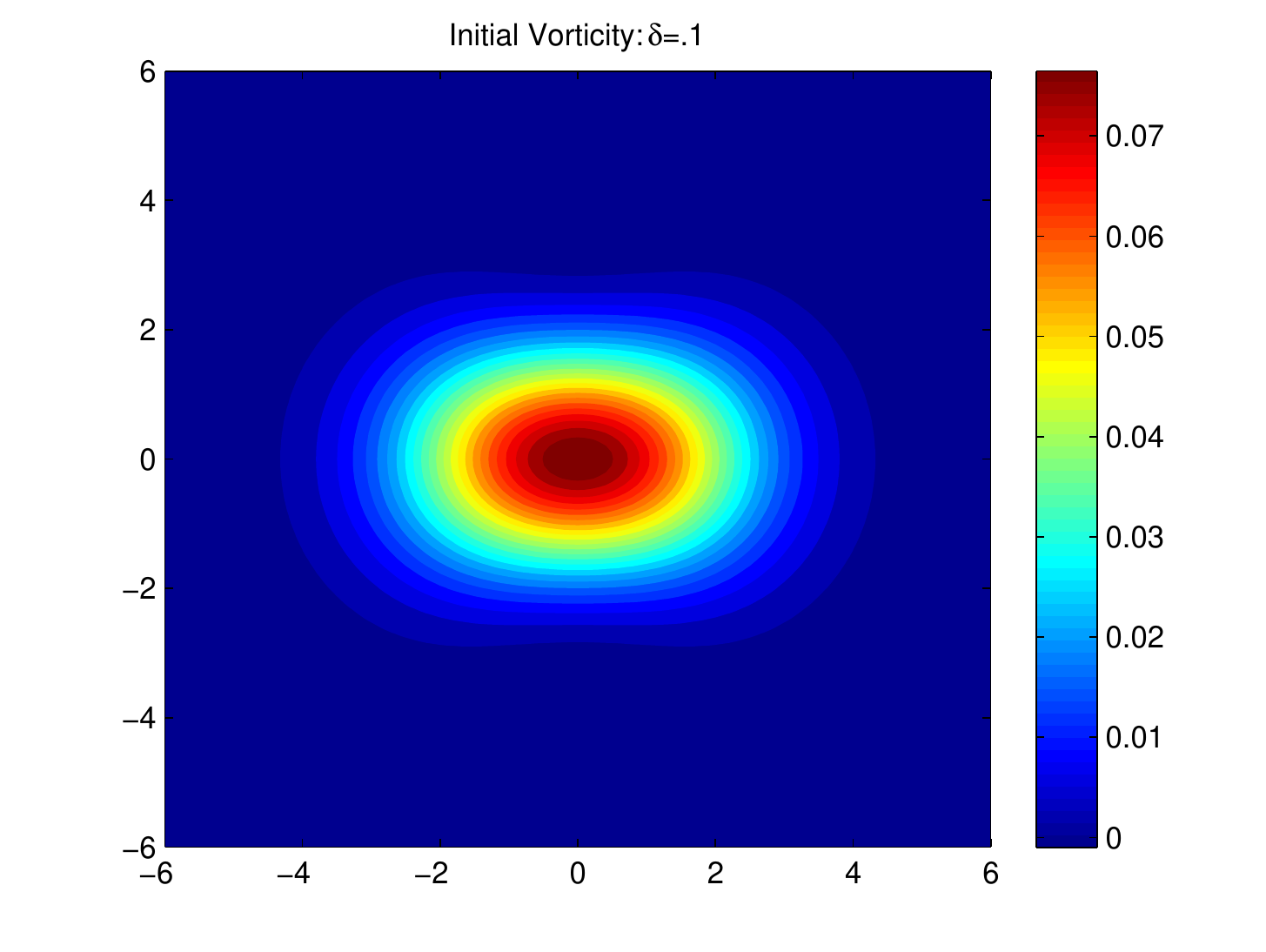}\\
\caption[fig:IC]{Initial elliptical perturbation of the Lamb-Oseen vortex of the form $\omega(\xx,0) =\phi_{00}(\xx,0)+ 4\delta(\phi_{20}(\xx,0)-\phi_{02}(\xx,0))$, for $\delta=.1$.}\label{fig:IC}
\end{center}
\end{figure}

\begin{figure}[!hbp]
\begin{center}
\includegraphics[width=.75\textwidth]{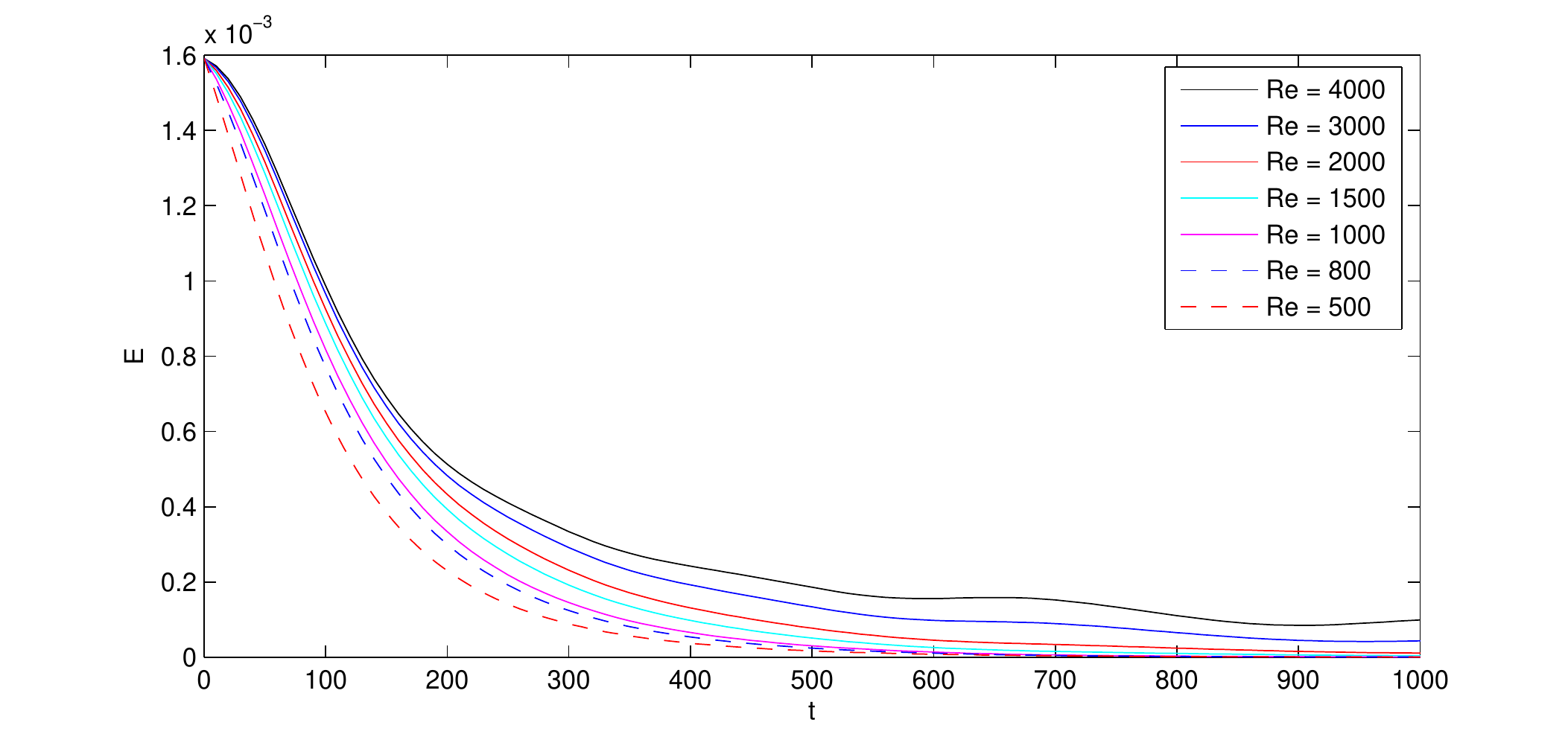}\\
\caption[fig:L2unscaled]{Plot of nonaxisymetric enstrophy (E) vs unscaled time $t$, showing relaxation of purely azimuthal perturbation (second mode)using $m=24$ moments. }\label{fig:L2unscaled}
\end{center}
\end{figure}

To capture and quantify axisymmetrization, we adopt the conventions of \cite{rossi:1997} and define the nonaxisymmetric enstrophy of the vortex to be
\begin{equation} \label{eq:nonaxi_enstrophy}
E  =  \int (\omega(\xx)-<\omega(|\xx|)>)^2 {\rm d}\xx
\end{equation}
where
\begin{equation*}
<\omega(|\xx|)>  =  \frac{1}{2\pi}\int_0^{2\pi} \omega(\xx){\rm d}\theta.
\end{equation*}
This $E$ represent the $L^2$ norm of the nonaxisymmetric portion of the solution and it is this quantity which we demonstrate decays on the $Re^{1/3}$ time scale. We solve the single-particle MMVM equations \eqref{eq:FinalSingleI}-\eqref{eq:moment_evolutionFinish} found in \ref{sec:Combinatorial} using the initial conditions \eqref{eq:IC_sd} with $\delta=.1$ for $Re = 500$, $800$, $1000$, $1500$, $2000$, $3000,$ and $4000$. In Figure \ref{fig:L2unscaled}, we plot $E$ against unscaled time and observe that each solution indeed goes through a rapid axisymmetrization.

\begin{figure}[!hbp]
\begin{center}
\includegraphics[width=.75\textwidth]{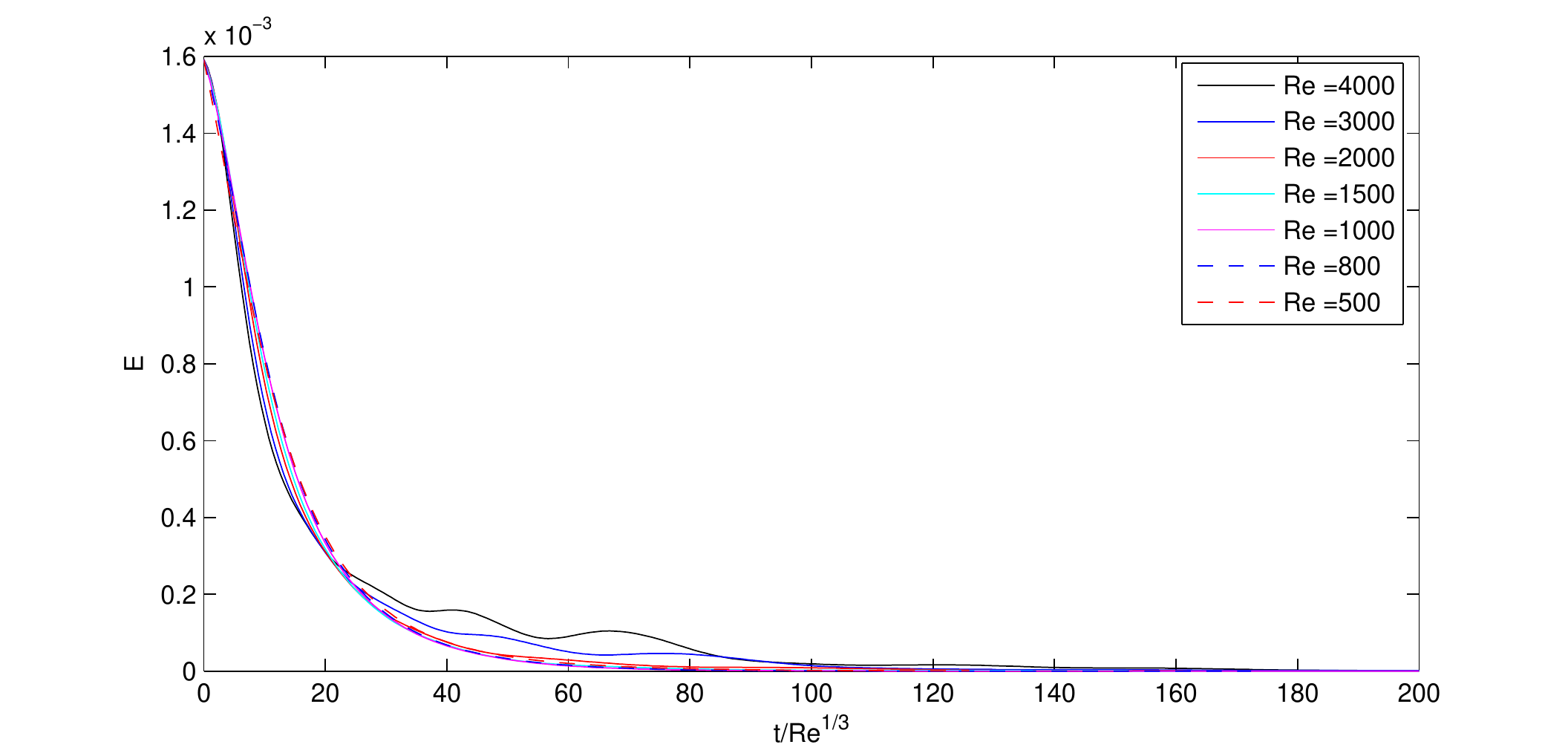}\\
\caption[fig:L2scaled]{Plot of nonaxisymetric enstrophy (E) vs $t/Re^{1/3}$, showing relaxation of purely azimuthal perturbation (second mode)using $m=24$ moments. }\label{fig:L2scaled}
\end{center}
\end{figure}

\begin{figure}[!hbp]
\begin{center}
\includegraphics[width=.75\textwidth]{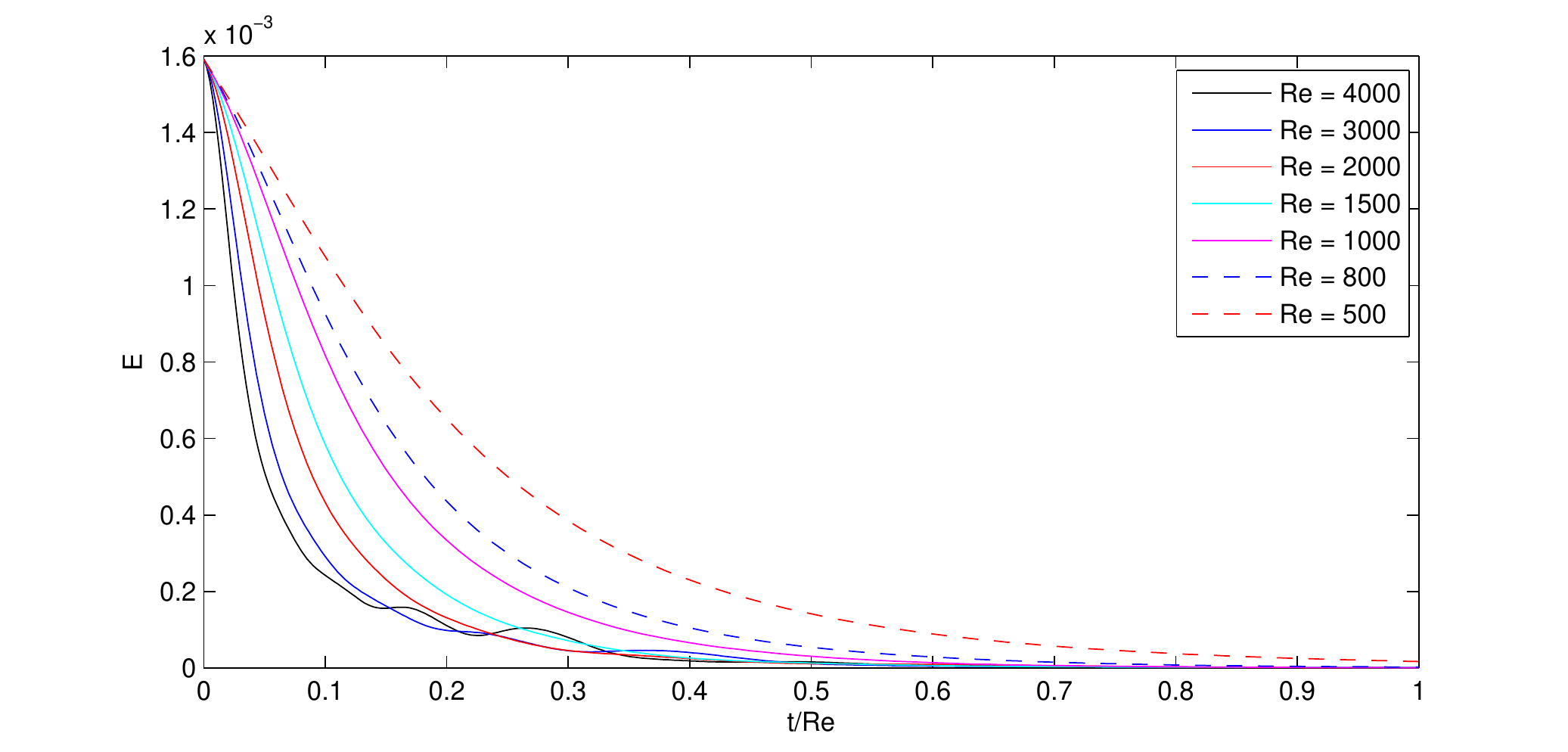}\\
\caption[fig:L2badscale]{Plot of nonaxisymetric enstrophy (E) vs $t/Re$, showing relaxation of purely azimuthal perturbation (second mode) using $m=24$ moments. }\label{fig:L2badscale}
\end{center}
\end{figure}

Notice in addition that in Figure \ref{fig:L2unscaled}, there are small oscillations after the initial relaxation in the cases of $Re = 3000$ and $4000$.  These oscillations capture the exchange of energy between the zeroth mode and the higher order modes that occurs before the diffusive time scale as discussed in \cite{barba:2006}. To confirm that the shear diffusion acts on the $Re^{1/3}$ time scale, we rescale time. Figure \ref{fig:L2scaled} provides clear evidence that the early rapid decay collapses on the $Re^{1/3}$ time scale, confirming that for early times, this is indeed the appropriate time scale for shear diffusion. Further note that in Figure \ref{fig:L2badscale}, when one tries to capture shear diffusion on the diffusive time scale $Re$, no such collapse occurs. In fact, the decay fans, out signaling a poor choice of time scale for the early decay.

\begin{figure}[!hbp]
\begin{center}
\includegraphics[width=.3\textwidth]{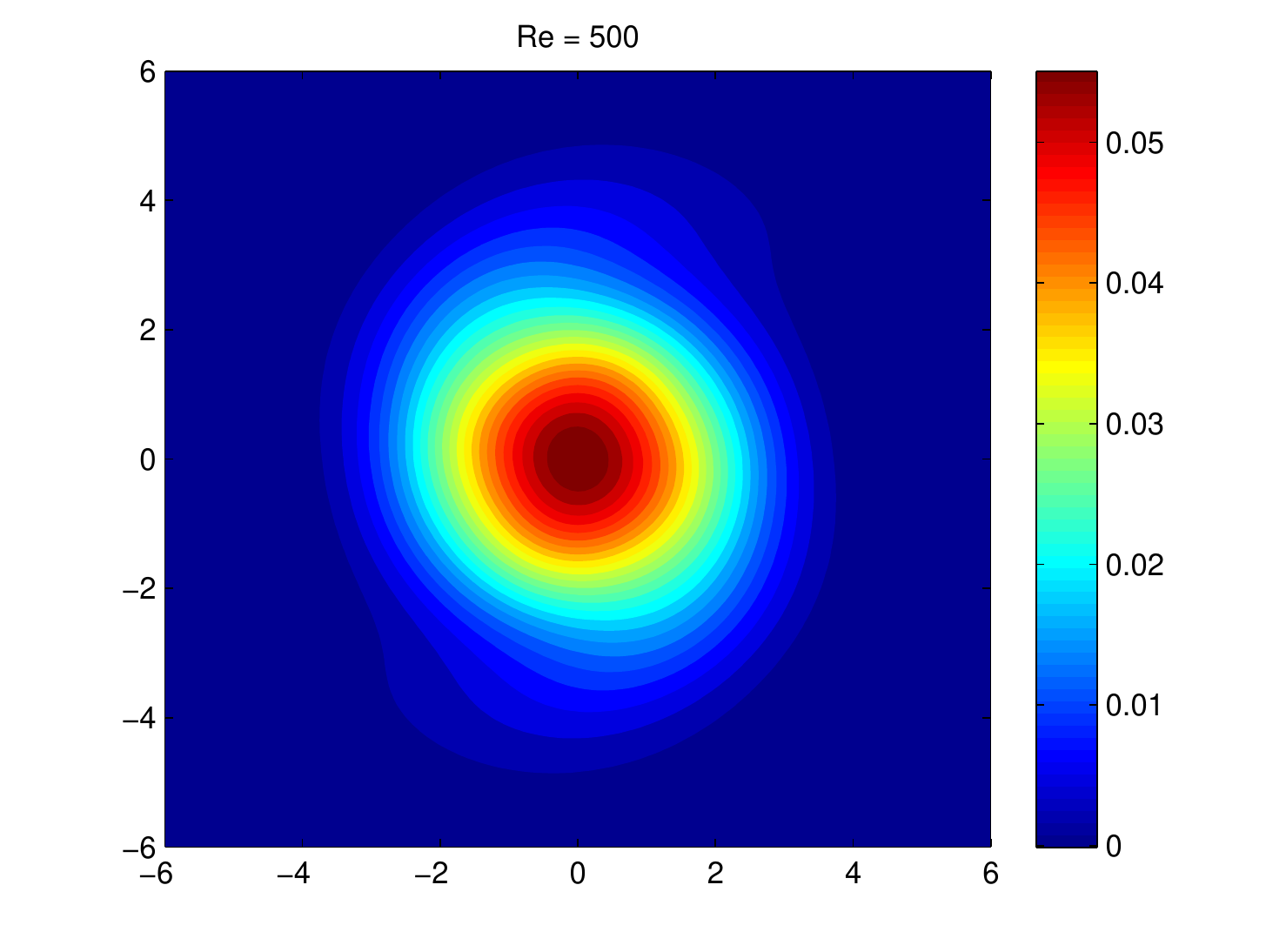}
\includegraphics[width=.3\textwidth]{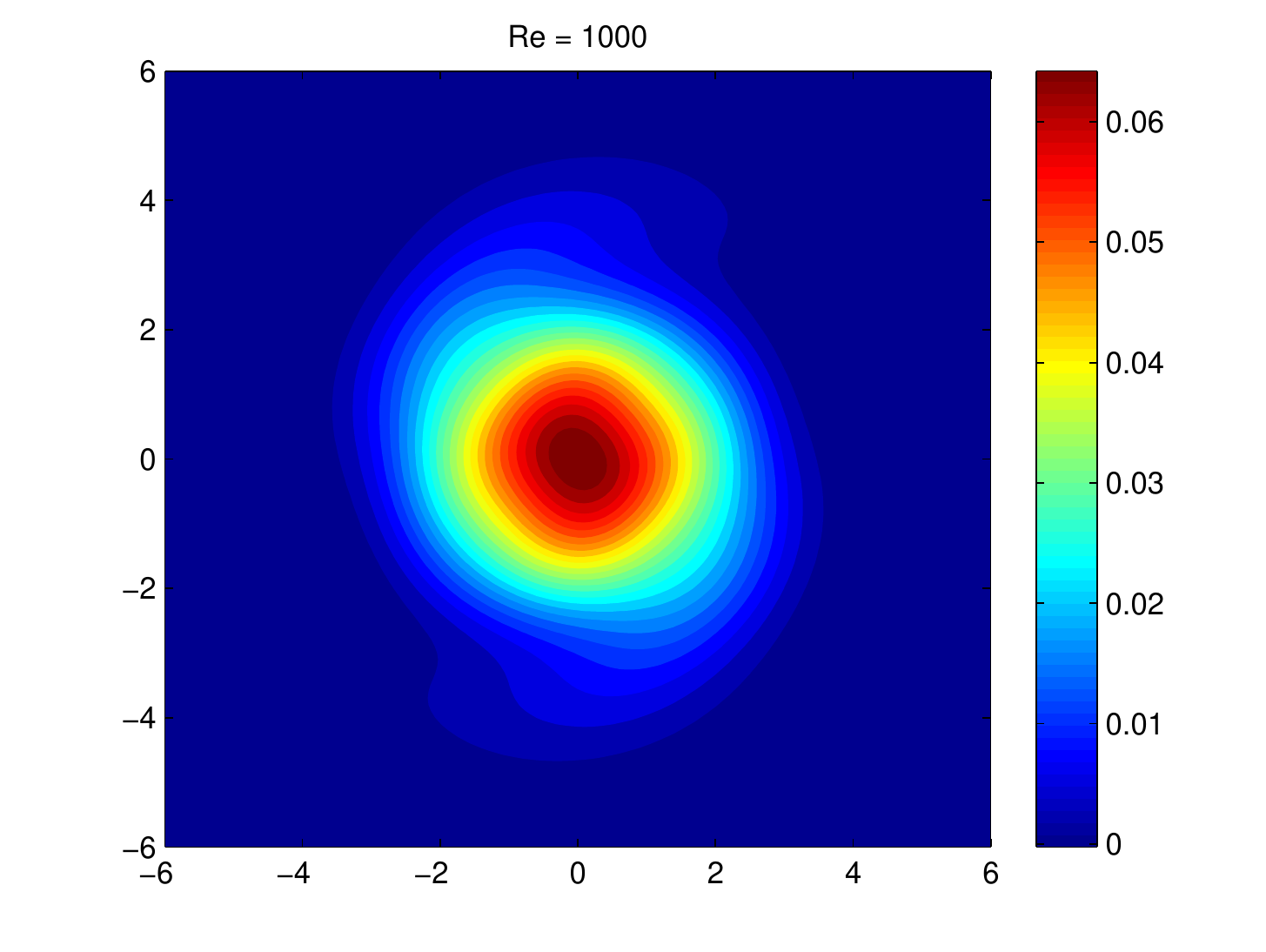}\\
\includegraphics[width=.3\textwidth]{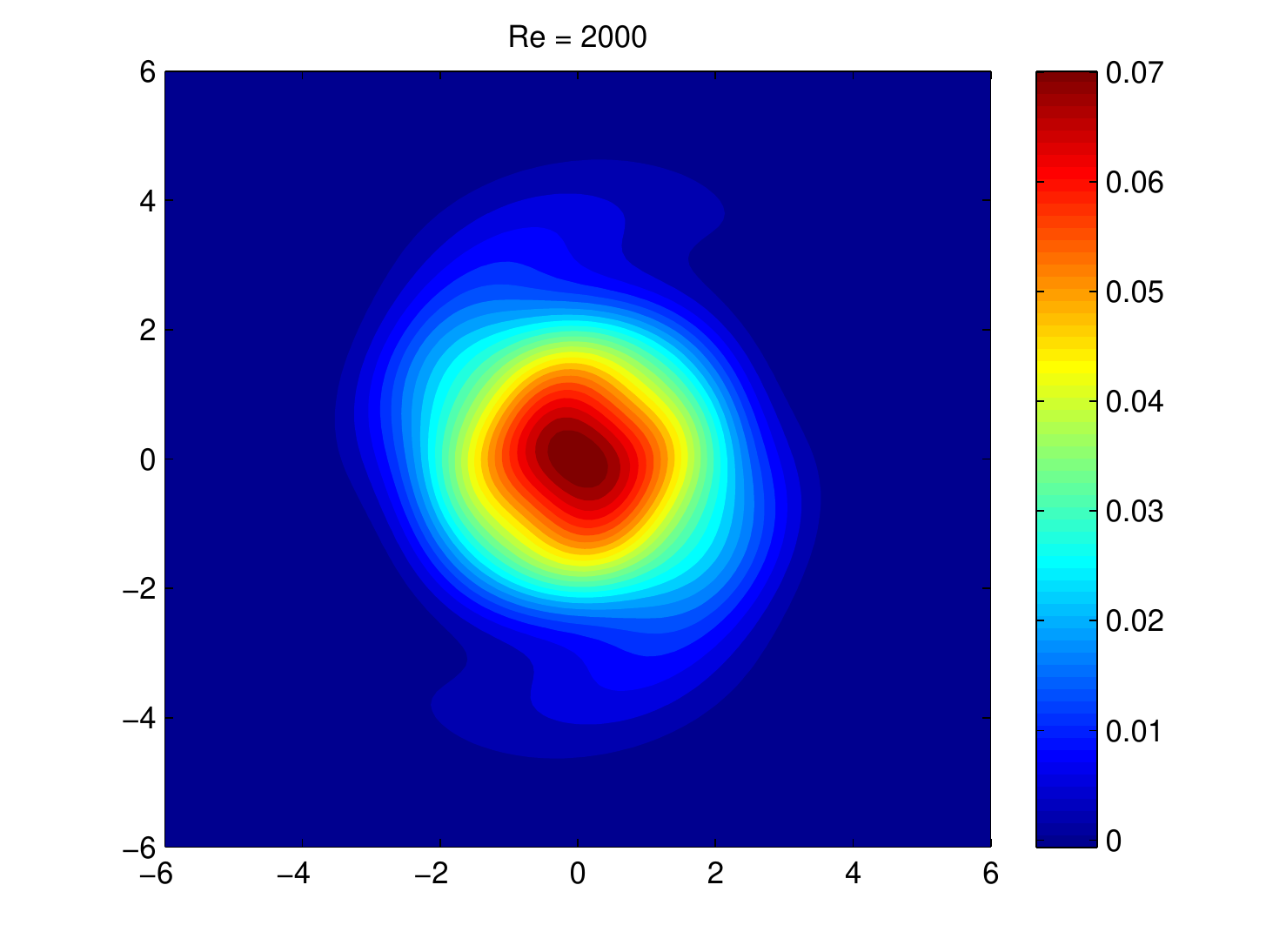}
\includegraphics[width=.3\textwidth]{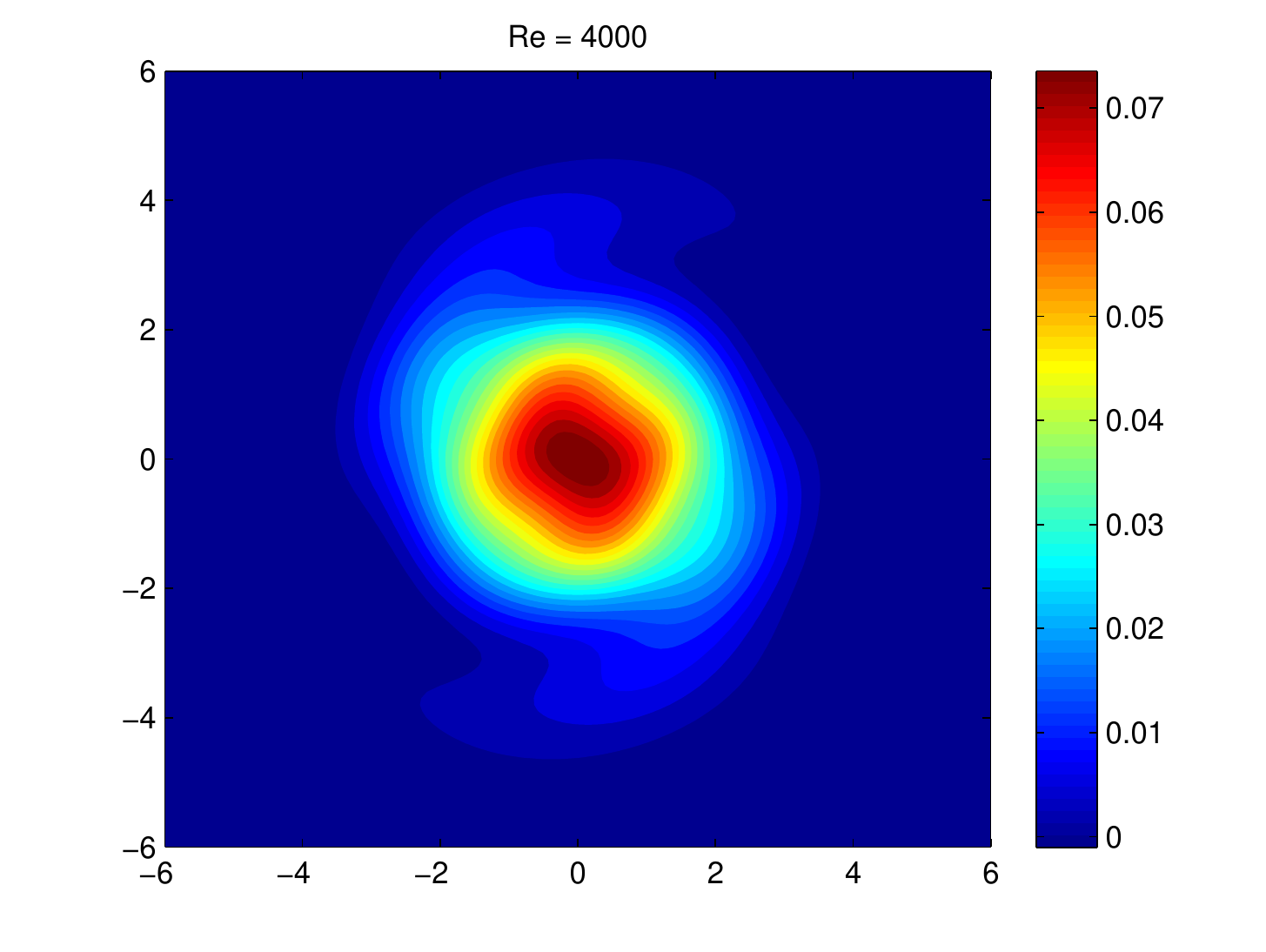}
\caption[fig:t200]{A plot of vorticity distribution at time $t=200$. From top to bottom and left to right the cases are $Re = 500,~1000,~2000,~4000$.}\label{fig:t200}

\end{center}
\end{figure}

To observe the actual vorticity distribution,  in Figure \ref{fig:t200} we reconstruct the 2D vorticity at $t=200$ for $Re = 500,$ $~1000,$ $~2000,$ and $4000$. Notice that diffusion dominates and smooths the vorticity for the cases of $Re = 500$ and $1000$ as expected. Notice that we also capture short spiral arms, characteristic of rapid axisymmetrization, in the higher Reynolds number cases of $Re = 2000$ and $4000$.

\subsubsection{A note on larger perturbations}\label{sec:SD_LargePerturb}

In the case where $\delta$ is large, relaxation back to a monopole is unlikely to occur.  Instead, it has been shown in \cite{Barba:2004,barba:2006,rossi:1997} that the vorticity distribution relaxes back to a rotating tripole (at least on an intermediate time scale). Although larger $\delta$ perturbations create four distinct regions of vorticity and thus may not be well suited to a single-particle MMVM, we nonetheless obtain excellent agreement to the work in \cite{Barba:2004} in both frequency of rotation and magnitude for the low Reynolds number calculation $Re=500$ and $\delta=.25$.

\begin{figure}[!hbp]
\begin{center}
\includegraphics[width=.3\textwidth]{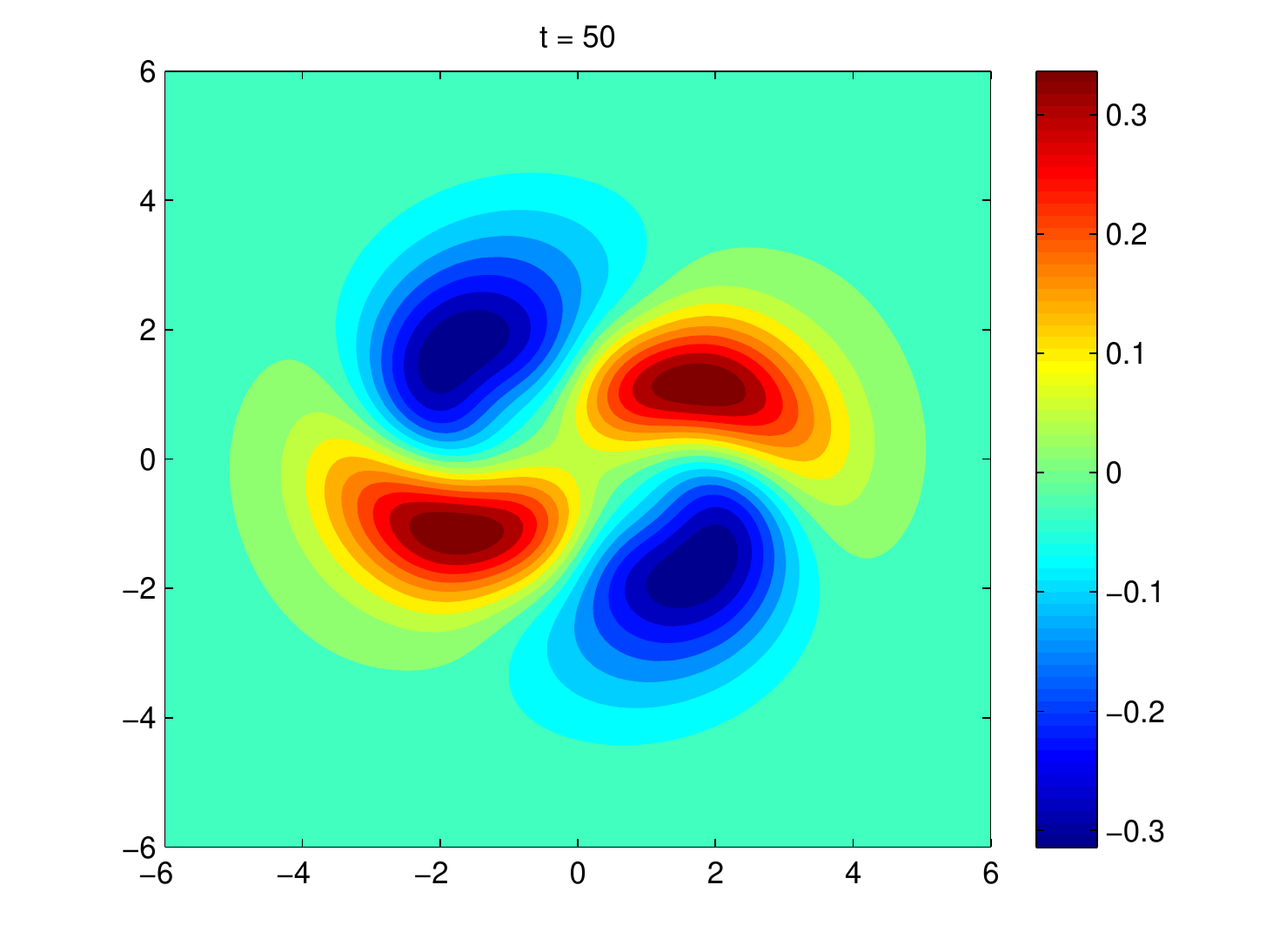}
\includegraphics[width=.3\textwidth]{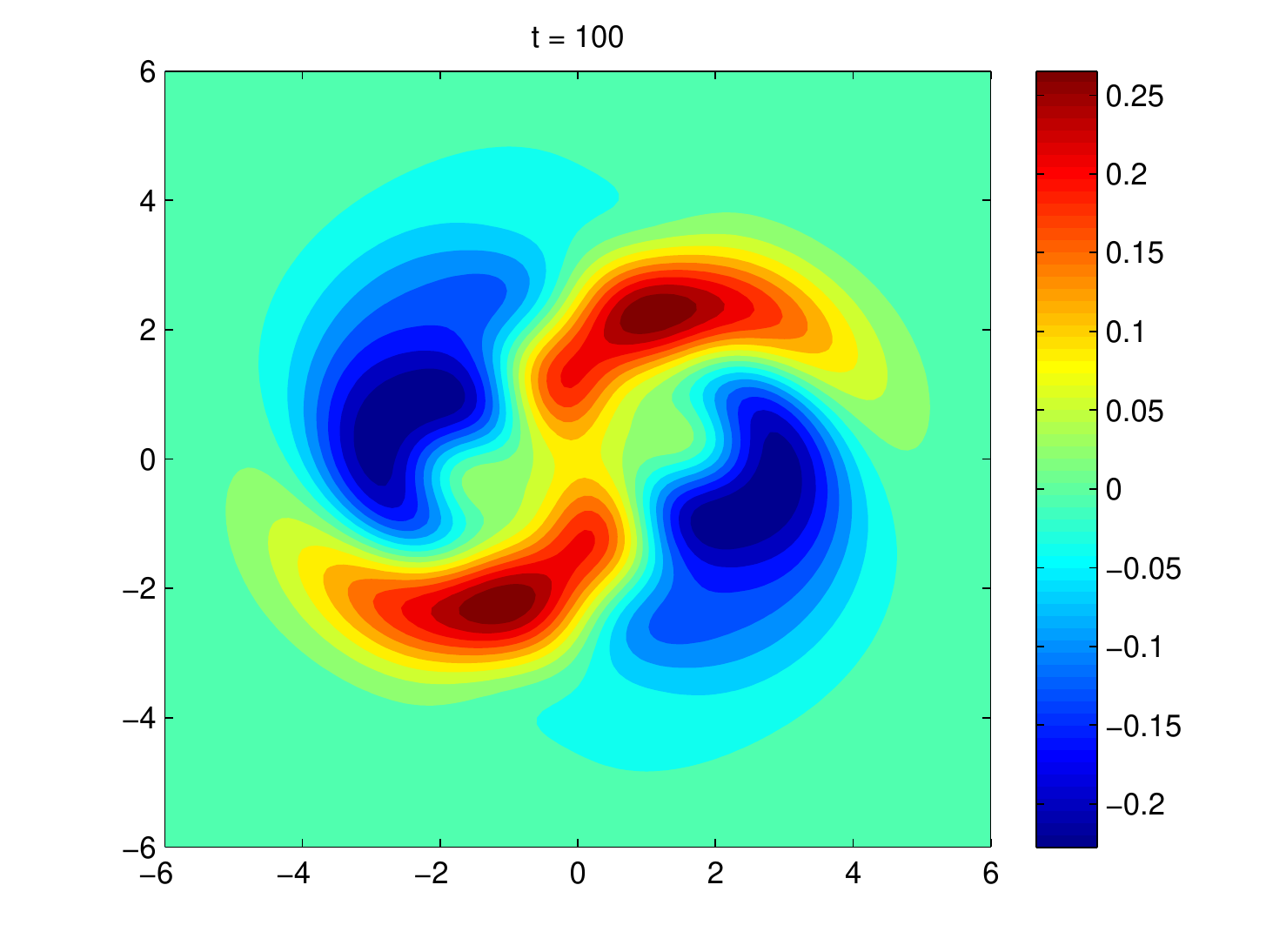}
\includegraphics[width=.3\textwidth]{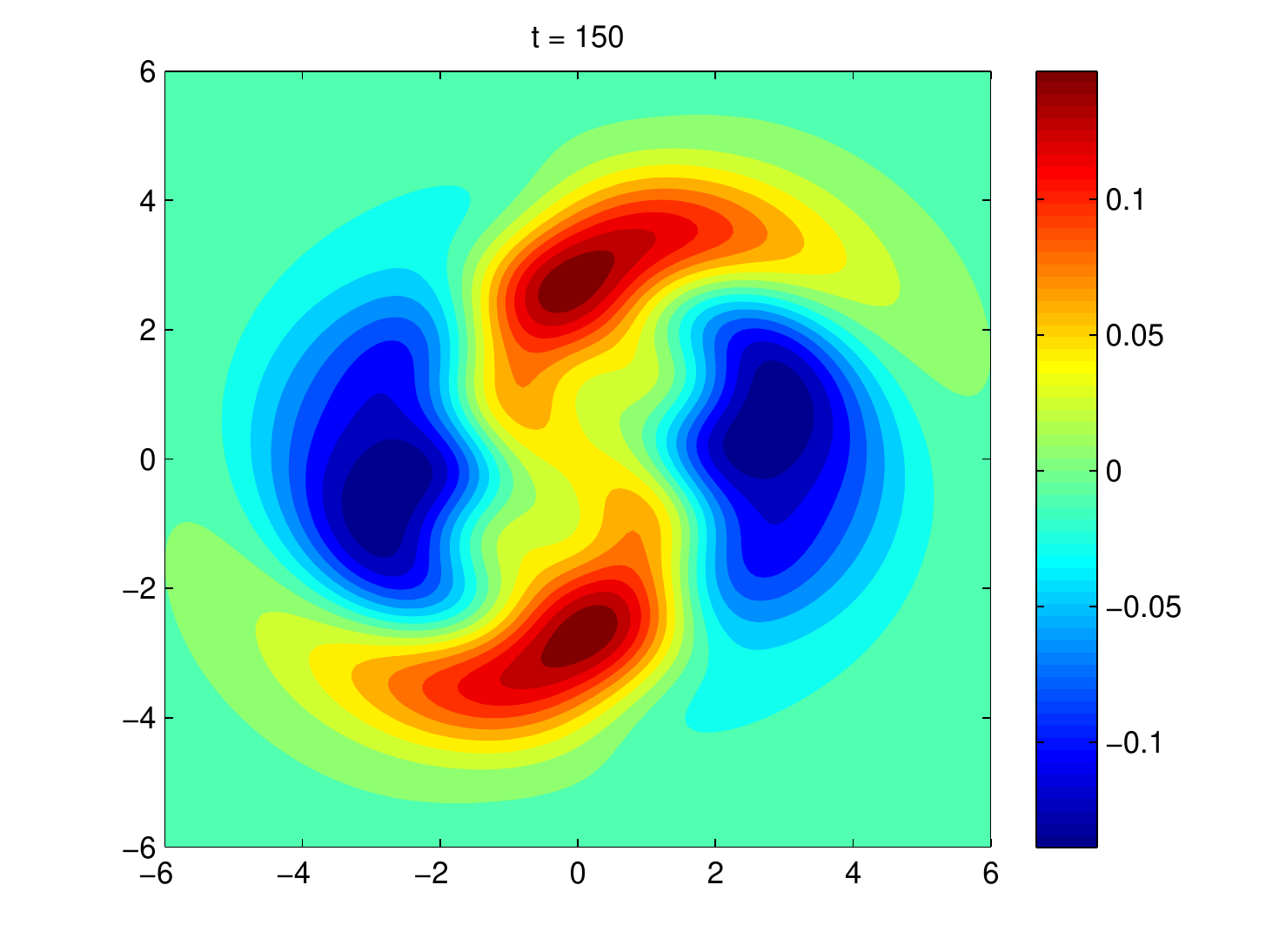}\\
\includegraphics[width=.3\textwidth]{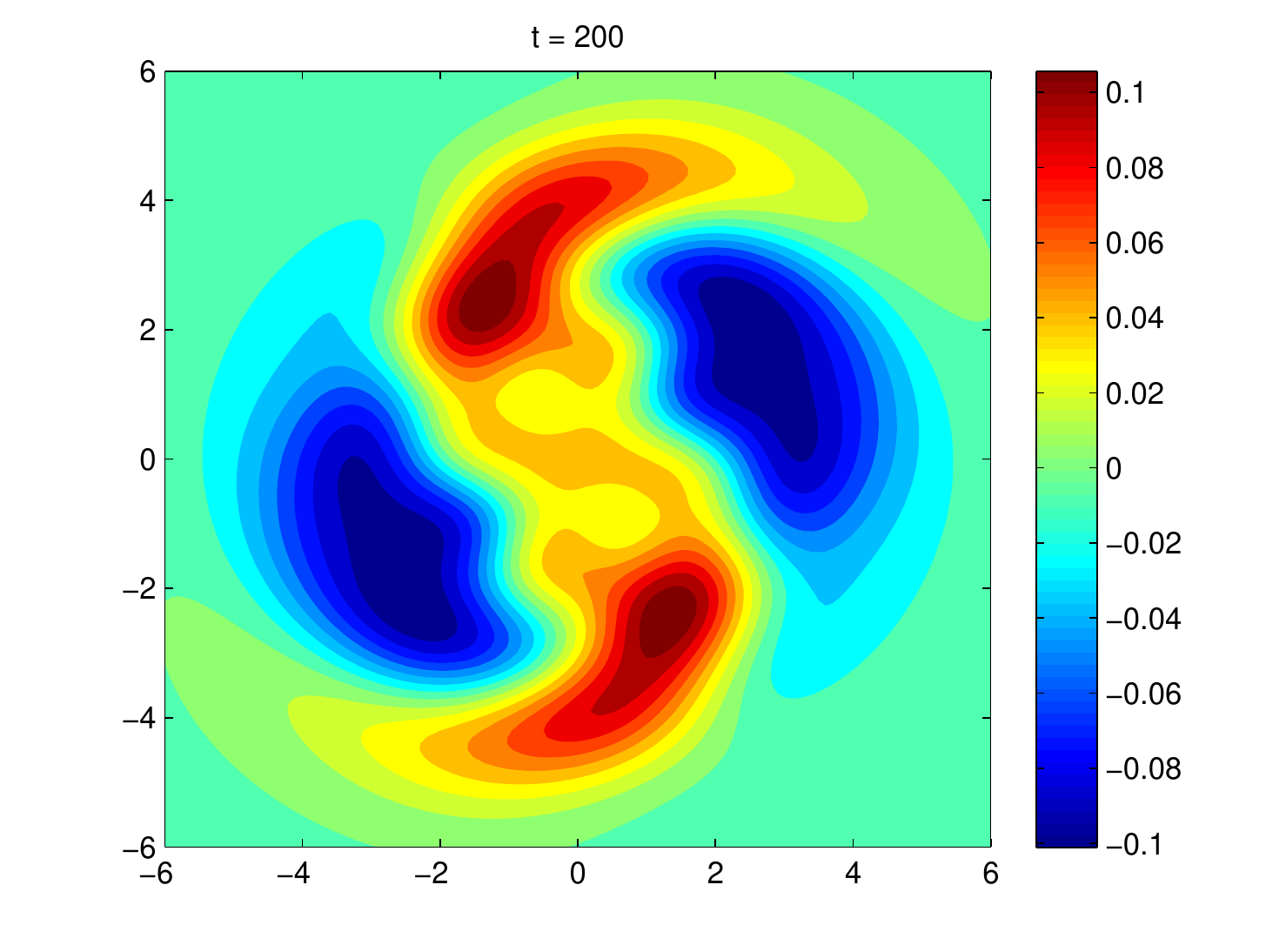}
\includegraphics[width=.3\textwidth]{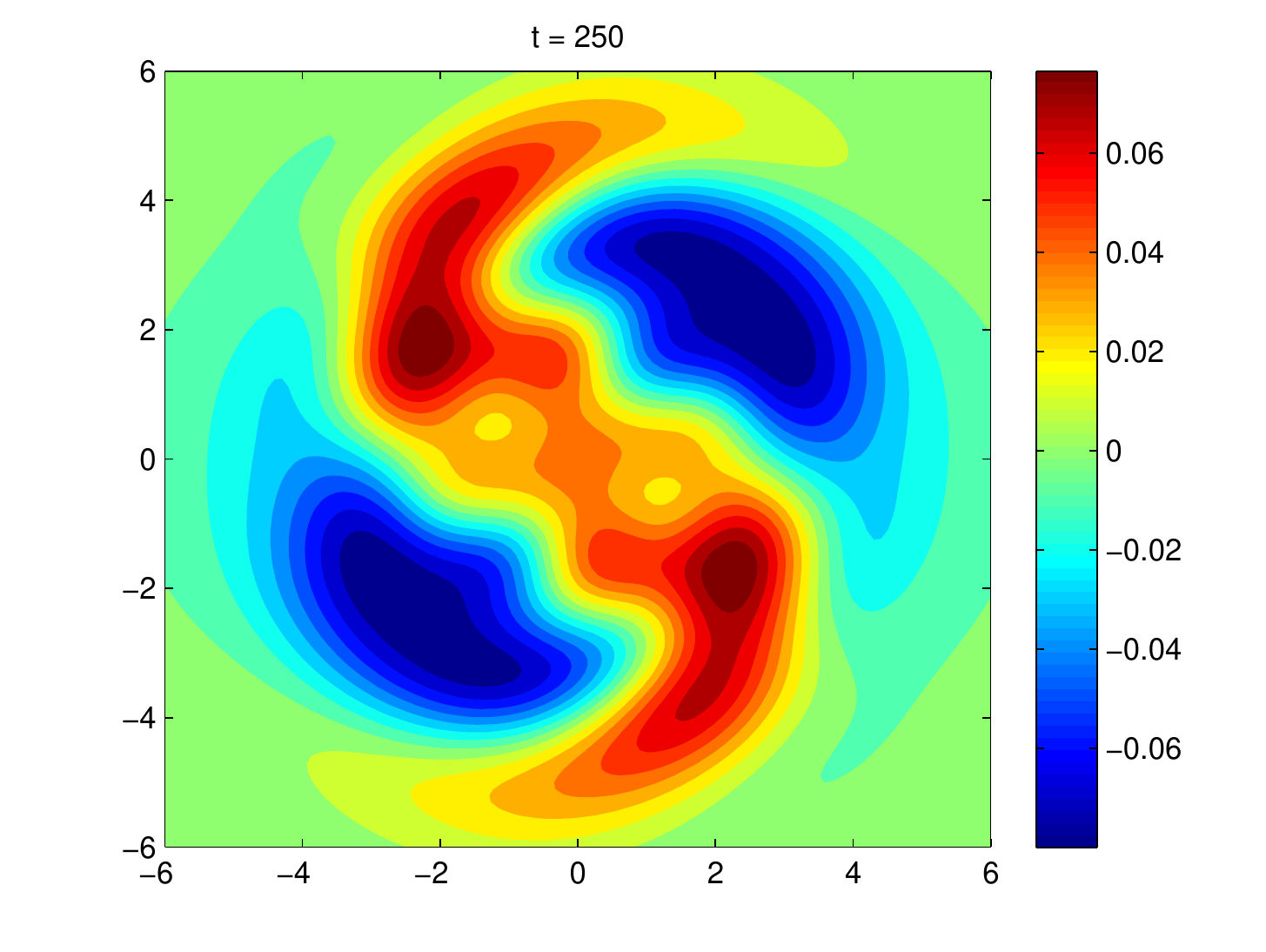}
\includegraphics[width=.3\textwidth]{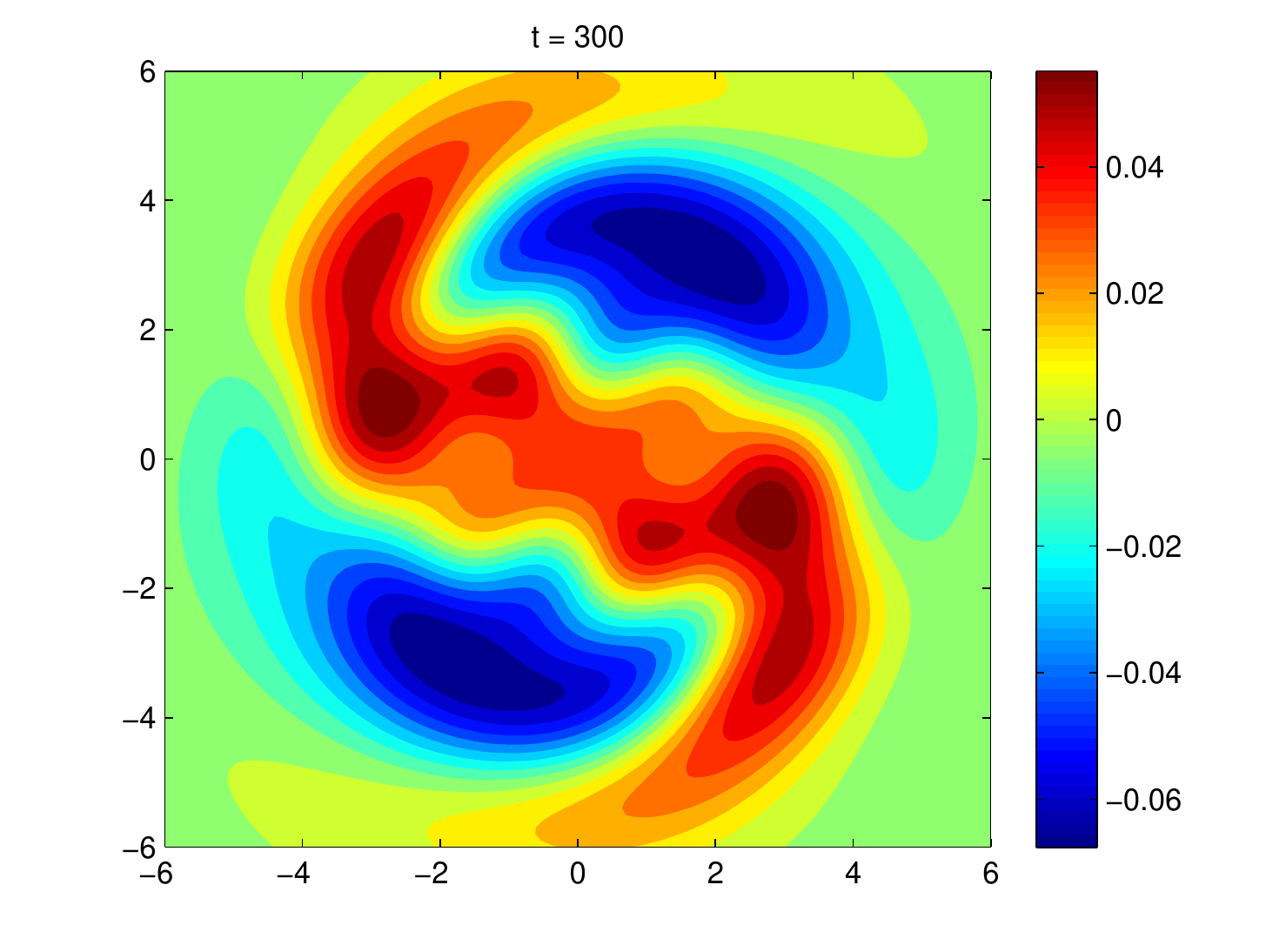}\\
\includegraphics[width=.3\textwidth]{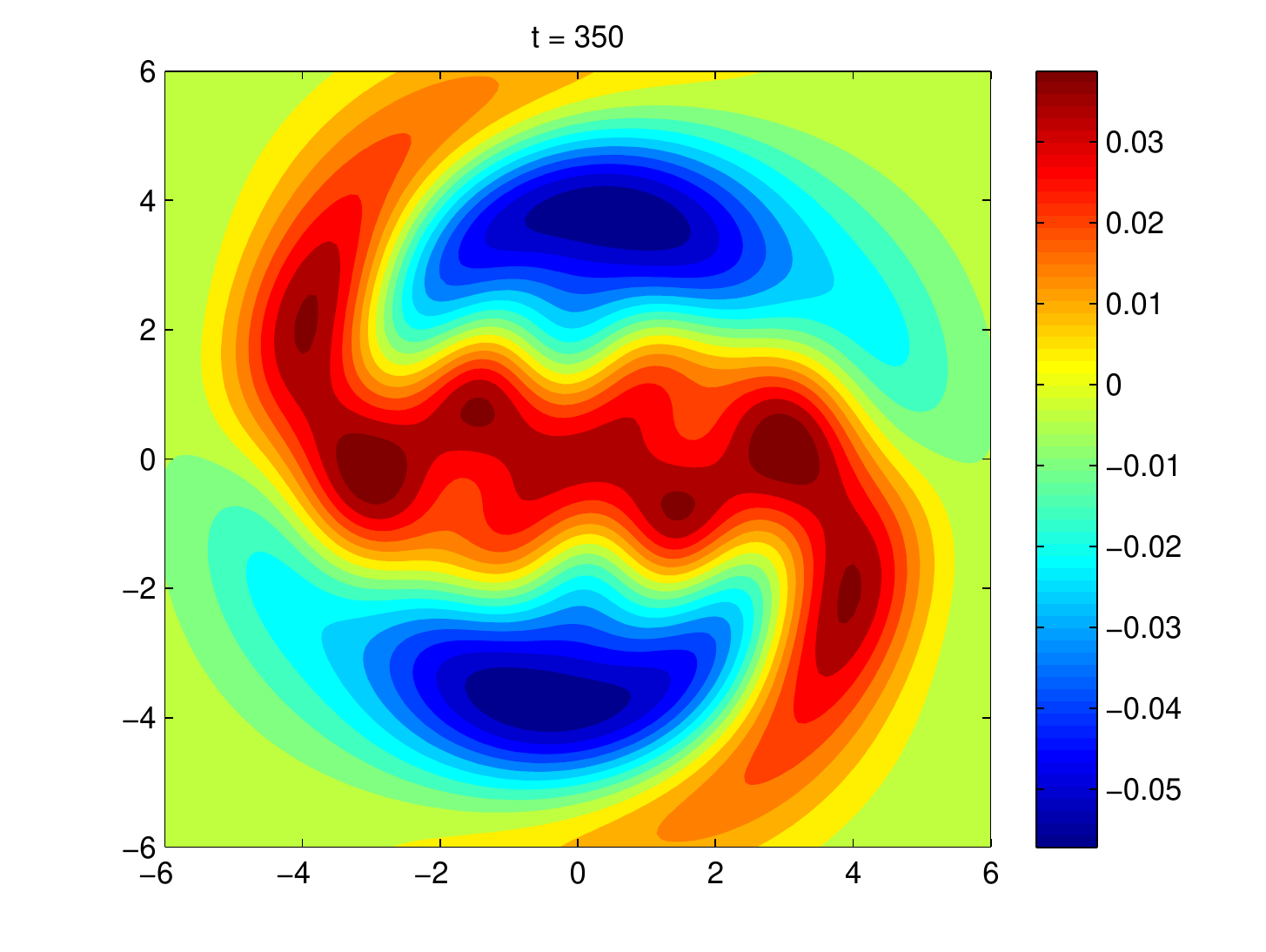}
\includegraphics[width=.3\textwidth]{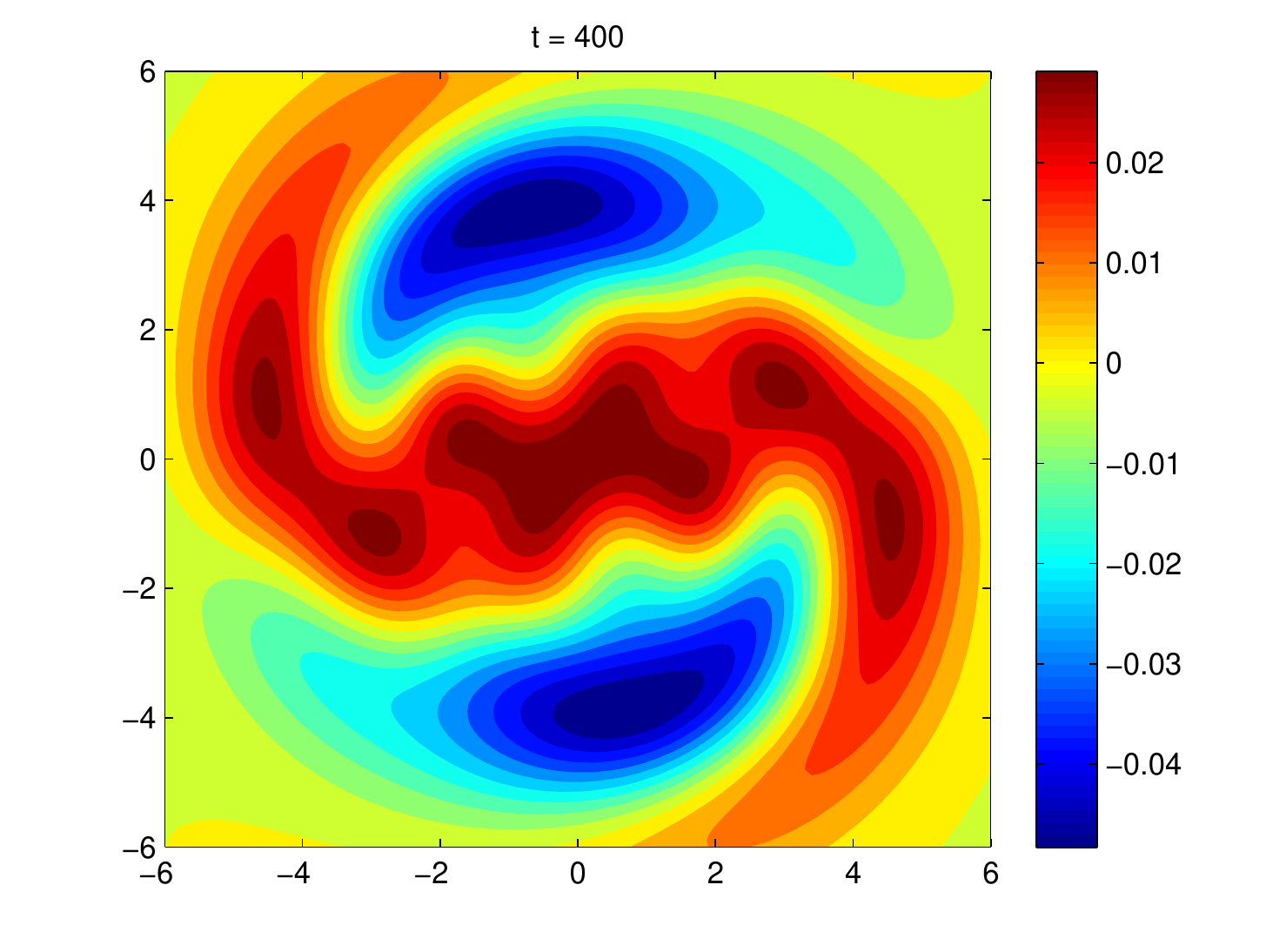}
\includegraphics[width=.3\textwidth]{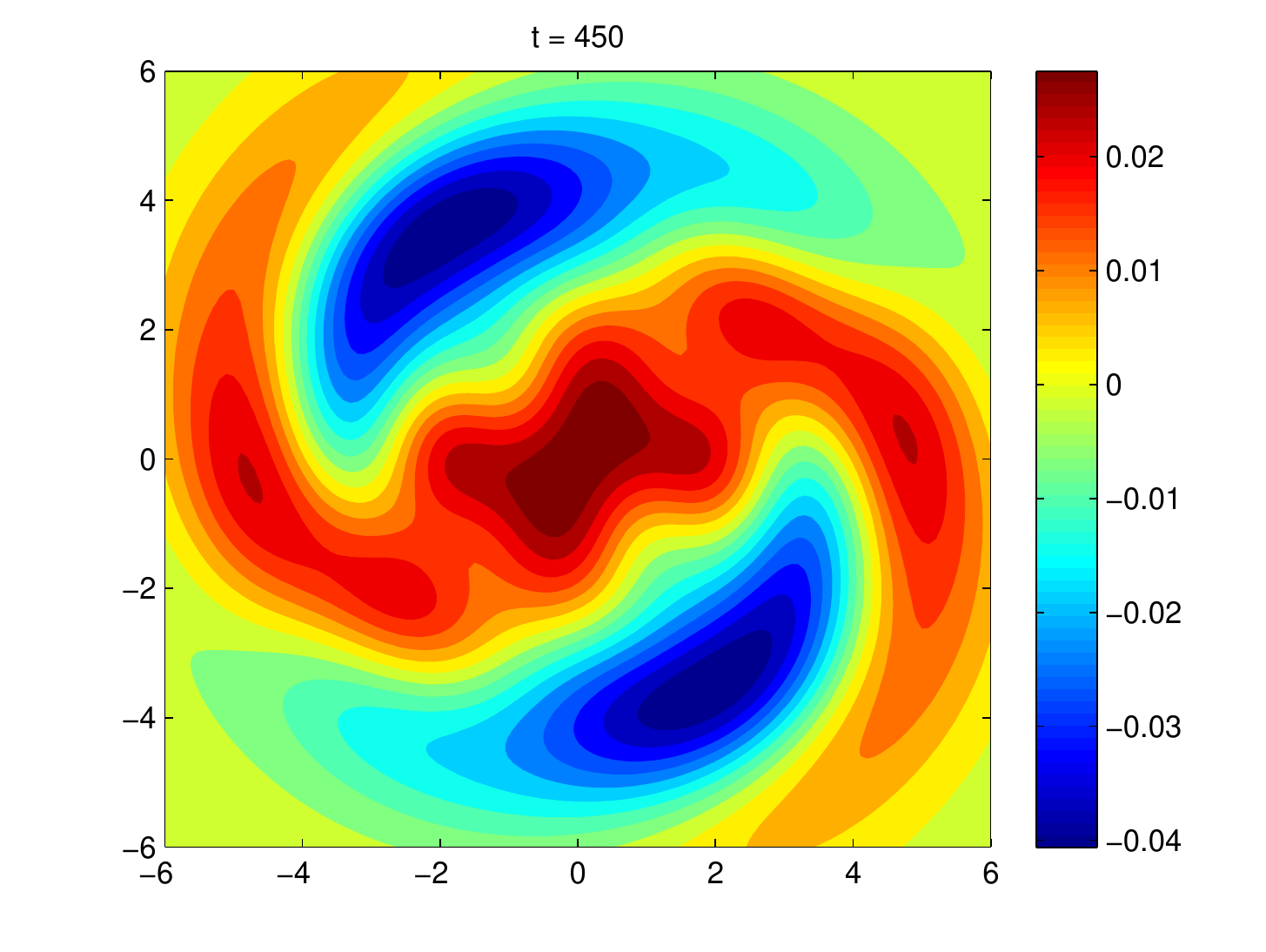}\\
\includegraphics[width=.3\textwidth]{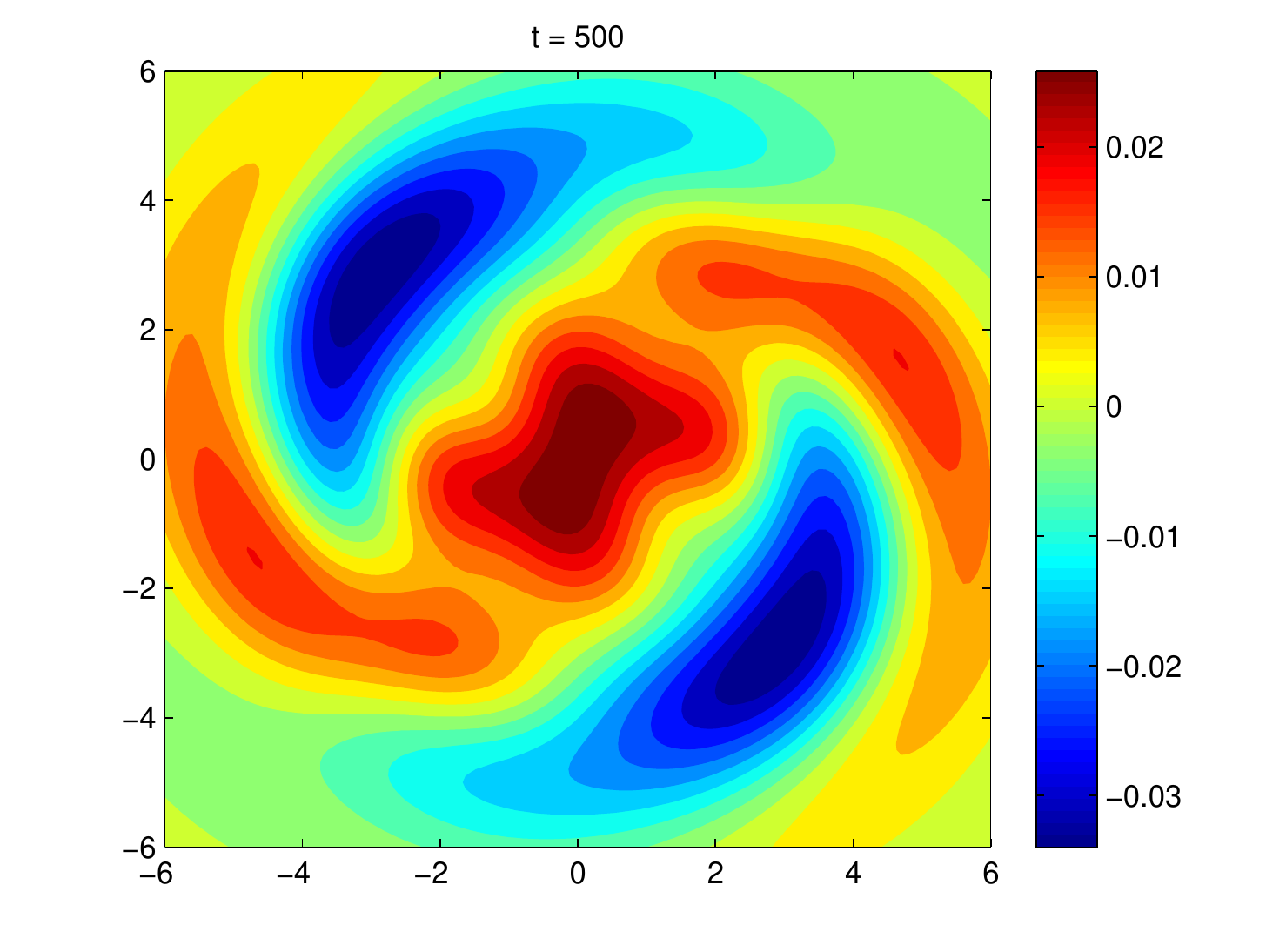}
\includegraphics[width=.3\textwidth]{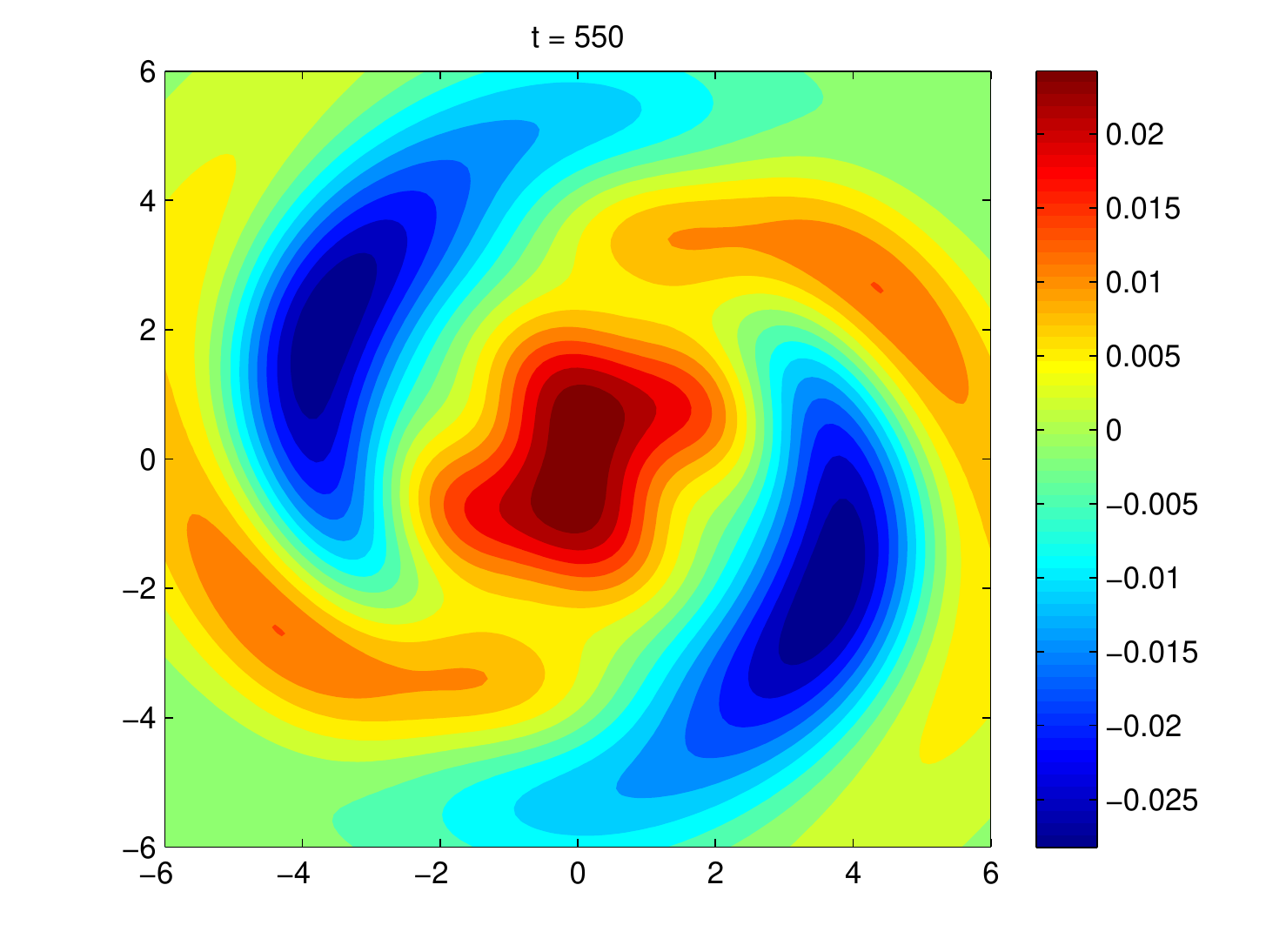}
\includegraphics[width=.3\textwidth]{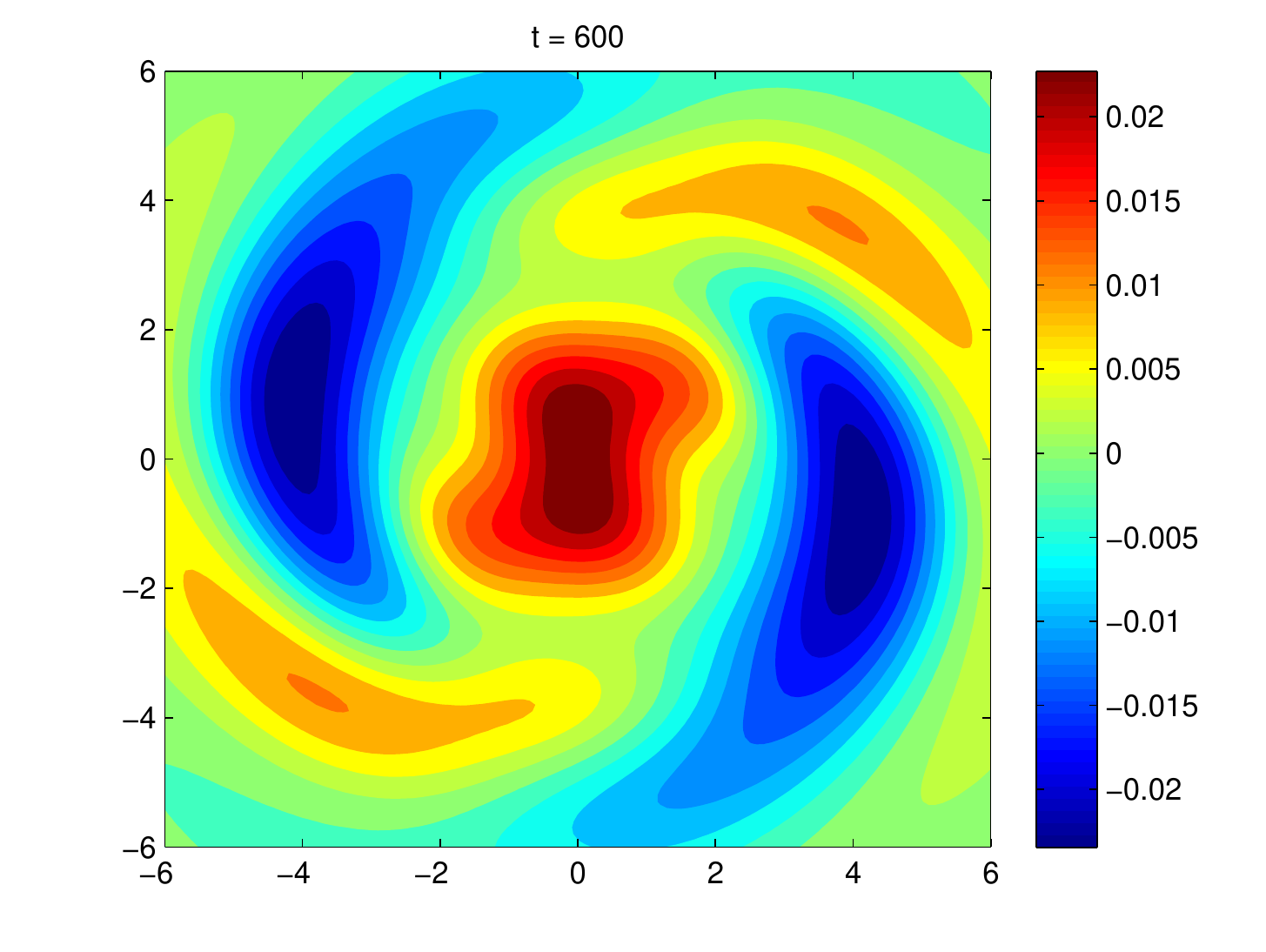}\\
\caption[Triple]{Perturbation vorticity normalized by $\omega_{\max}(0)$ with $Re=500$ and $\delta=.25$. The frequency of roation and distribution of perturbation vorticity has very good agreement with Figure 6.8 in \cite{Barba:2004}. The most significant difference is the more {``blobby''} and less elliptical center for later times.}\label{fig:Re500Perturbation}
\end{center}
\end{figure}

Figure \ref{fig:Re500Perturbation} can be directly compared to Figure 6.8 in \cite{Barba:2004}. It is important to note that Barba's calculations were done using a very high resolution an adaptive Lagrangian scheme with thousands of vortex elements.  The main differences in the results occur at the center of the tripole. We do not find a smooth elliptical center vortex and instead have a slightly ``blobby" approximation.  For higher Reynolds numbers, we do not find good agreement with the work of \cite{Barba:2004}.  This is likely due to our inability to resolve very small scale behavior without retaining many more moments, see Section \ref{sec:Numerical} for further discussion.

\subsection{Full MMVM Examples}
As we saw in Section \ref{sec:Sheardiffusion}, a large single vortex element with many Hermite moments can closely capture the shear diffusion mechanism. One difficulty the method encounters is in resolving the finest scales, which includes filamentation. By dividing the initial vorticity distribution into many smaller vortex ``particles" and keeping perhaps only a few moments for each particle, one could reasonably hope to better resolve the vortex dynamics similarly to the approach of classical vortex methods.  In this section we present two examples of the full MMVM,  the first being a simple $n=2$ model for vortex merger which we now discuss.

\subsubsection{Modeling vortex merger}
In this calculation, we use the MMVM to model a classic problem of vortex merger which has been well studied \cite{DizeVerga:2002,AgulloVerga:2001,meunier:2002,Meunier:2005,Josserand:2007}. In this example we use $n=2$ vortex elements, initialized with non-overlapping Gaussian profiles centered at $(-1,0)$ and $(1,0)$.  Our goal of this example is to observe the effect on the dynamics of merger produced by  including the Hermite moments of each vortex. We specifically focus our attention on the effect that the number of Hermite moments has on the motion of the centroid of each vortex.

Whether two axisymmetric, equal, and like-signed vortices will merge is determined largely by the ratio between the core size of the vortices, $a$, and the distance between them, $b$. For vortices with Lamb-Oseen or Gaussian profiles, the critical ratio which determines whether the co-rotating vortices destabilize and begin to merge is in the range of $a/b =0.22$-$0.24$, see \cite{meunier:2002,DizeVerga:2002,CerretelliWilliamson:2003}. If the ratio of the vortices are below this range, then they are in the well separated regime and the vortices will co-rotate, diffuse, and their Gaussian profiles may begin to elliptically deform as they co-rotate. On the other hand, if Gaussian vortices are initialized above this critical ratio, the vortices rapidly distort,  begin to merge, and one might observe ejection of spiral arms of vorticity.

\begin{figure}[!hbp]
\begin{center}
\includegraphics[width=.5\textwidth]{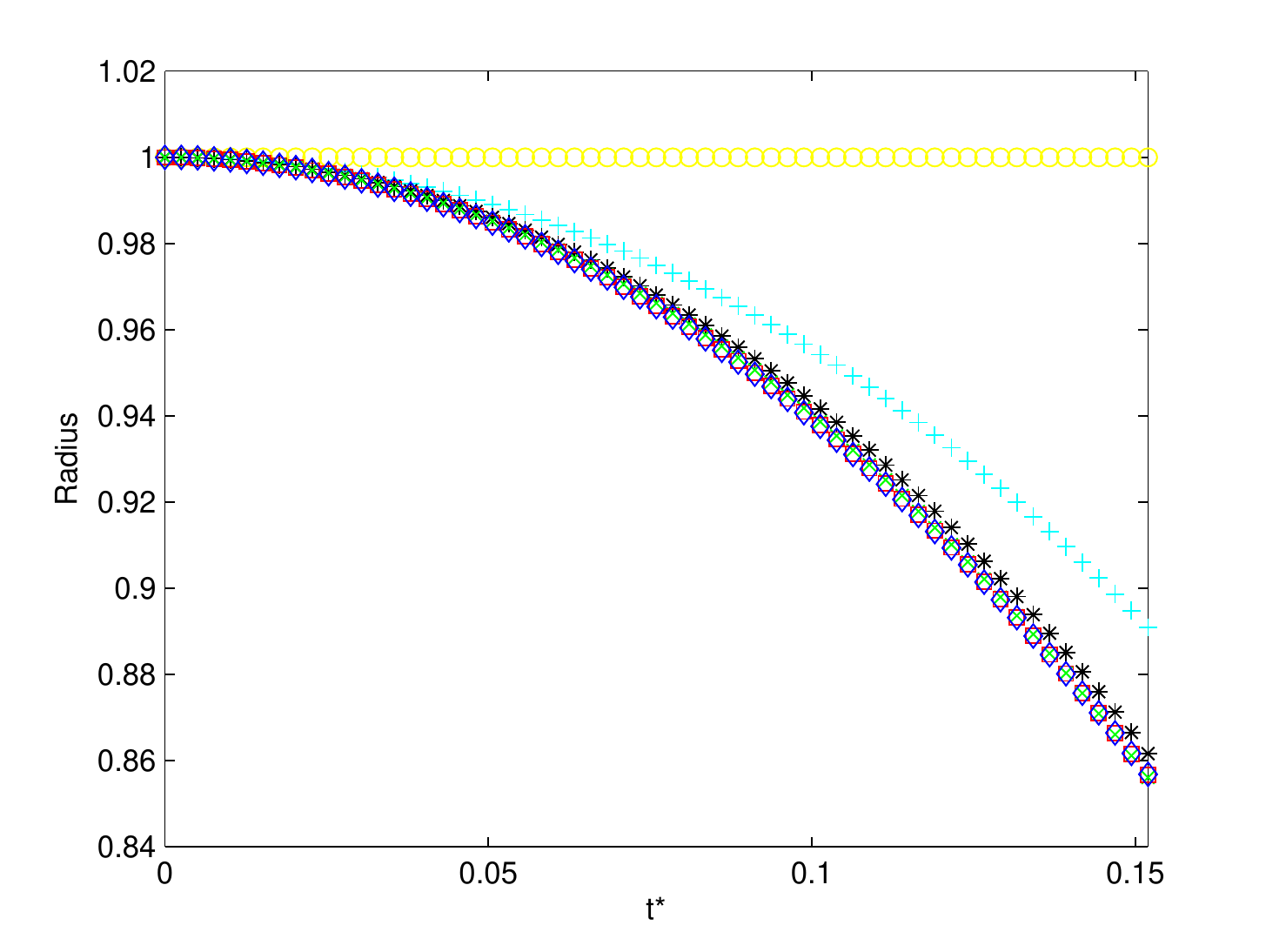}\includegraphics[width=.5\textwidth]{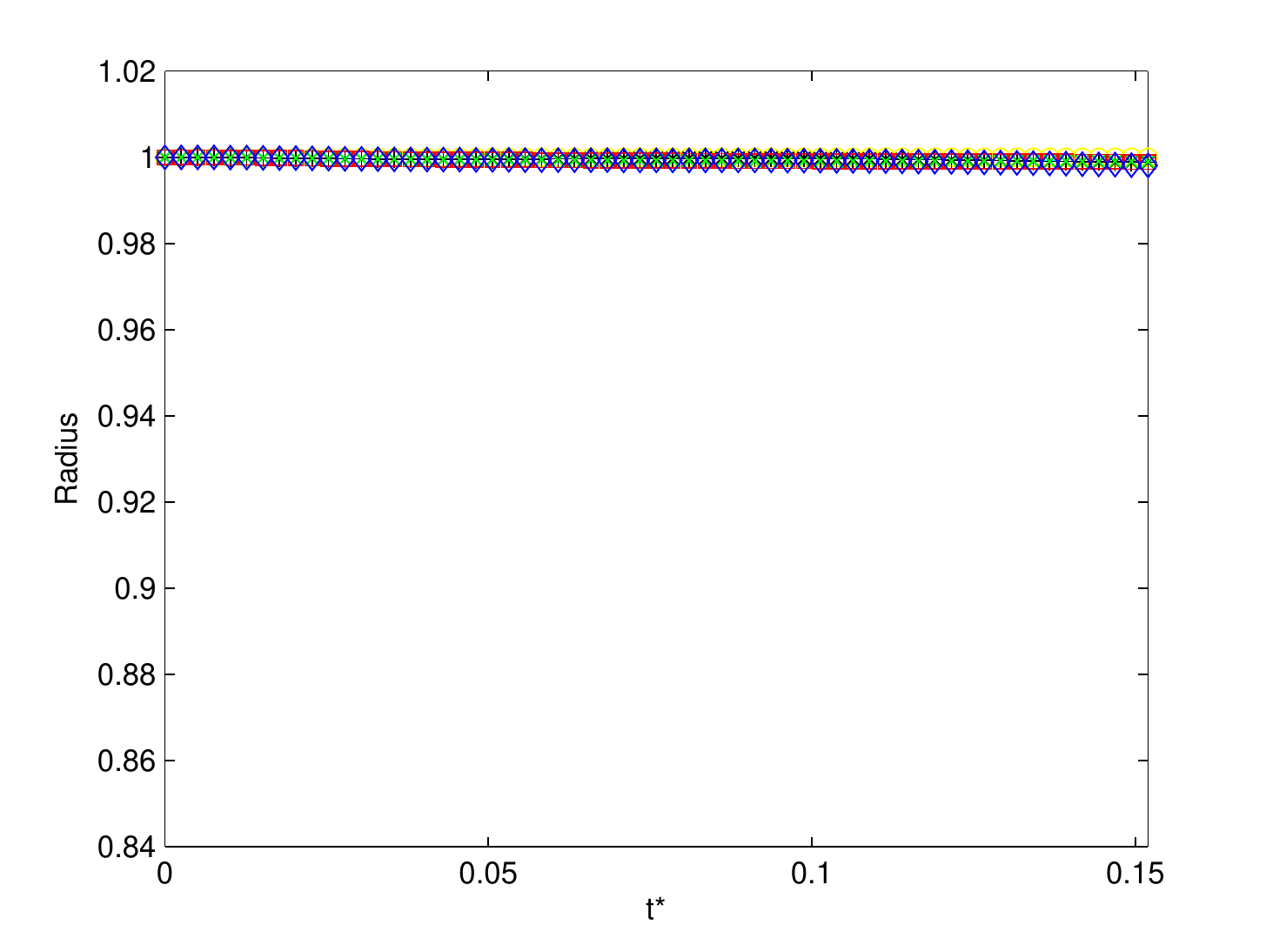}
\caption[Triple]{In this figure we plot the distance from the origin to the centroid of the vortices. On left $a/b=.375$ and on the right $a/b=.125$. For both plots the $m=0$ plot is ({\color{yellow} $\circ$}), the $m=2$ plot is ({\color{cyan}+}), the $m=3$ plot is ({\color{black} ${*}$}), the $m=4$ plot is ({\color{green}$\times$}), the $m=5$ plot is ({\color{red} $\Box$}), and the $m=6$ plot is  (\color{blue}$\diamond$}).\label{fig:radiusPlot}
\end{center}
\end{figure}

Here, we consider two different initial conditions for vortex merger: the first is well above the critical ratio at $a/b = 0.375$ and second is well below the critical ratio at $a/b=
0.125$. Each vortex is initialized with total circulation of $M[i,0,0] =1$ and $\nu=0.001$, and thus our Reynolds number is $Re = M[i,0,0]/\nu = 1000$.  These two examples allow us to study the effect that increasing the number of moments for each vortex has on the motion of the centroid of vorticity. To compare both examples we non-dimensionalize time using the turnover time $t^* = \frac{1}{2\pi^2b^2}t$ and implement both examples with Hermite moments of order $m=0,~2,~3~,4~,5,$ and $6,$ from $t^*=0$ to $t^*=.152$, to focus on the early time behavior of merger. In the first case ($a/b =0.375$), we expect the vortices to rapidly begin to merge.  In the left plot in Figure \ref{fig:radiusPlot}, we plot the distance from the origin to the centroid of the vortices. In the context of the merger regime we expect to observe a rapid decline of this distance.

\begin{figure}[!hbp]
\begin{center}
\includegraphics[width=.33\textwidth]{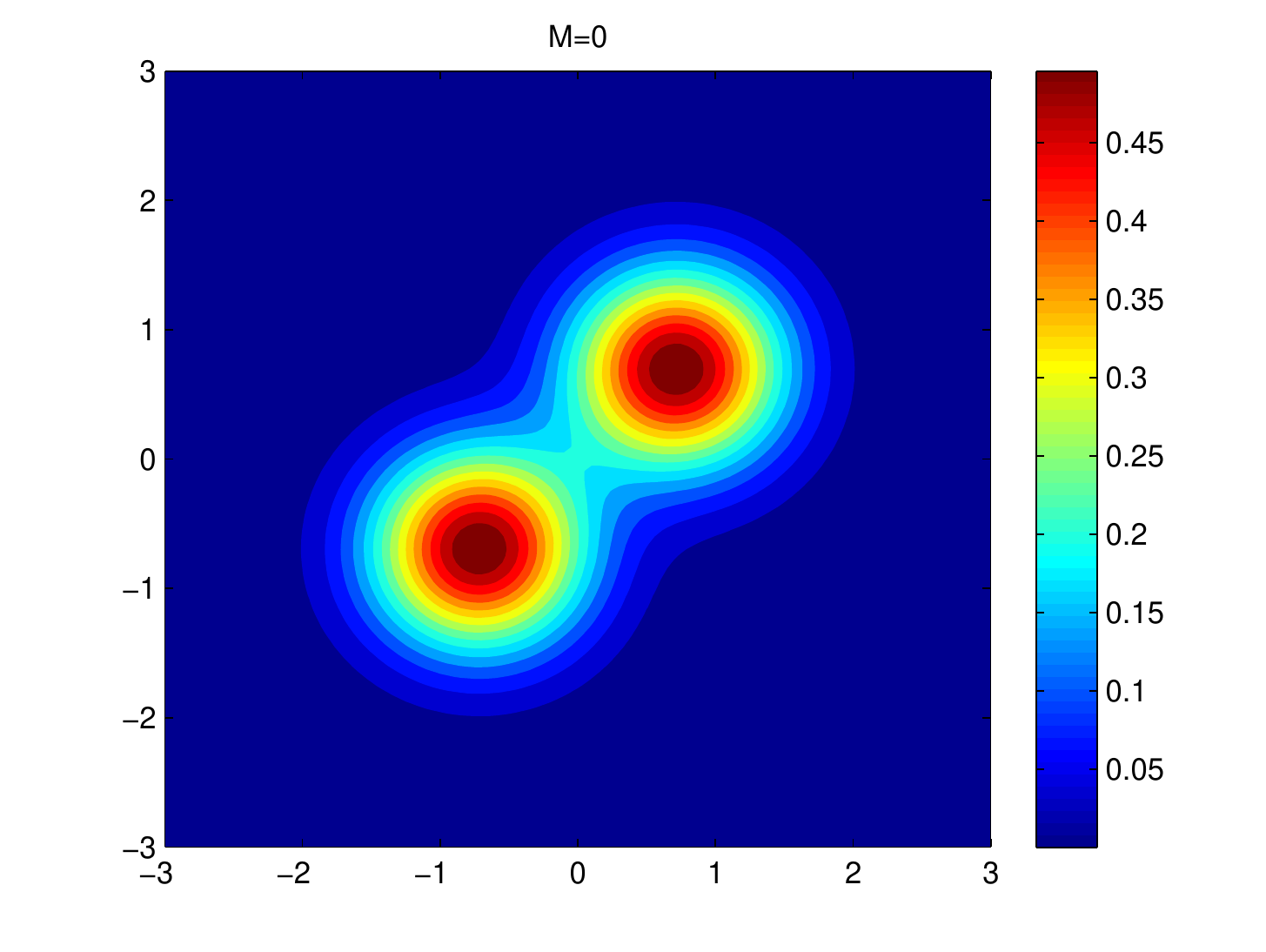}\includegraphics[width=.33\textwidth]{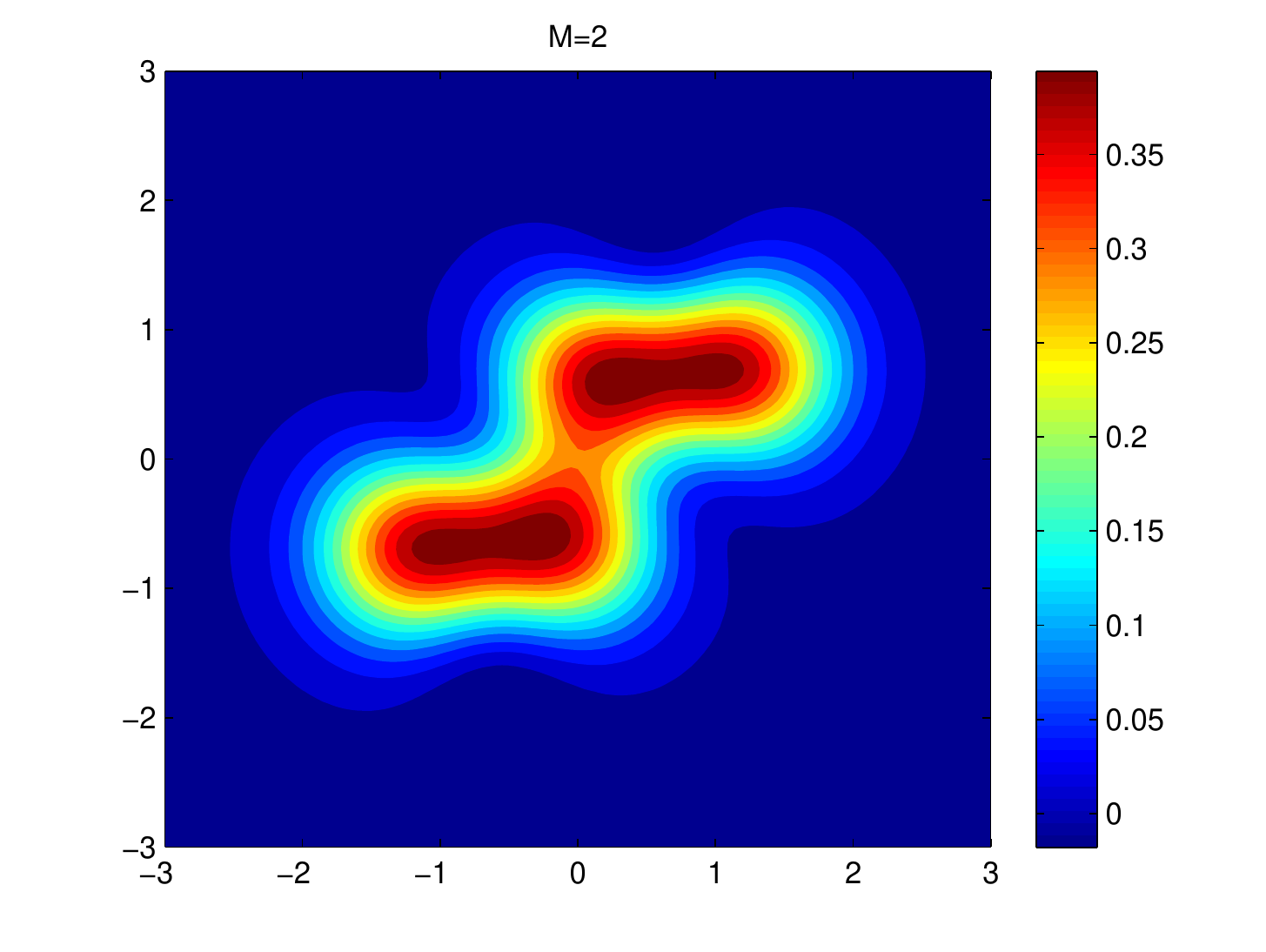}\includegraphics[width=.33\textwidth]{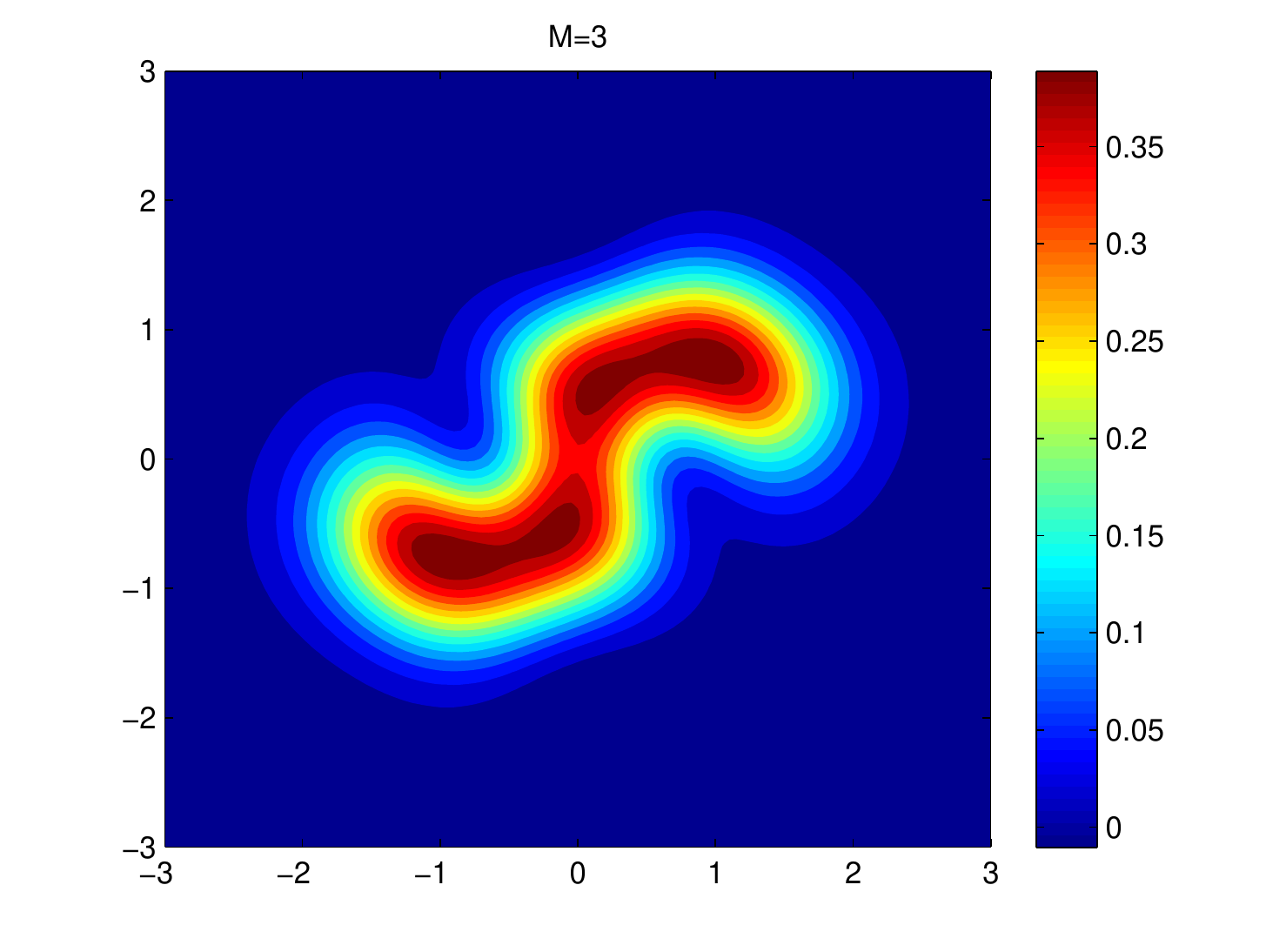}\\
\includegraphics[width=.33\textwidth]{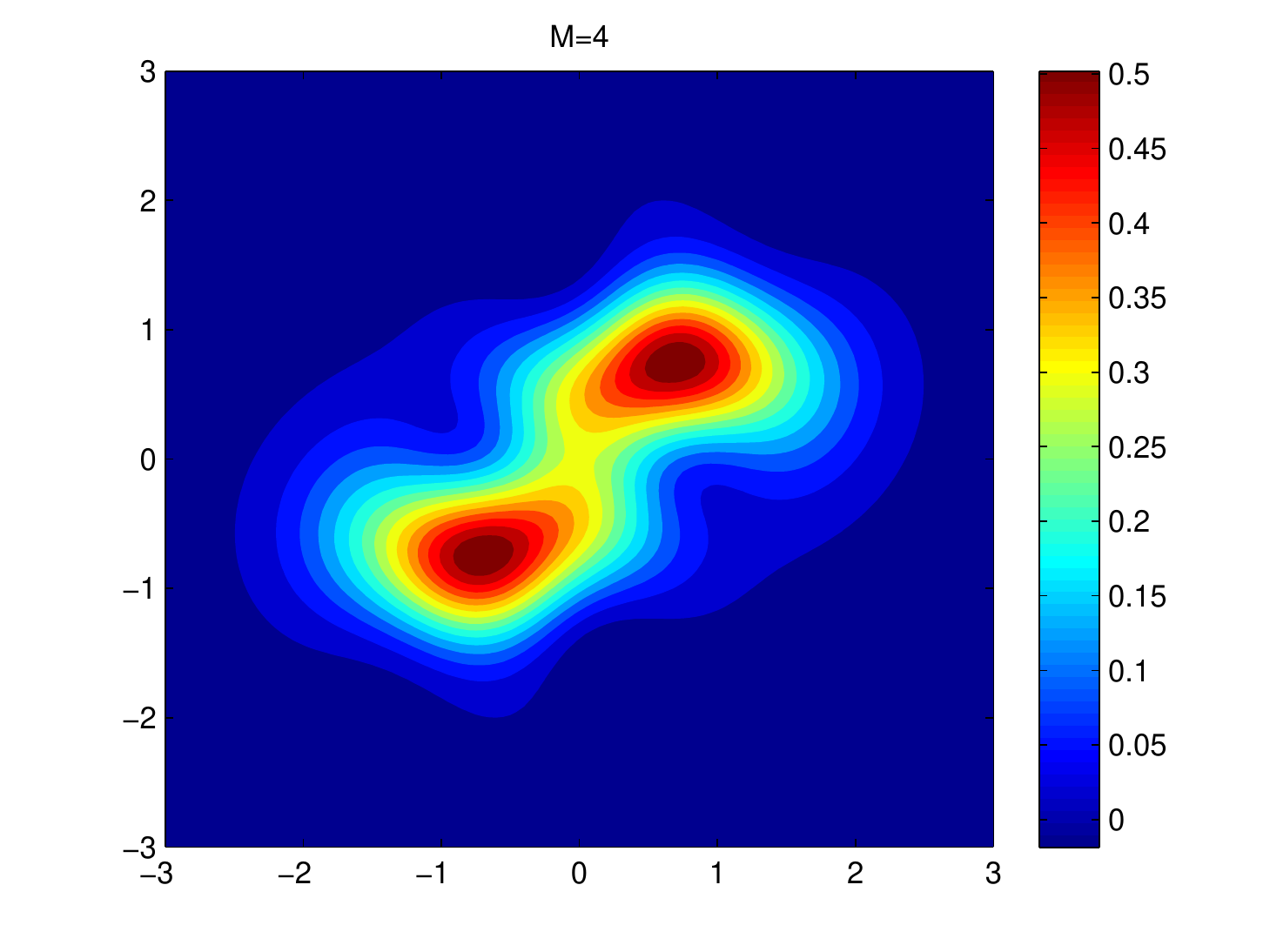}\includegraphics[width=.33\textwidth]{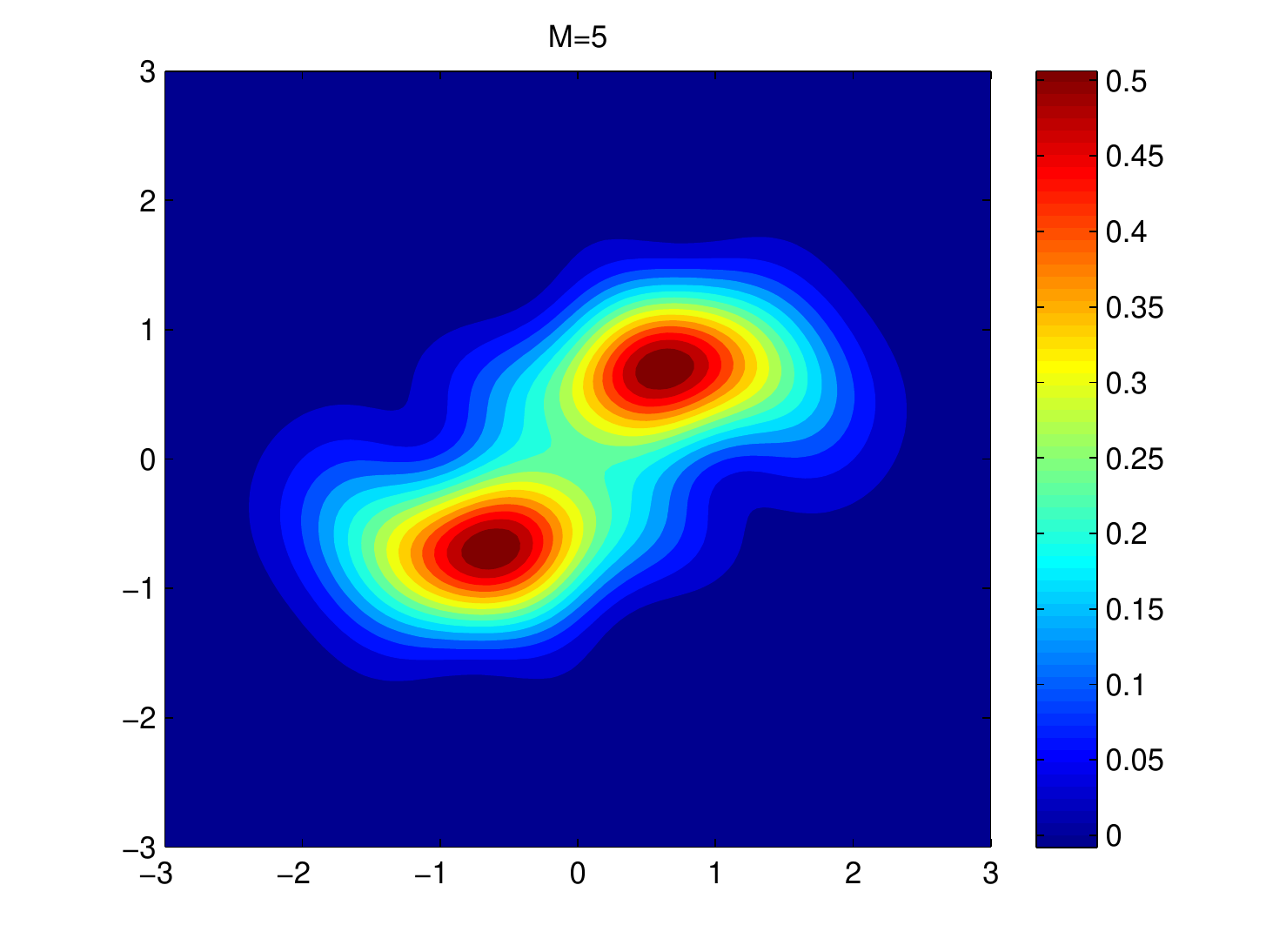}\includegraphics[width=.33\textwidth]{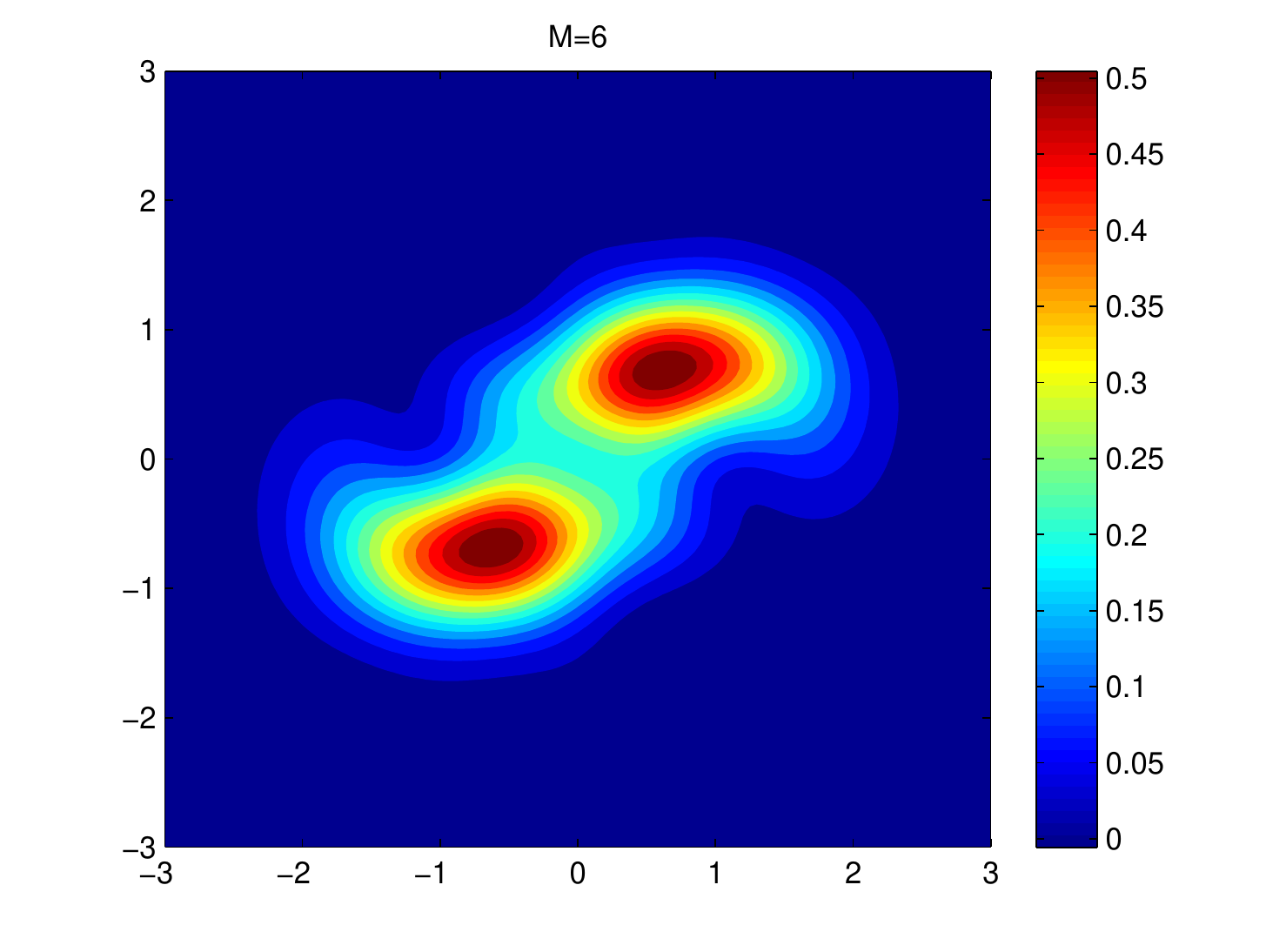}
\caption[Triple]{The plots of the vorticity in the merger regime at $t^*=.127$. The plots (from left to right and top to bottom) are $m=0,2,3,4,5,6$.}\label{fig:vorticityMerg}
\end{center}
\end{figure}

Let us first consider the $m=0$ trajectory in Figure \ref{fig:radiusPlot}.  This is the case of a pure Gaussian basis element, and the constant trajectory shows that vortex continues to rotate at a strict distance of $1$ from the origin and this is clearly inaccurate for this regime. As $m$ increases, we see that the distance to the origin rapidly begins to decline and for early times,  the trajectory of the centroid of vorticity's early descent to the origin seems to stabilize in $m$ to a limiting trajectory. In particular we can see that in the time the vortices rotate by $.152$ radians the distance of the centroid to the origin has declined by $20\%$. We include a plot of the vorticity at $t^*=.127$ for each $m$ in Figure \ref{fig:vorticityMerg}.

There are several points worthy of comment in Figure \ref{fig:vorticityMerg}. First, notice that the top left plot is the case of pure Gaussian vortices ($m=0$). As expected, they simply turn and diffuse without any deformation, entirely unaffected by being inside the vortex merger regime. The next two plots on the top line, $m=2$ and $3$, show large deformations and merger as the low order approximations attempt to exhibit qualitatively correct behavior. Notice that only elongated deformations occur for $m=2$, since shearing and filamentation are not yet captured  by only retaining up to $m=2$ moments. For $m=3$, shearing and merging have both begun but are only qualitatively correct. By $m=4$, the vorticity distribution is beginning to settle into the approximately correct vorticity distribution. Focusing on the $m=6$ vorticity plot, we see all the characteristics expected in the early onset of merger: shearing and the early development  of spiral arms of vorticity, deformation of the initial circular core near the centroid of vorticity and mutual, rapid attraction of each centroid of vorticity toward the other. This example, while only a model for vortex merger, clearly demonstrates that it is precisely the inclusion of higher Hermite moments that allow us to more accurately capture the correct fluid dynamics of vortex merger.

\begin{figure}[!hbp]
\begin{center}
\includegraphics[width=.33\textwidth]{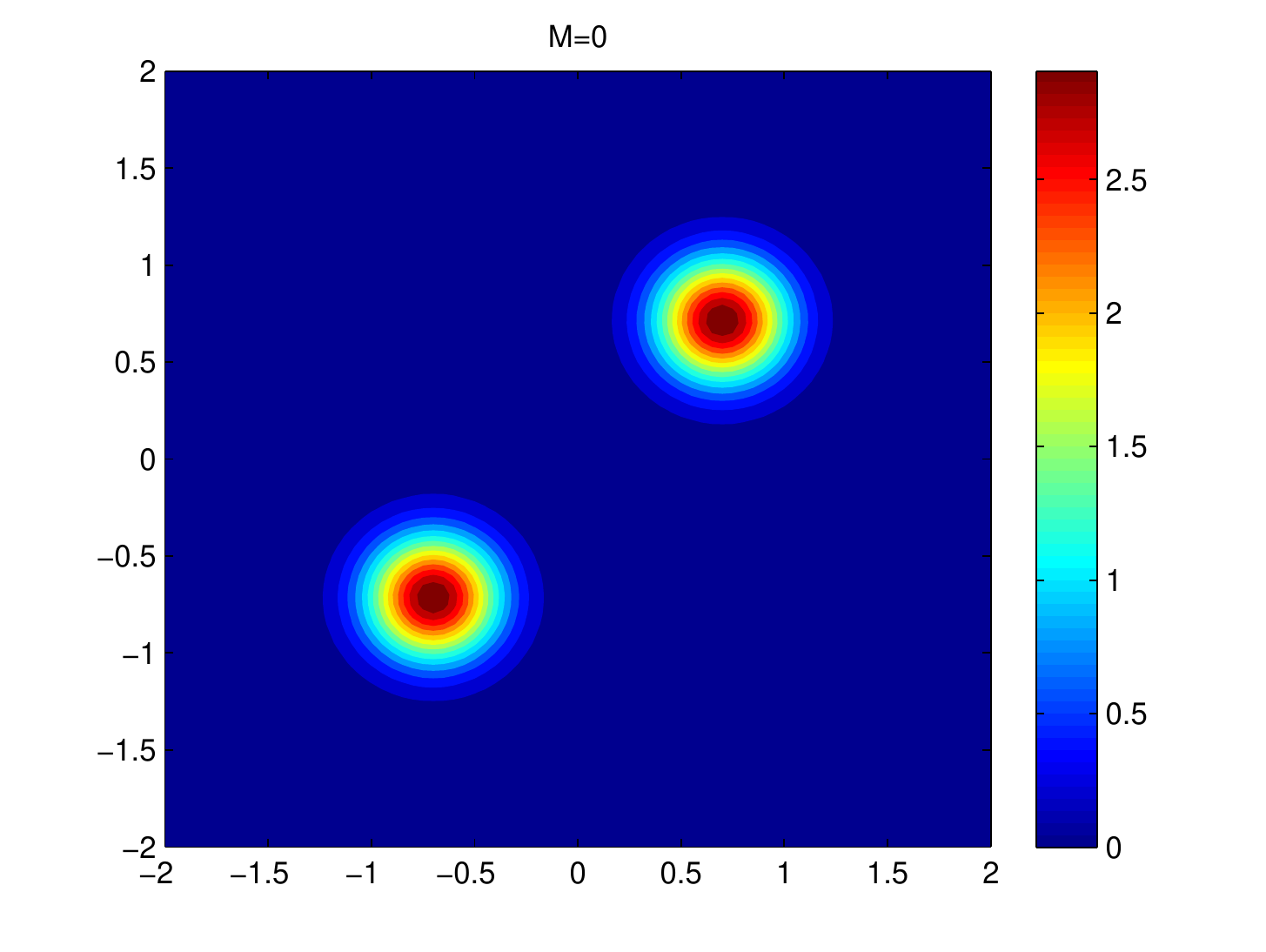}\includegraphics[width=.33\textwidth]{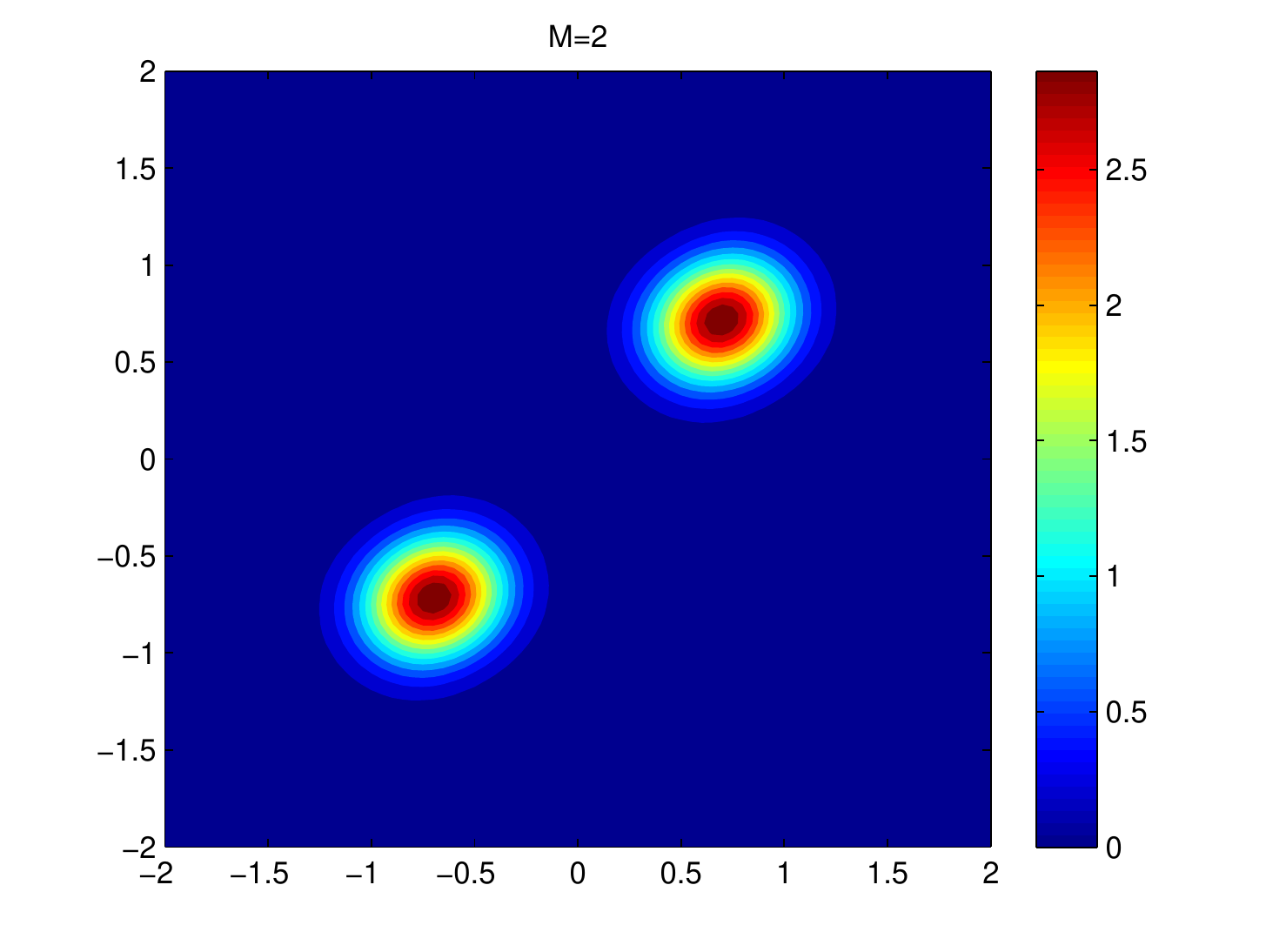}\includegraphics[width=.33\textwidth]{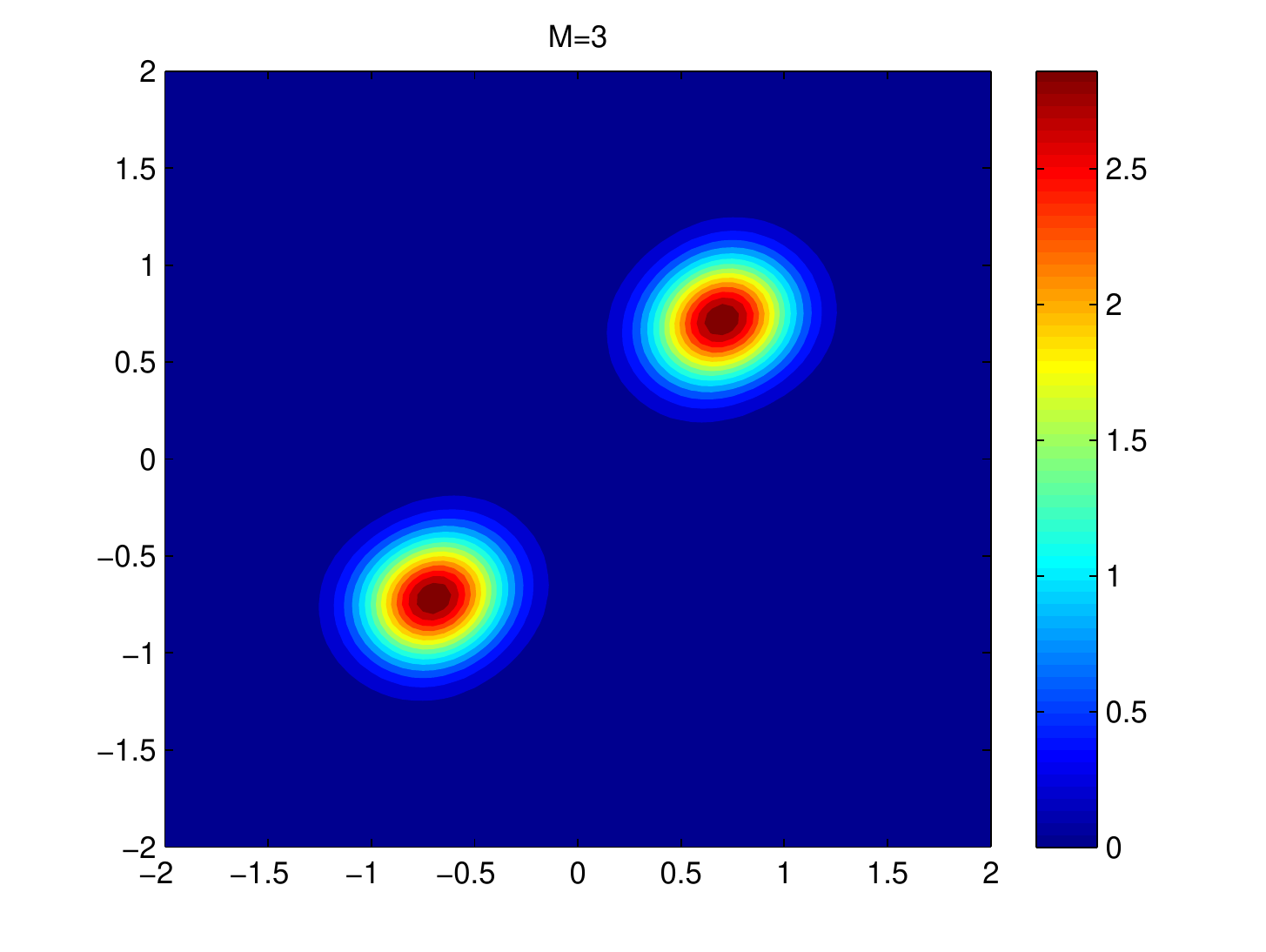}\\
\includegraphics[width=.33\textwidth]{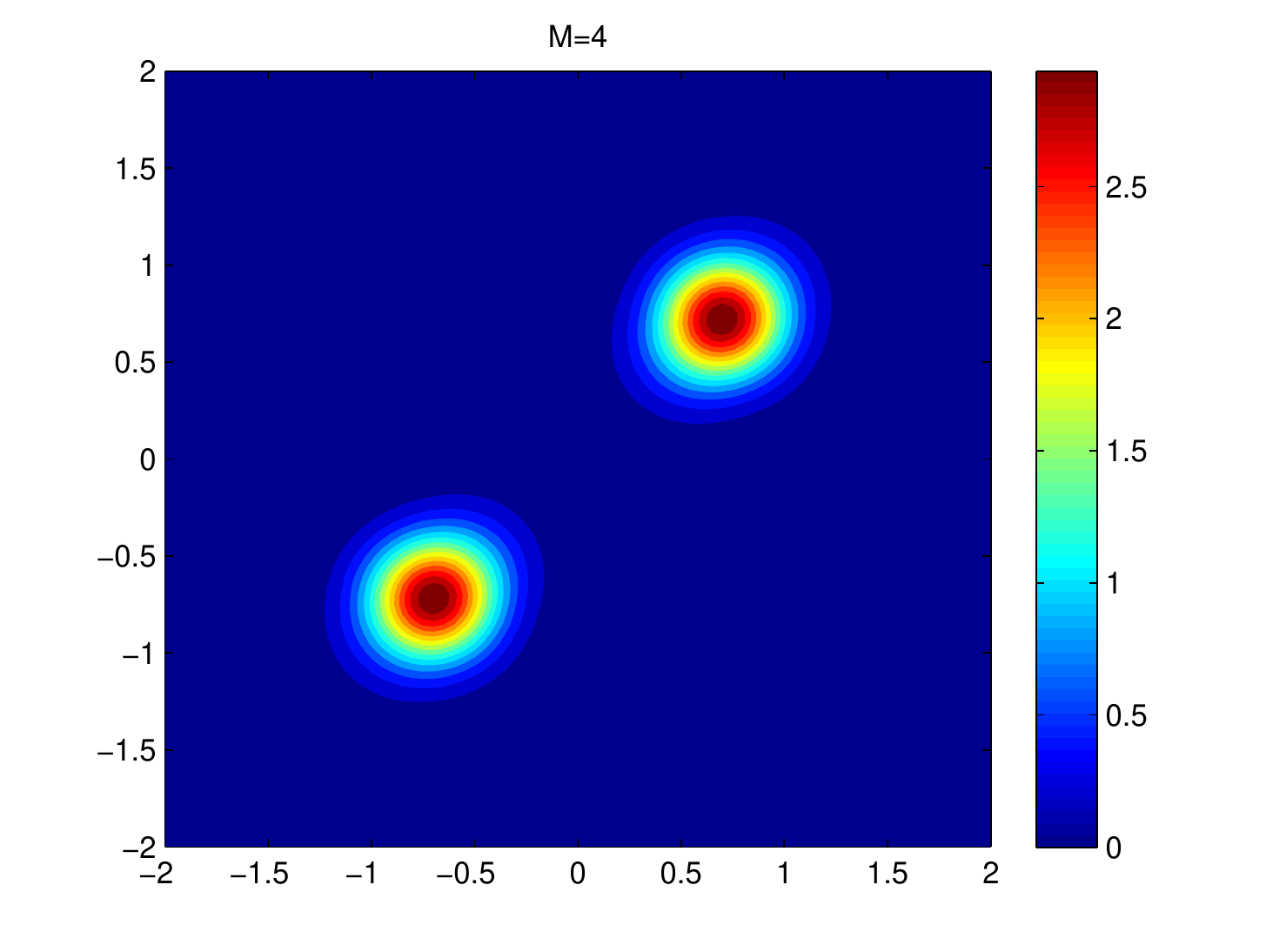}\includegraphics[width=.33\textwidth]{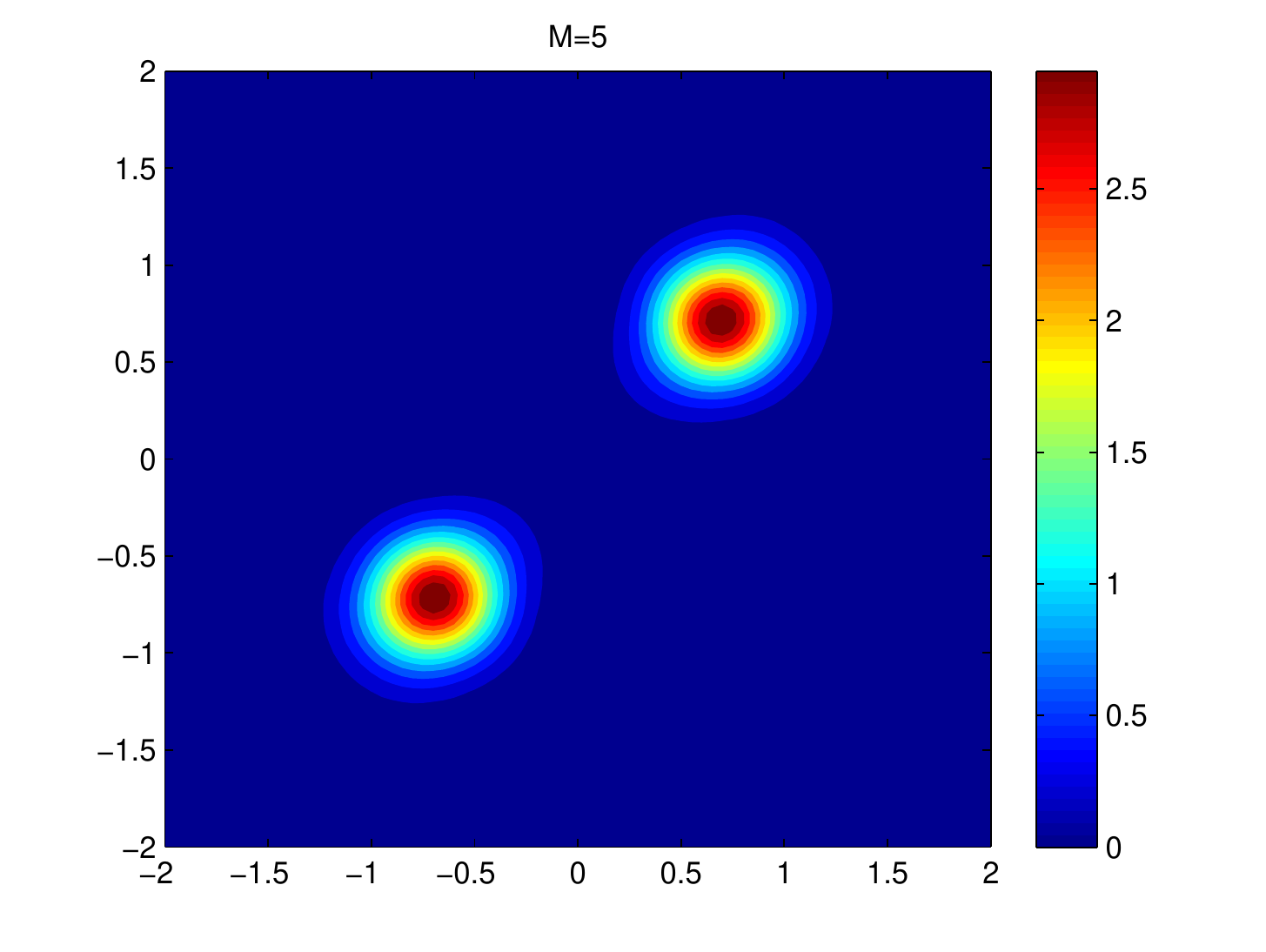}\includegraphics[width=.33\textwidth]{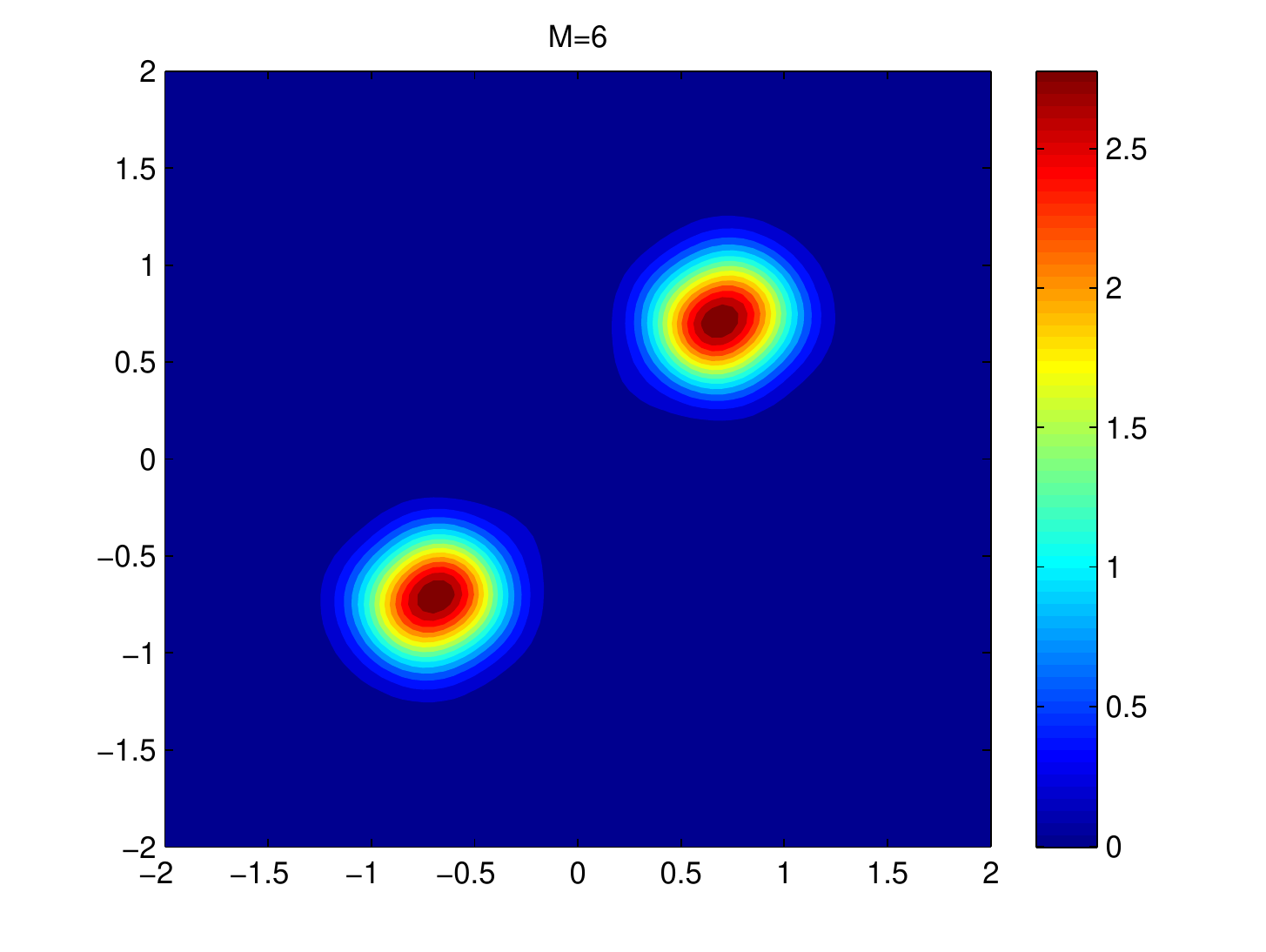}
\caption[Triple]{The plots of the vorticity in the well separated regime at $t^*=.127$. The plots (from left to right and top to bottom) are $m=0,2,3,4,5,6$.}\label{fig:vorticityRotate}
\end{center}
\end{figure}

In our second example, we consider the case where the ratio of the co-rotating vortices is $a/b= .125$, well outside the critical ratio. Here, we expect that the centroid of vorticity should maintain the distance from the origin or perhaps only slightly decline, and this is precisely what we observe in the right plot in Figure \ref{fig:radiusPlot}. In addition we expect the well separated vortices to maintain there circular shape and perhaps begin to a exhibit slightly elliptical deformations. We  plot the vorticity distribution at $t^*=.127$ in Figure \ref{fig:vorticityRotate}.
First notice that in each plot of Figure \ref{fig:vorticityRotate} the rotation angle is the same. There is also very little deformation of the vortices with perhaps the slightest elliptical deformation being exhibited in the distributions with higher Hermite moments as expected.

These two examples demonstrate that simply through the inclusion of higher order Hermite moments, our model can capture significantly more accurate physics. In the $m=0$ case of diffusing Gaussians, the governing equations for short times are essentially equivalent to the point vortex theory which states that two point vortices of equal circulation will rotate on a circle. Thus, it is precisely the inclusion of the higher order Hermite moments that allows two initially Gaussian vortex elements to exhibit {\it convective} merger. Moreover, both examples demonstrate that the Hermite moments correctly select the regime of the vortices (co-rotation or merger). Of course, vortex merger is an extremely complicated behavior and to continue to exhibit accurate solutions for longer times, more Hermite moments are ultimately needed to capture the dynamics.

\subsubsection{Tripole Relaxation}\label{sec:CourseGrid}

In our final example, we present a coarse grid direct numerical simulation of large quadrupole perturbations of the Lamb-Oseen vortex of the form \eqref{eq:IC_sd} found in Section \ref{sec:Sheardiffusion}. As noted in Section \ref{sec:Sheardiffusion}, larger values of $\delta$ cause the vorticity distribution to relax to a rotating tripole, see \cite{Barba:2004,barba:2006,rossi:1997} for careful studies of this phenomena. We have already shown in Figure \ref{fig:Re500Perturbation} that a single vortex element with $m=22$ moments successfully captures this relaxation for lower Reynolds number ($Re=500$). We now revisit this problem, but at higher Reynolds number($Re=1000$), and use a coarse grid ($6$ x $6$) approximation of the initial vorticity distribution. For this example, we use an initial Lamb-Oseen vortex with core size $\lambda_0=1$ and our initial vorticity distribution once again takes the form:
\begin{equation}\label{eq:perturbM2}
\omega_0(\xx)= \omega(\xx,0) = \phi_{00}(x,0) + 4\delta(\phi_{20} -\phi_{02}).
\end{equation}

\begin{figure}[!hbp]
\begin{center}
\includegraphics[width=.45\textwidth]{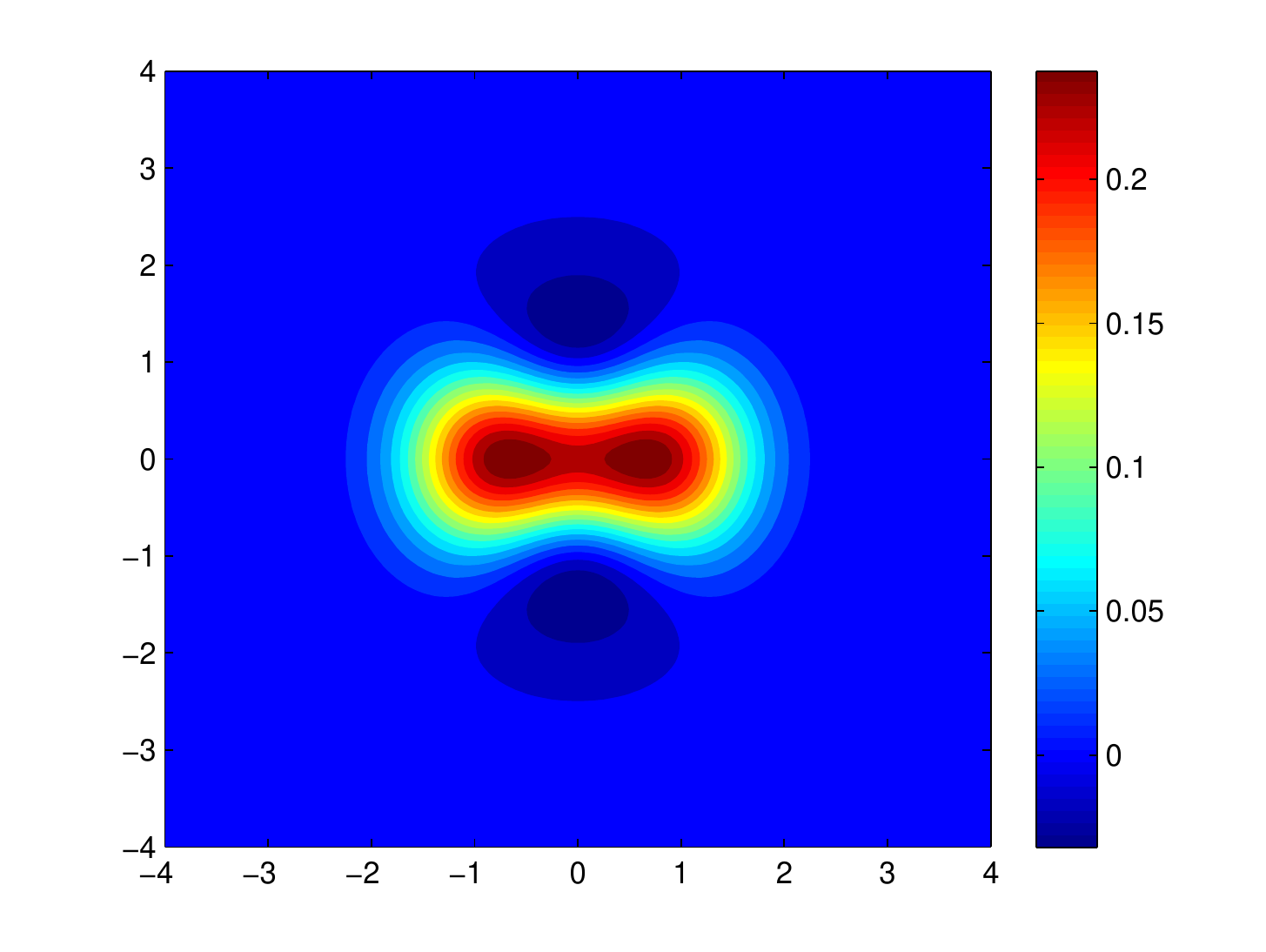}\\
\caption[Triple]{The initial vorticity distribution constructed from equation \ref{eq:blurredIC}.}\label{fig:tripoleIC}
\end{center}
\end{figure}

For this experiment, we select $\delta =.25$, a large quadrapole perturbation of the Lamb-Oseen vortex (note that this perturbation is even larger than in Section \ref{sec:SD_LargePerturb} since $\lambda_0$ is smaller). We overlay our $6\times 6$ grid of vortex centers on $[-1,1] \times [-1,1]$ and approximate the initial condition \eqref{eq:perturbM2} by initializing the circulation of each initial Gaussian vortex element using the standard approach
\begin{equation}\label{eq:blurredIC}
M^j[0,0;0] = \omega_0(\xx_j)\Delta\xx_j
\end{equation}
where $\Delta\xx_j$ is the area of each node.  We first compute a pure Gaussian simulation ($m=0$), which is equivalent to a standard core-spreading vortex method, and then for our second simulation we re-run the identical initial conditions but include quadrupole moments ($m=2$) for each vortex element. Since we are using core-spreading and a coarse grid, we focus our study on the early relaxation dynamics. A plot of the coarse grid approximation of \eqref{eq:perturbM2} is presented in Figure \ref{fig:tripoleIC}.

\begin{figure}[!hbp]
\begin{center}
\includegraphics[width=.33\textwidth]{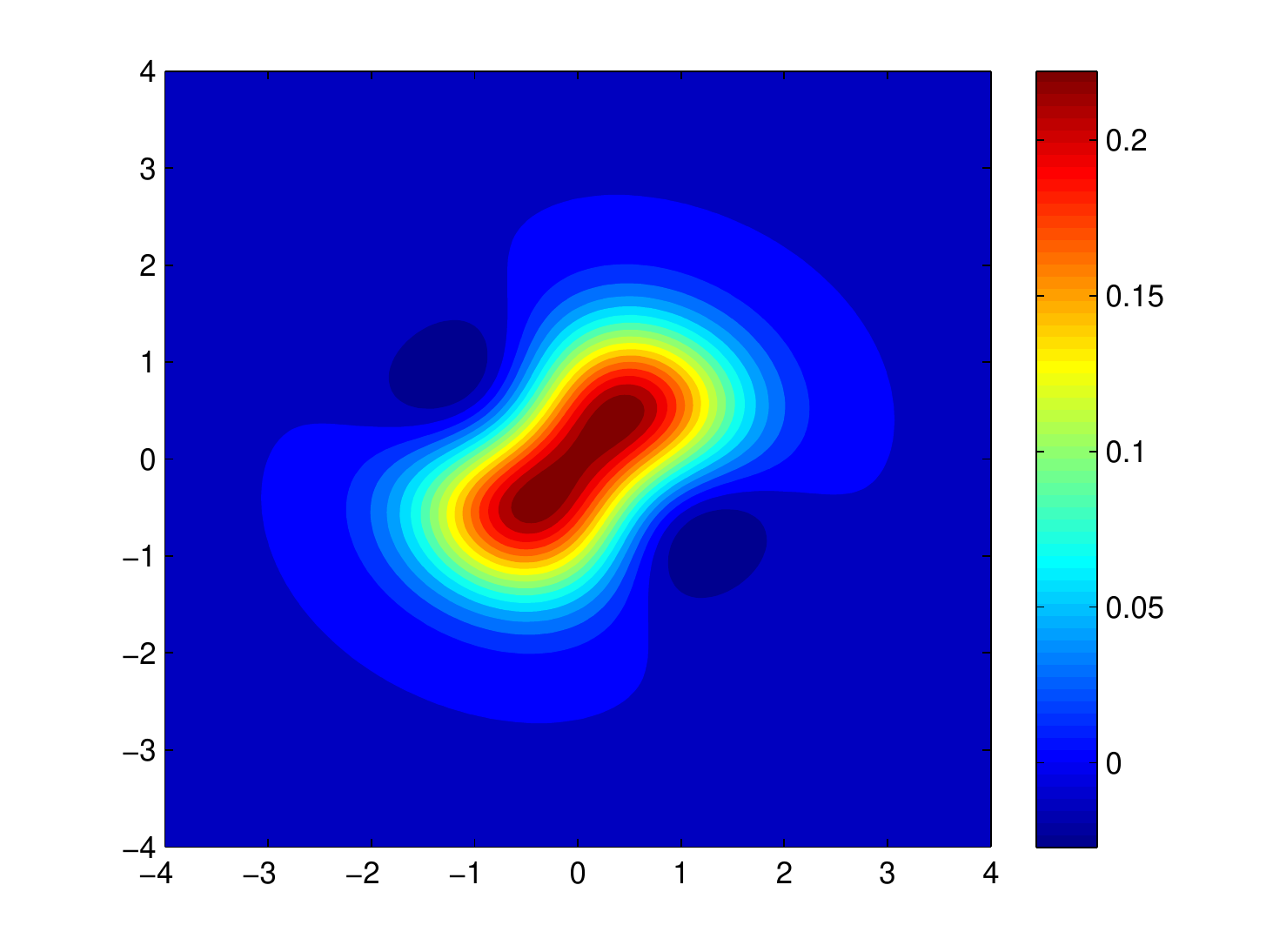}\includegraphics[width=.33\textwidth]{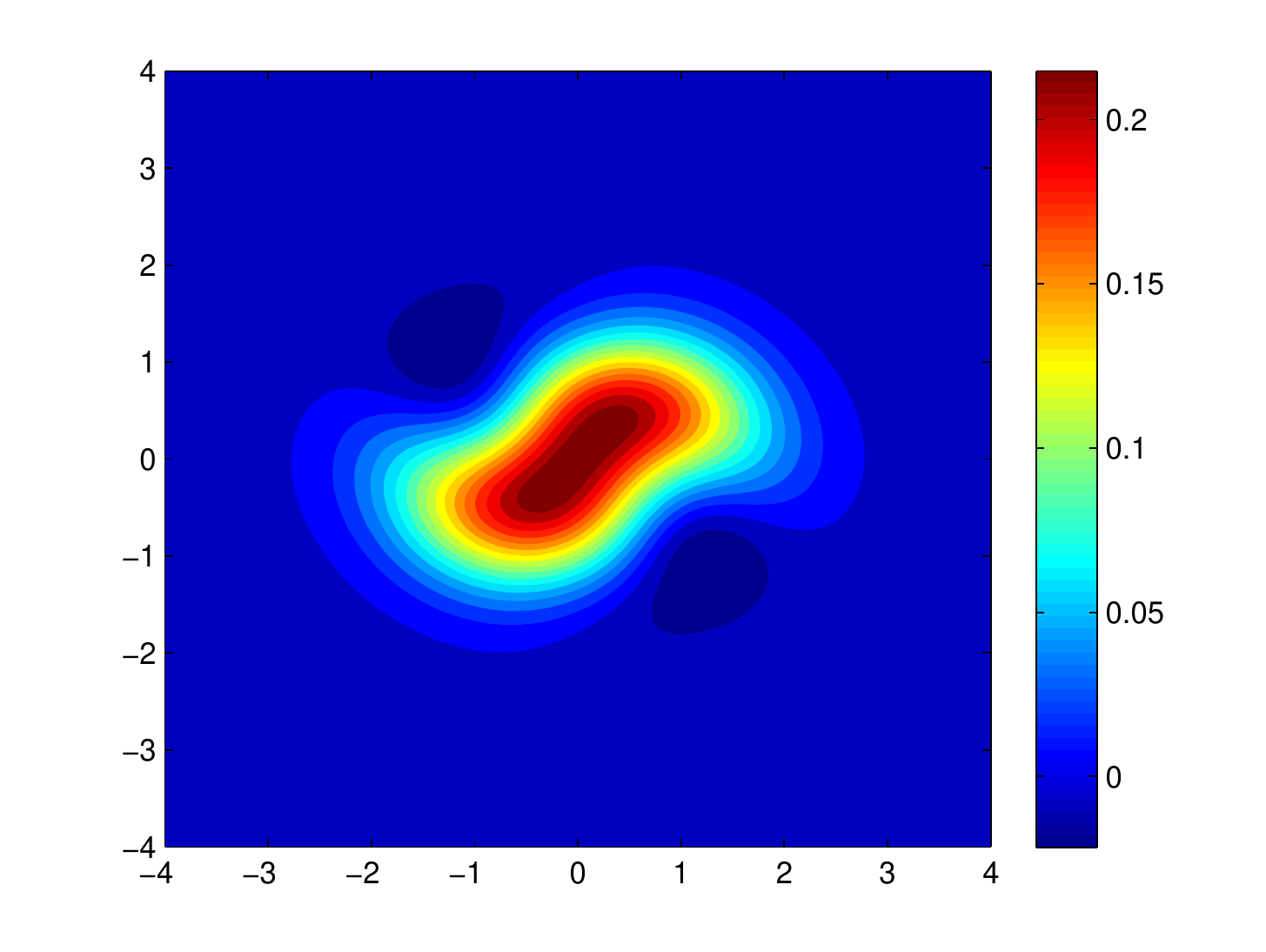} \includegraphics[width=.33\textwidth]{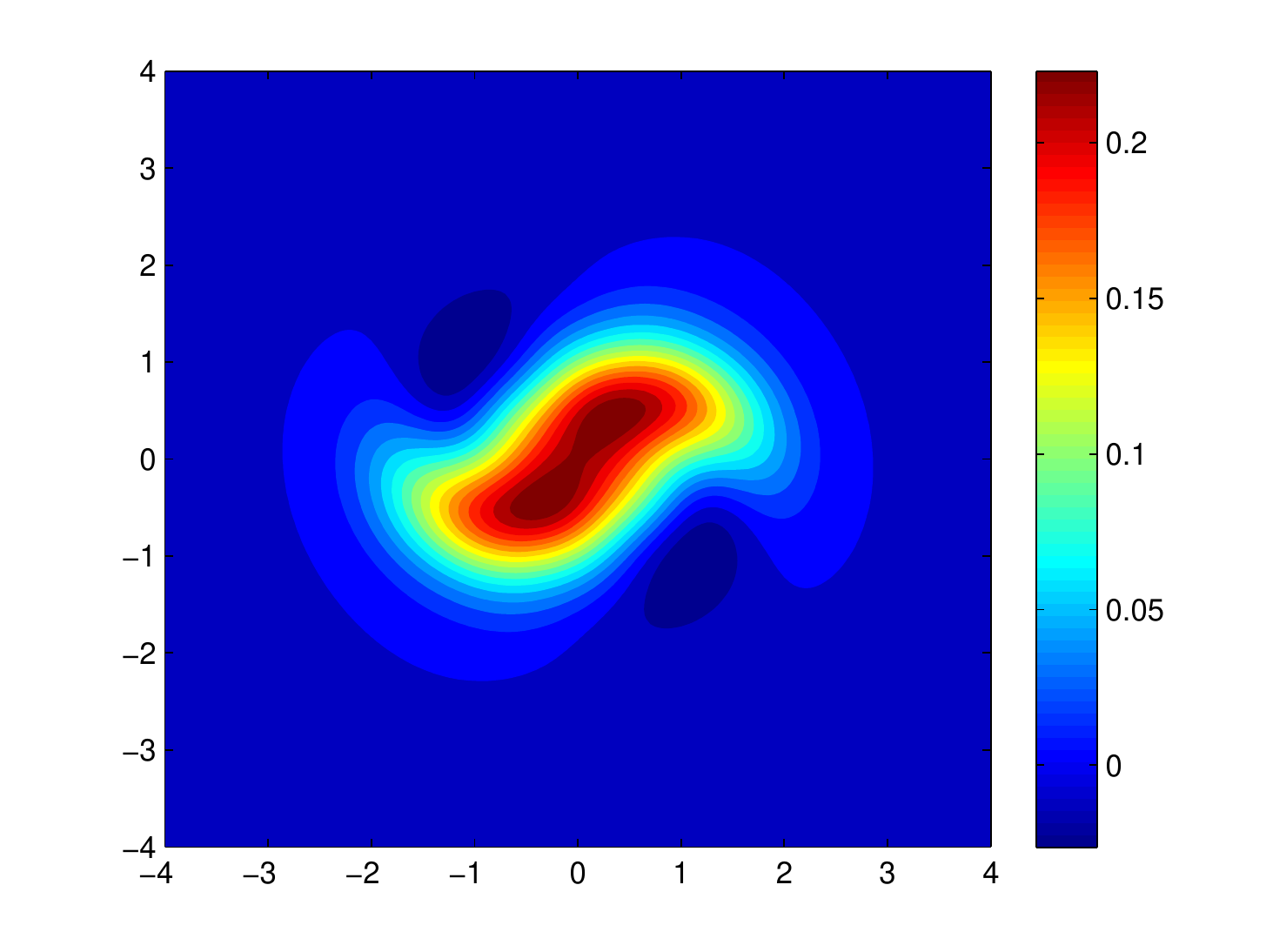}\\
\caption[Triple]{The plot at $t=15$ on the left is the solution using MMVM with $m=0$ and the plot on the right is the solution using MMVM with $m=2$.}\label{fig:tripolet15}
\end{center}
\end{figure}

In Figure \ref{fig:tripolet15}, we juxtapose the implementations of the full MMVM at $t=15$ for $m=0$ and $m=2$ using the same initial conditions \ref{eq:blurredIC}. The left plot is the pure Gaussian $m=0$ computation. In the center plot is the result of including the quadrupole Hermite moments $m=2$ into the calculation. The right plot is high order ($m=24$) single-particle MMVM simulation of the same initial conditions.

The results shown in Figure \ref{fig:tripolet15} are quite striking.  The right plot is taken to be a high accuracy solution of the identical initial conditions at $t=15$ and captures the early time tripole relaxation observed in the work of Barba et al. \cite{Barba:2004,Barba:2005} and Rossi et al. \cite{rossi:1997}.  In this plot we observe shearing and relaxation of the small satellite vortices. In addition, at each contour level we see characteristic spiral arm wind up occurring as well.

With such a coarse grid, we expect to miss much of these important characteristics of the early tripole relaxation.  Indeed, in the case of classical vortex methods with Gaussian basis functions, i.e. MMVM with $m=0$, the approximation is poor. There is barely any spiral arm wind up occurring in the distribution of vorticity, and very little shearing and relaxation is observed. The simulation with $m=0$ looks more like a rotation and diffusion of the initial condition than the characteristic nonlinear rapid relaxation to a tripole.

In contrast, for $m=2$, the MMVM calculation shown in the middle plot of Figure \ref{fig:tripolet15} shows a dramatically more accurate approximation to the high accuracy solution on the right as compared to the $m=0$ example.  In this case, we observe spiral wind up at each contour level of vorticity  and the small satellites of negative vorticity have started to shear the outer arms of positive vorticity  which is also characteristic of the early relaxation dynamics. Since these two experiments are initialized with identical initial vorticity, we can  attribute the improvement of accuracy in the early relaxation dynamics of a tripole solely to the inclusion of the higher moments. We will quantify precisely this observed improvement of accuracy in Section \ref{sec:Numerical}.


\section{Convergence studies and performance}\label{sec:Numerical}
In this section we will present two convergence studies for the single-particle MMVM: The first test case demonstrates the ability of our method to accurately approximate the Lamb-Oseen Vortex, an exact solution to \eqref{eq:2DIVP}. The second example is a tripole relaxation example where no exact solution is known. In each example exponential convergence is demonstrated numerically. We will next quantify the  improvement of accuracy in the coarse grid approximation example presented in Section \ref{sec:CourseGrid}.  Finally we will briefly discuss the computational implementation used to compute these examples.

\subsection{The single-particle MMVM}

\subsubsection{First test case: the Lamb-Oseen vortex}

A standard test example for $2$D vortex methods is the Lamb-Oseen monopole which has initial conditions:
\begin{equation}\label{eq:LambIC}
\omega(r,0) = \frac{\Gamma_0}{4\pi \lambda_0^2} e^{-\frac{r^2}{\lambda_0^2 }}
\end{equation}
where $r = (x^2 + y^2)^{\frac{1}{2}}$. Initial conditions (\ref{eq:LambIC}) are particularly useful because it results in the only  known nontrivial, analytic solution to system (\ref{eq:2DIVP}) known as the Lamb-Oseen vortex. This solution takes the form
$$\omega(r,t) = \frac{\Gamma_0}{4\pi (\lambda_0^2 + 4\nu t)}e^{-\frac{r^2}{(\lambda_0^2+4\nu t )}},$$
 and this example is often a first benchmark to compare the spatial accuracy of numerical solutions, \cite{Barba:2005,Rossi:2006A}.


Provided that we select our core size $\lambda_0$ in our expansion to be equal to the initial core size of the initial data \eqref{eq:LambIC} then, by construction, the Lamb-Oseen monopole represents the zeroth order moment in our expansion $M[0,0]$, which is constant because the total vorticity is a conserved quantity \cite{majda:2002}. Thus, no nontrivial dynamics will occur if we were to start a single-particle MMVM simulation using initial conditions $M[0,0](0) = 1$ and $M[i,j](0)=0$ for all $i,j$; in this case, our method predicts that for all time $t$,
$M[0,0](t) = 1$ and $M[i,j](t) = 0$ for all $(i,j)\ne (0,0)$, in agreement with the exact solution.  Hence, to perform a more meaningful convergence study for MMVM, we must approximate the Lamb-Oseen vortex in a way which tests the dynamics of the higher moments as well. We therefore choose to approximate a Lamb-Oseen vortex with core size $\lambda_0 = 2.1$, which remains an exact solution, with a Hermite expansion using a Hermite expansion with core size $\lambda_0=2$, thus allowing the higher moments to accommodate the error between the smaller core size Gaussian and the true solution. For our convergence study  we select $Re = 1000$.

\begin{figure}[!hbp]
\begin{center}
\includegraphics[width=.6\textwidth]{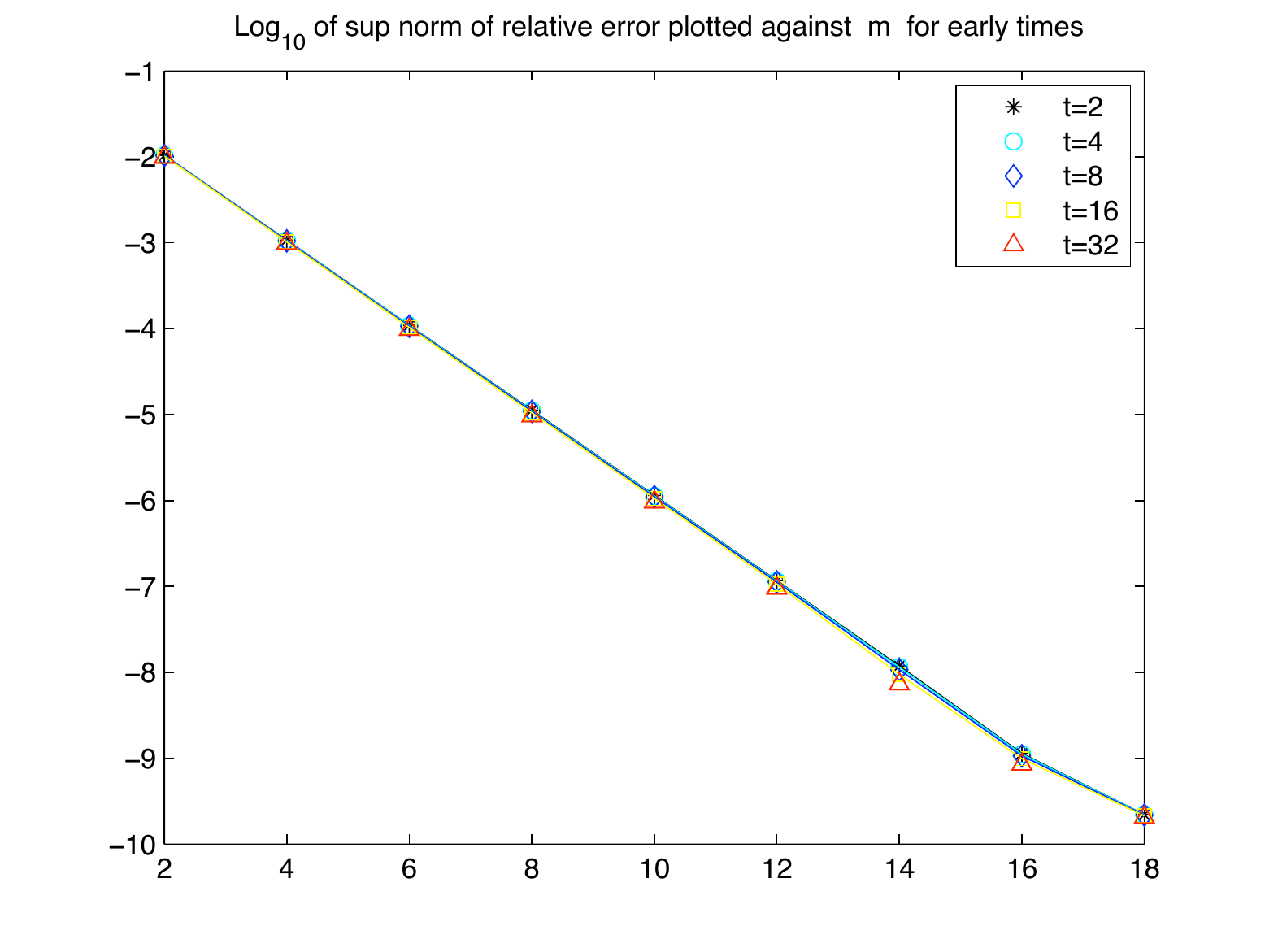}\\
\caption[Triple]{The $\log_{10}$ of the $L^\infty$ norm of the relative error between the Lamb-Oseen monopole with core size $2.1$ and our test approximation with $\lambda_0 = 2$ plotted against $m$, the number of moments retained in the simulation.  The different lines in the figure correspond to times $2,~4, ~8, ~16$ and $32$.}\label{fig:coresizeError}
\end{center}
\end{figure}

We calculate the error between the true solution for the Lamb-Oseen monopole and the $m$ moment approximation obtained with the single-particle MMVM for a range of values of $m$.  In Figure \ref{fig:coresizeError}, we show the $L^\infty$ norm of the relative error plotted against $m$. The linear nature of these plots confirms that, as one expects from a spectral method, the error between the true solution and the simulated solution converges exponentially as $m \to \infty$. The overlap of the four lines corresponding to times $2,~4,~8,~16$ and $32$ indicate that the error varies extremely slowly in time.  This is likely due to the axisymmetry of the initial data and the fact that diffusion is the sole component of the dynamics.

\subsubsection{Test Case $2$: A Larger Quadrupole Perturbation}

The previous test case numerically demonstrated that the convergence rate is exponential
in the number of moments, in the case of axisymmetric initial data. We now test our method for non-axisymmetric initial conditions which result in nontrivial dynamics for the evolution of the coefficients in our expansion. To this end, we select as a second test case, large quadrupole perturbations of the Lamb-Oseen vortex with initial conditions of the form \eqref{eq:IC_sd} which were studied in Section \ref{sec:Sheardiffusion}. We select large enough perturbations such that the solutions do not relax to an axisymmetric state and instead evolves toward a metastable tripole solution, see \cite{barba:2006,rossi:1997}, which has no known analytic solution.   Therefore, in order to measure the convergence of the single particle MMVM, we take a self consistent approach by selecting a high-order simulation ($m=24$ moments) in lieu of an exact solution and compare our lower-order simulations, in increasing order, against this solution.


\begin{figure}[!hbp]
\begin{center}
\includegraphics[width=.6\textwidth]{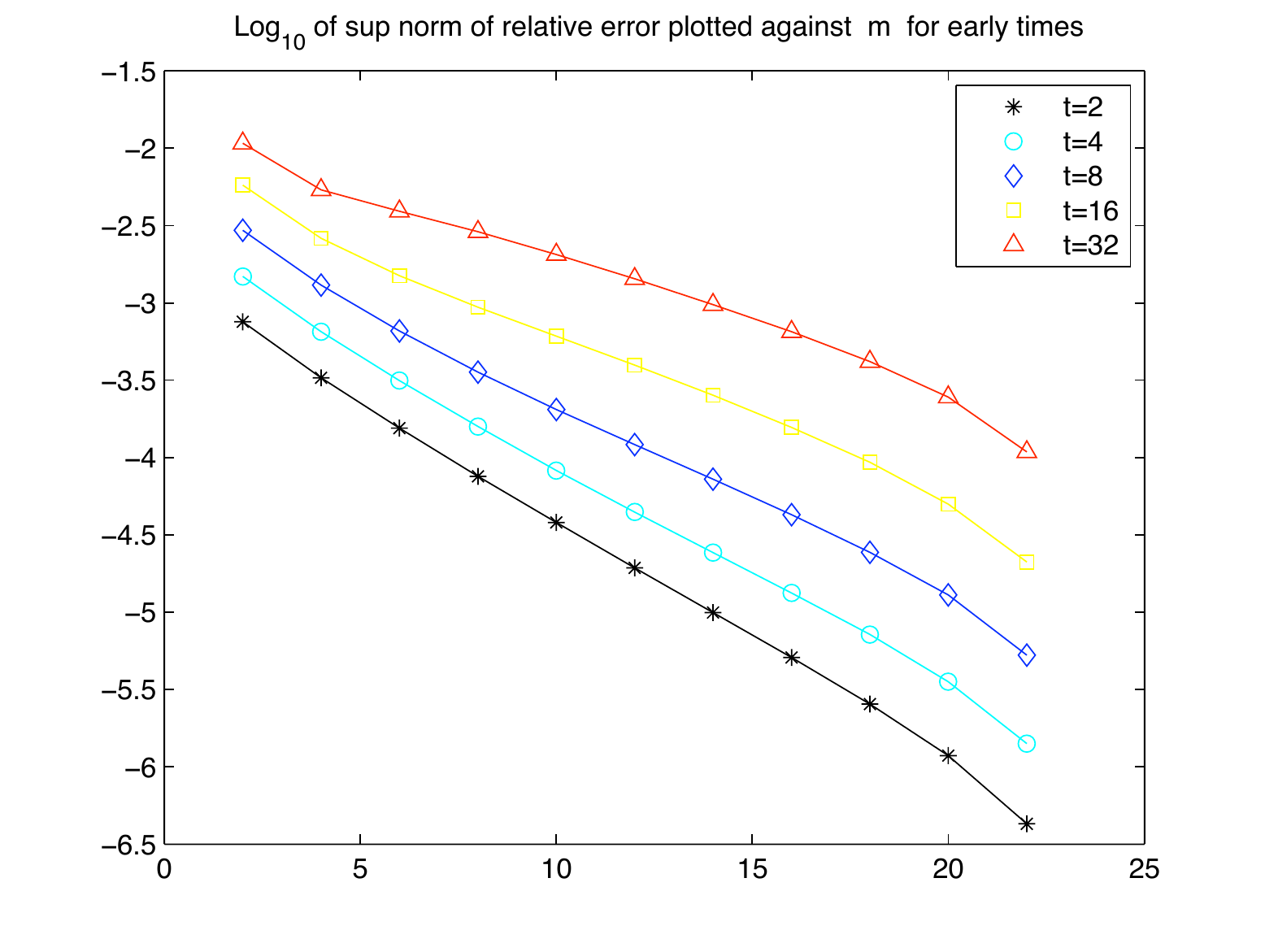}
\caption{Here, $\log_{10}$ of the $L^\infty$ norm of the relative error of the solution is plotted against the number of moments $m$ for $Re=1000$ and $\delta = 0.25$. The different lines correspond to times $2,~4,~8,~16$ and $32$ (see legend). }\label{fig:supnorm}
\end{center}
\end{figure}

In Figure \ref{fig:supnorm}, we examine the $\log_{10}$ of the relative difference between the simulated solution with $24$ moments and the simulations with fewer moments for $\delta = 0.25$ and $Re=1000$. The different lines in the plot correspond to times $t= 2,~4,~8,~16$ and $32$. The linear nature of the plot once again indicates that the $L^\infty$ norm of the relative error decreases exponentially in $m$.  The rate of exponential convergence is higher for earlier times. In addition, the error is very small at early times but increases over time. As expected, for any fixed number of moments, the dynamics will eventually diverge from the ``true solution," however, it is clear from this figure that the rate at which this divergence occurs depends heavily on the number of moments we retain in the simulation. Thus, in practice, one can tune the number of moments kept in order to approximate the solution at the desired level of accuracy.  

\begin{figure}[!hbp]
\begin{center}
\includegraphics[width=.30\textwidth]{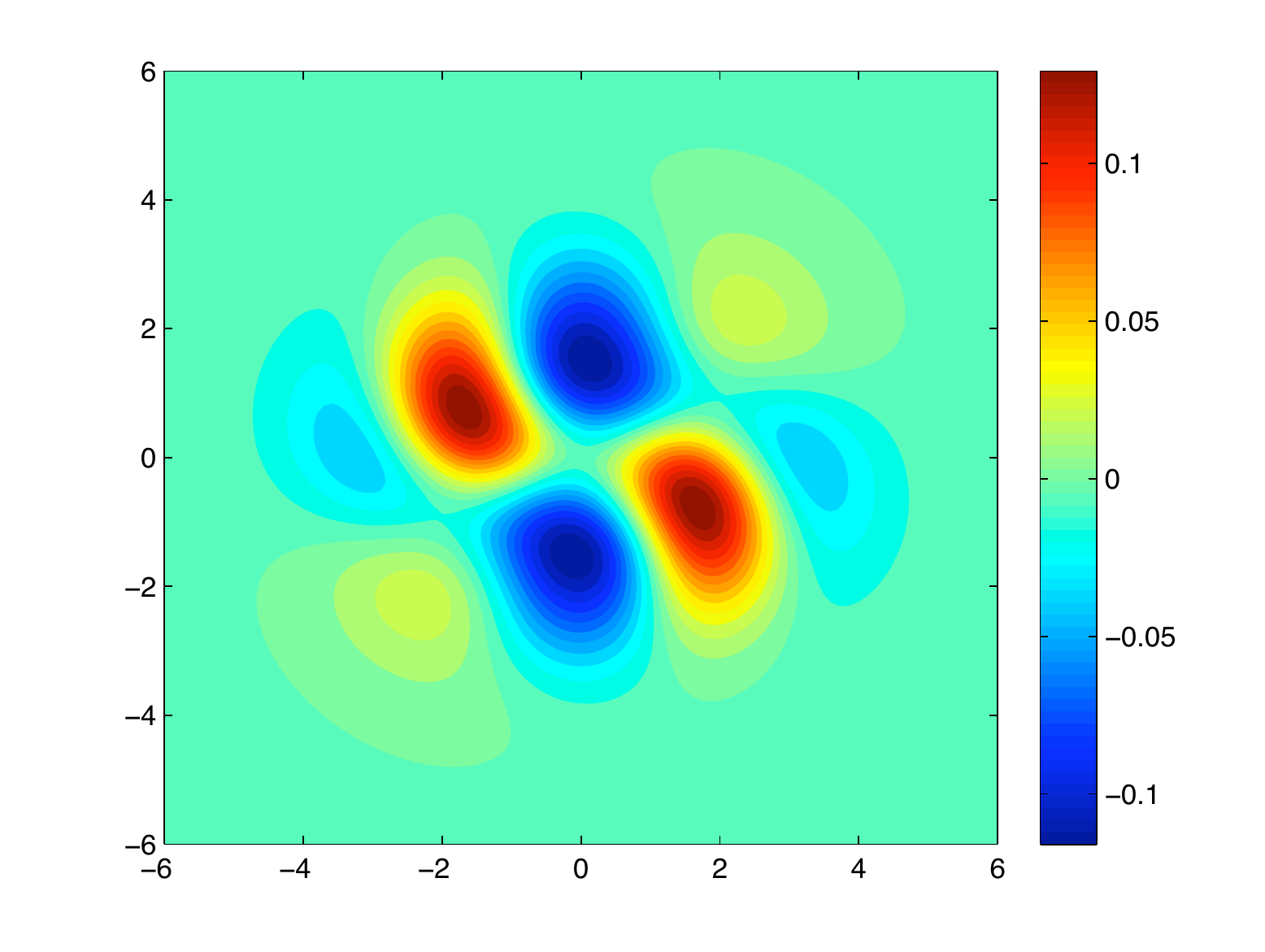}
\includegraphics[width=.30\textwidth]{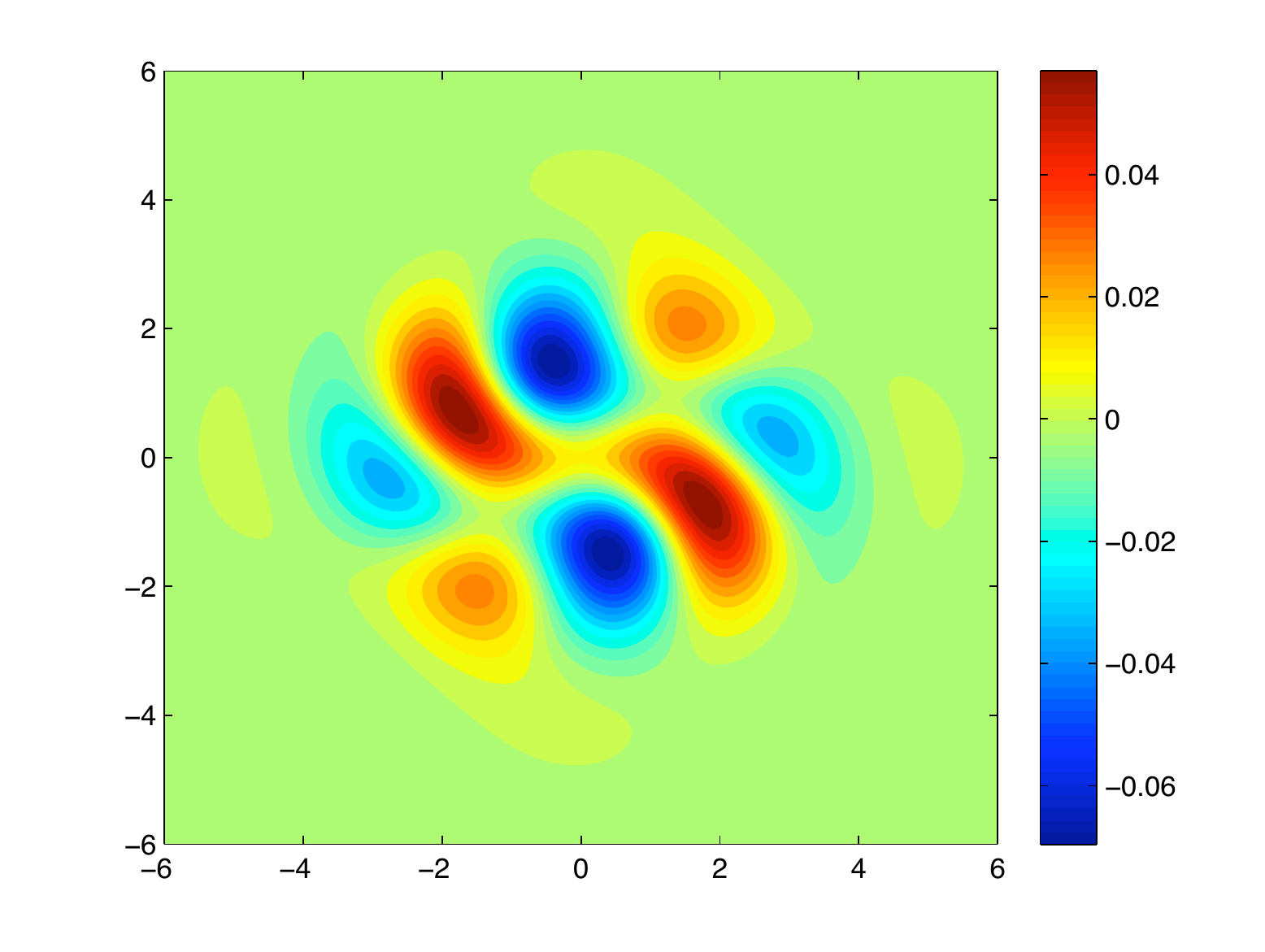}
\includegraphics[width=.30\textwidth]{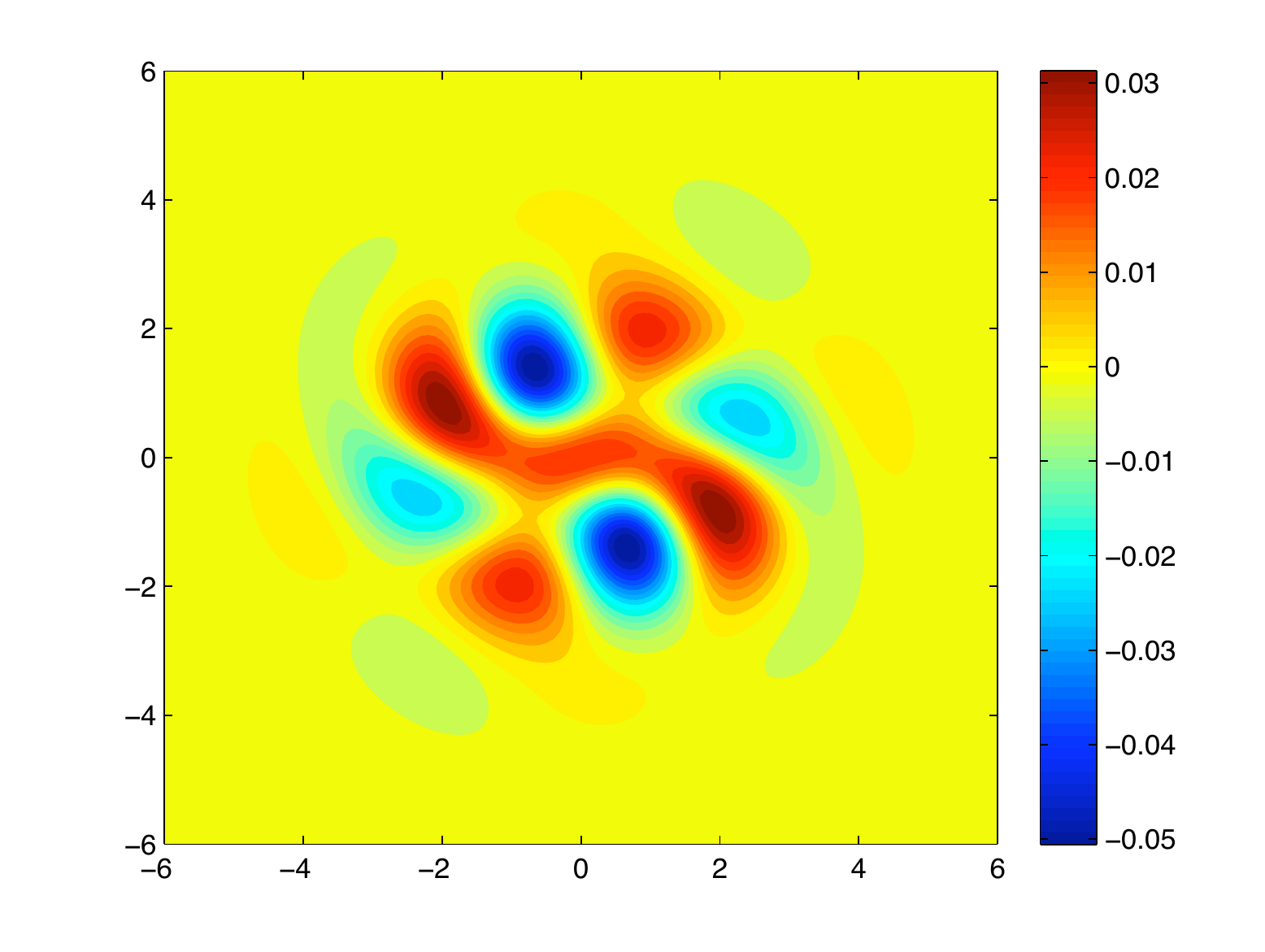}\\
\includegraphics[width=.30\textwidth]{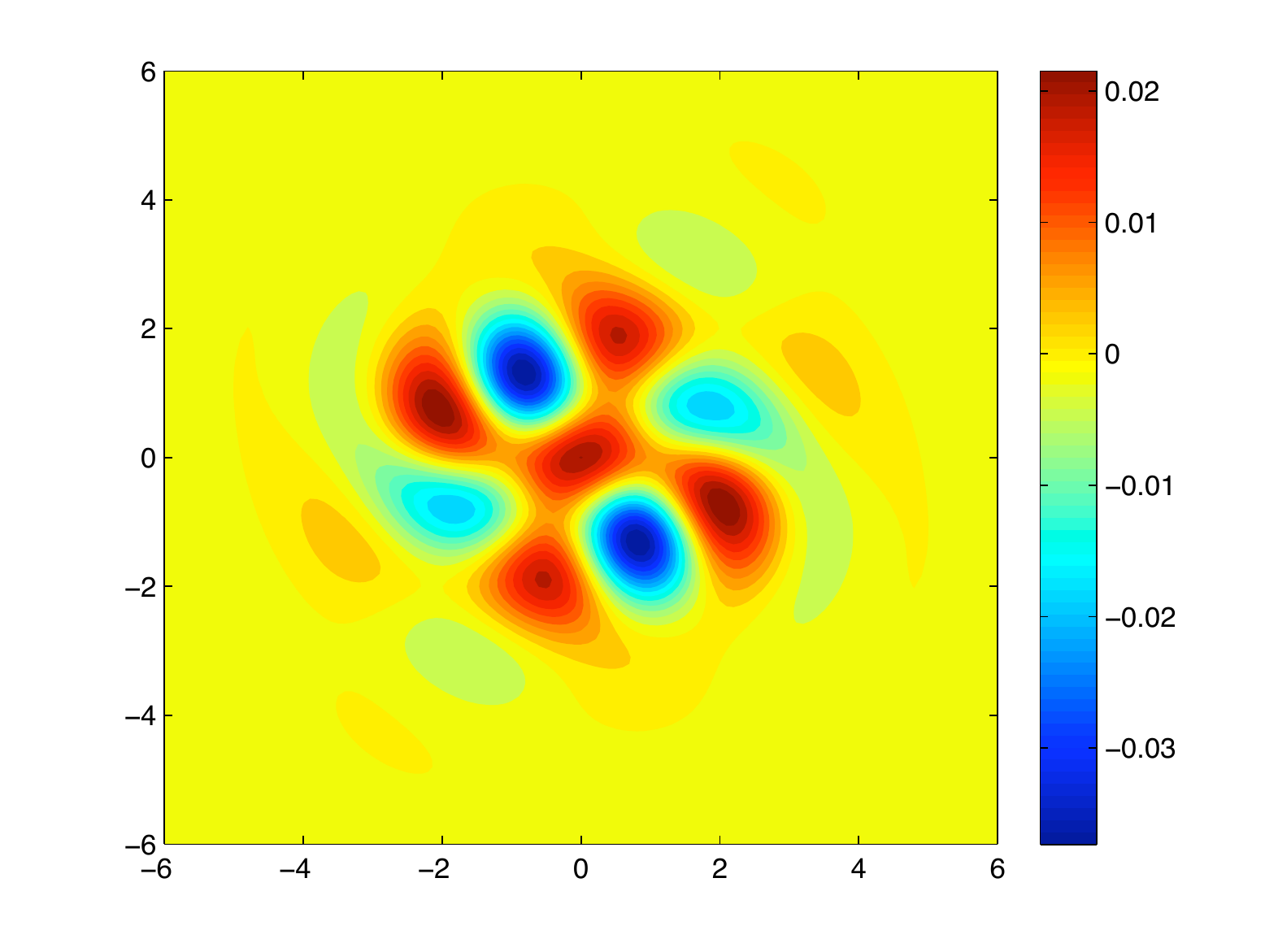}
\includegraphics[width=.30\textwidth]{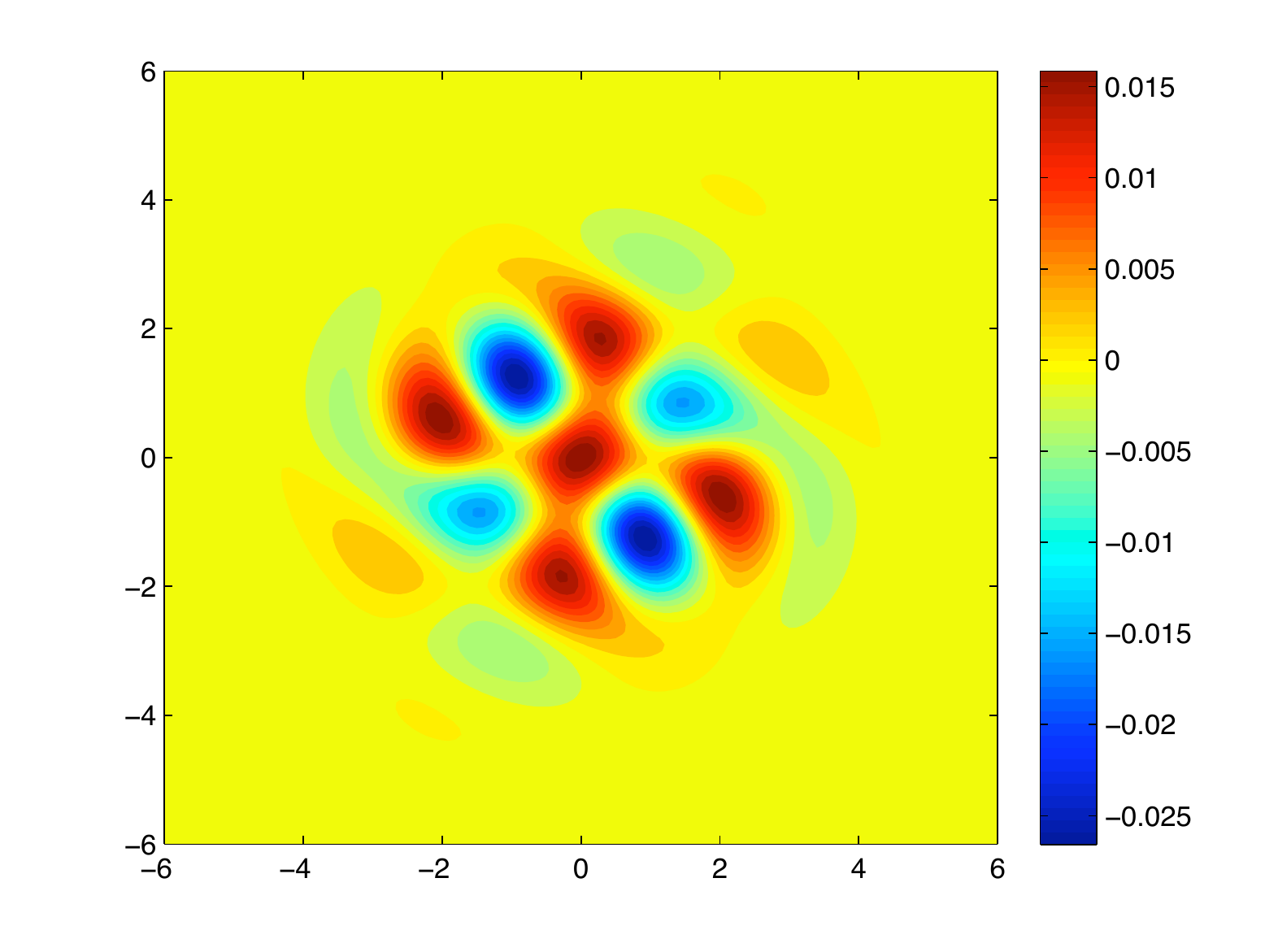}
\includegraphics[width=.30\textwidth]{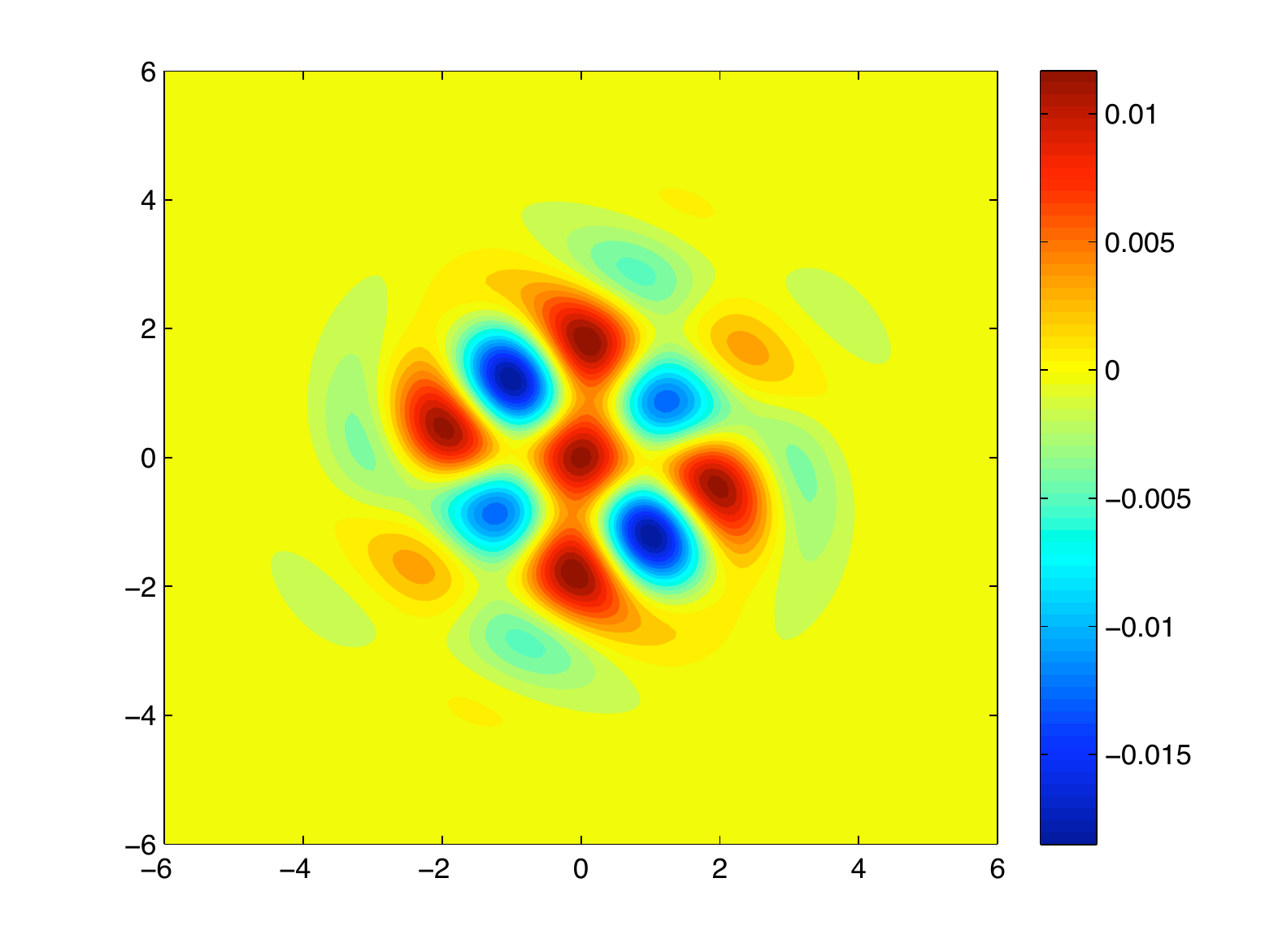}\\
\includegraphics[width=.30\textwidth]{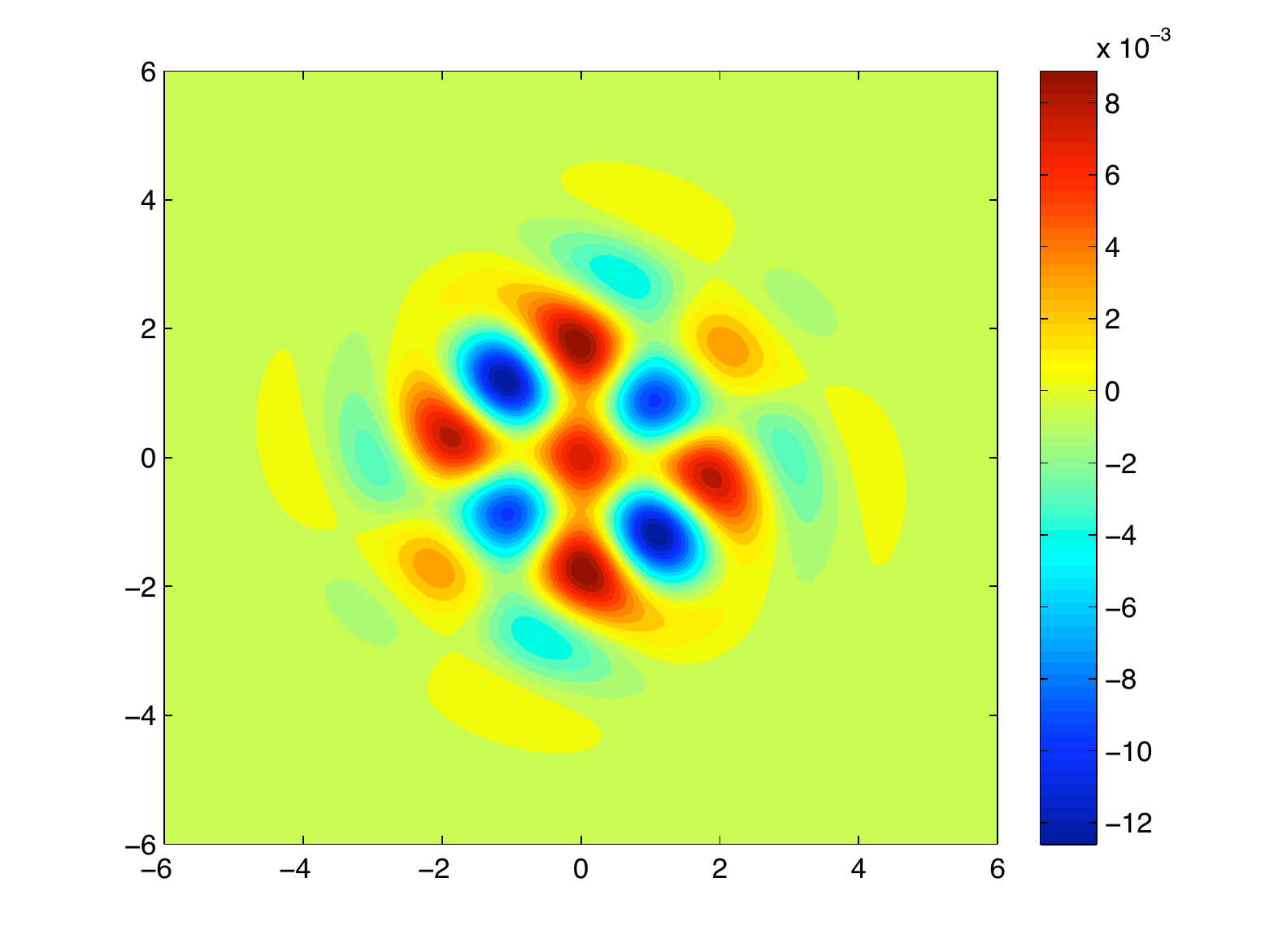}
\includegraphics[width=.30\textwidth]{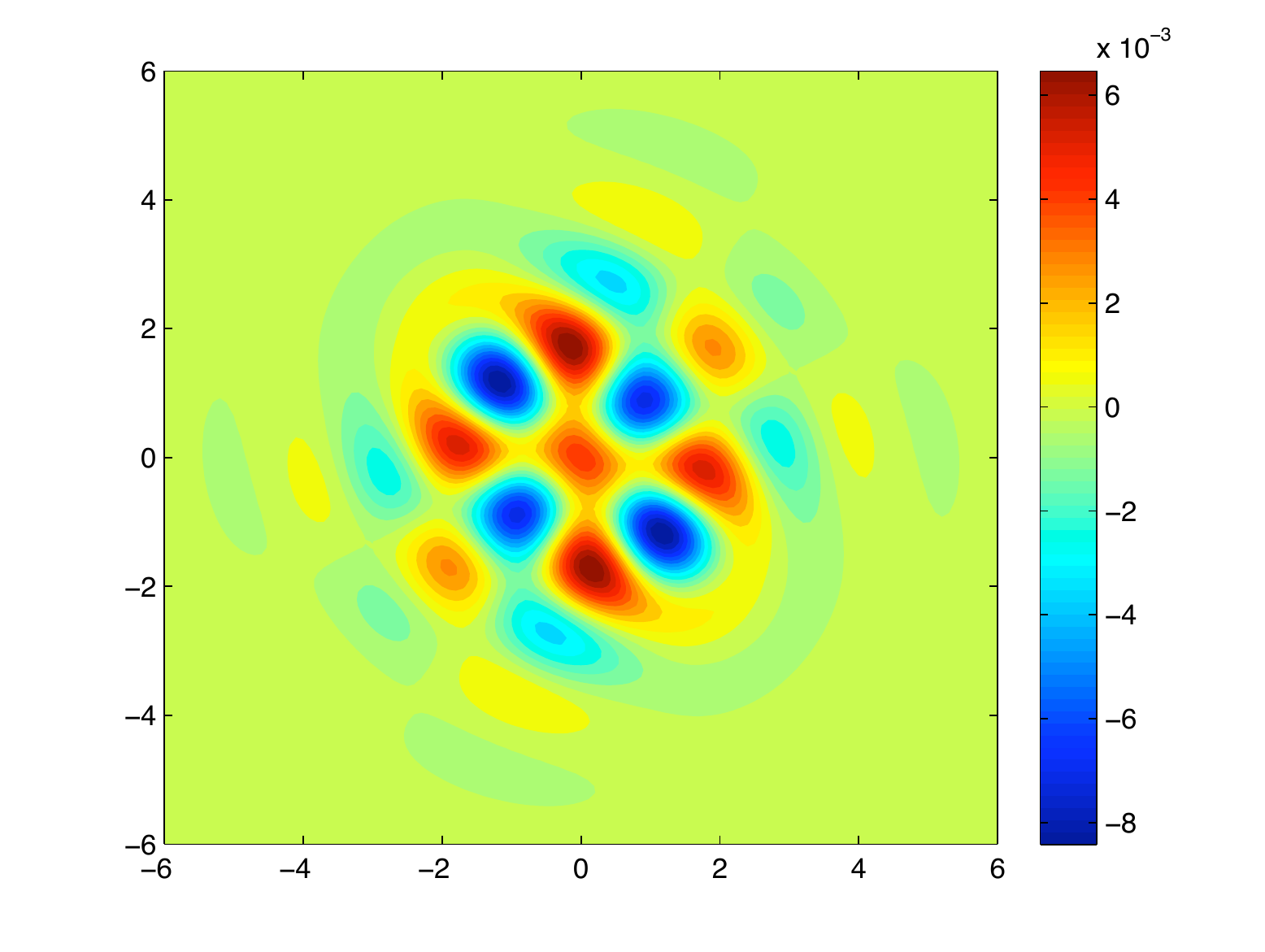}
\includegraphics[width=.30\textwidth]{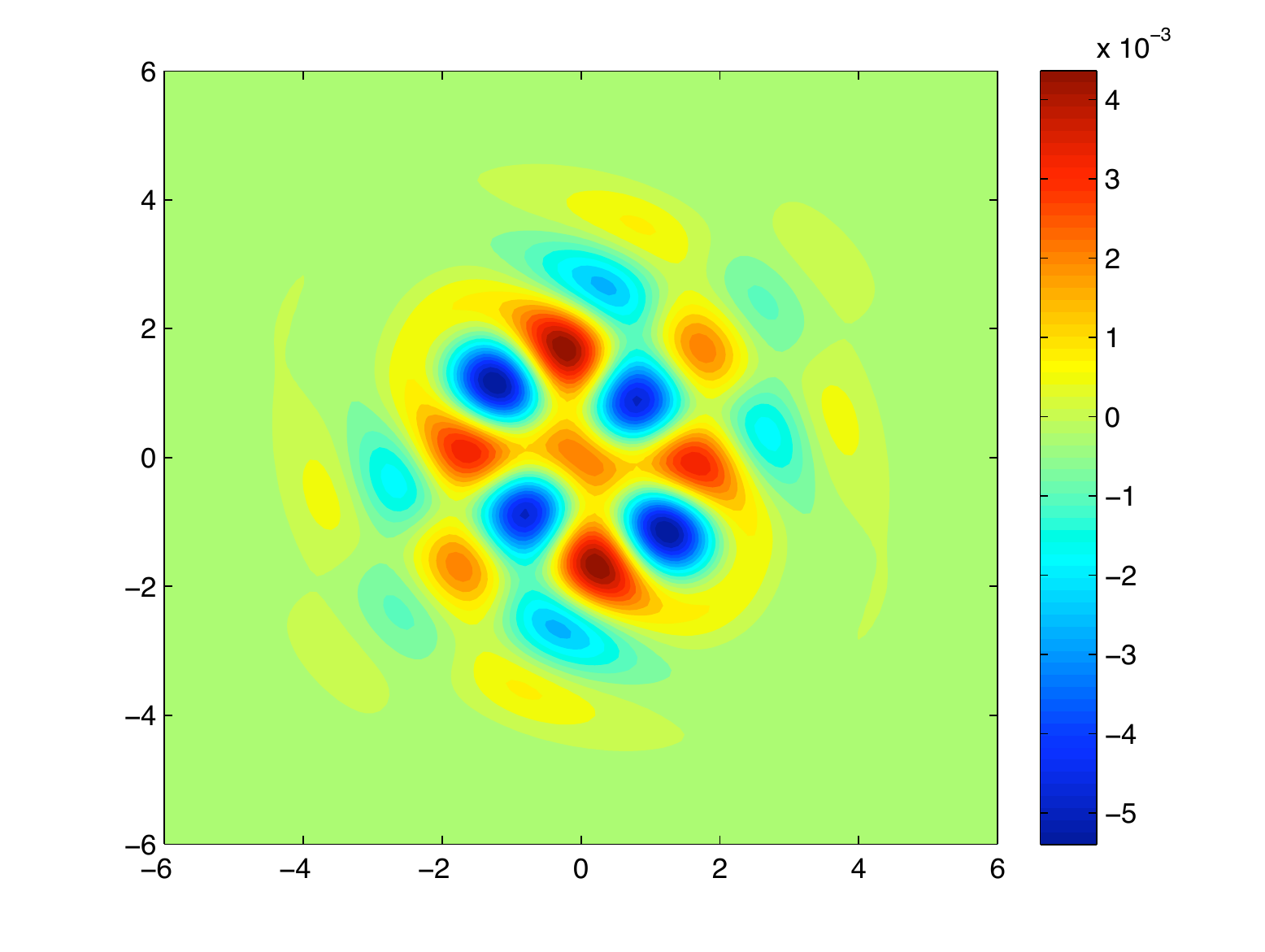}\\
\includegraphics[width=.30\textwidth]{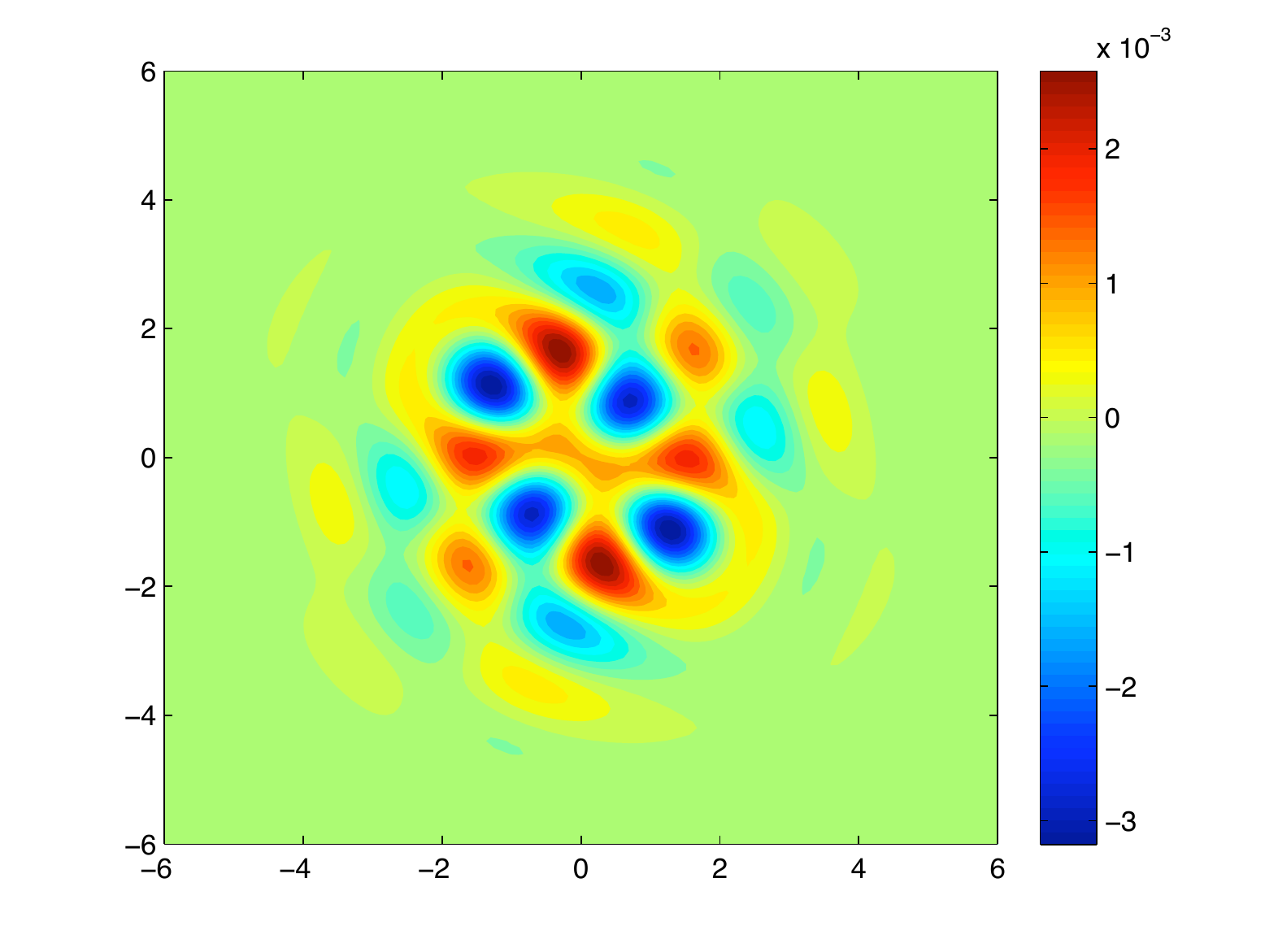}
\includegraphics[width=.30\textwidth]{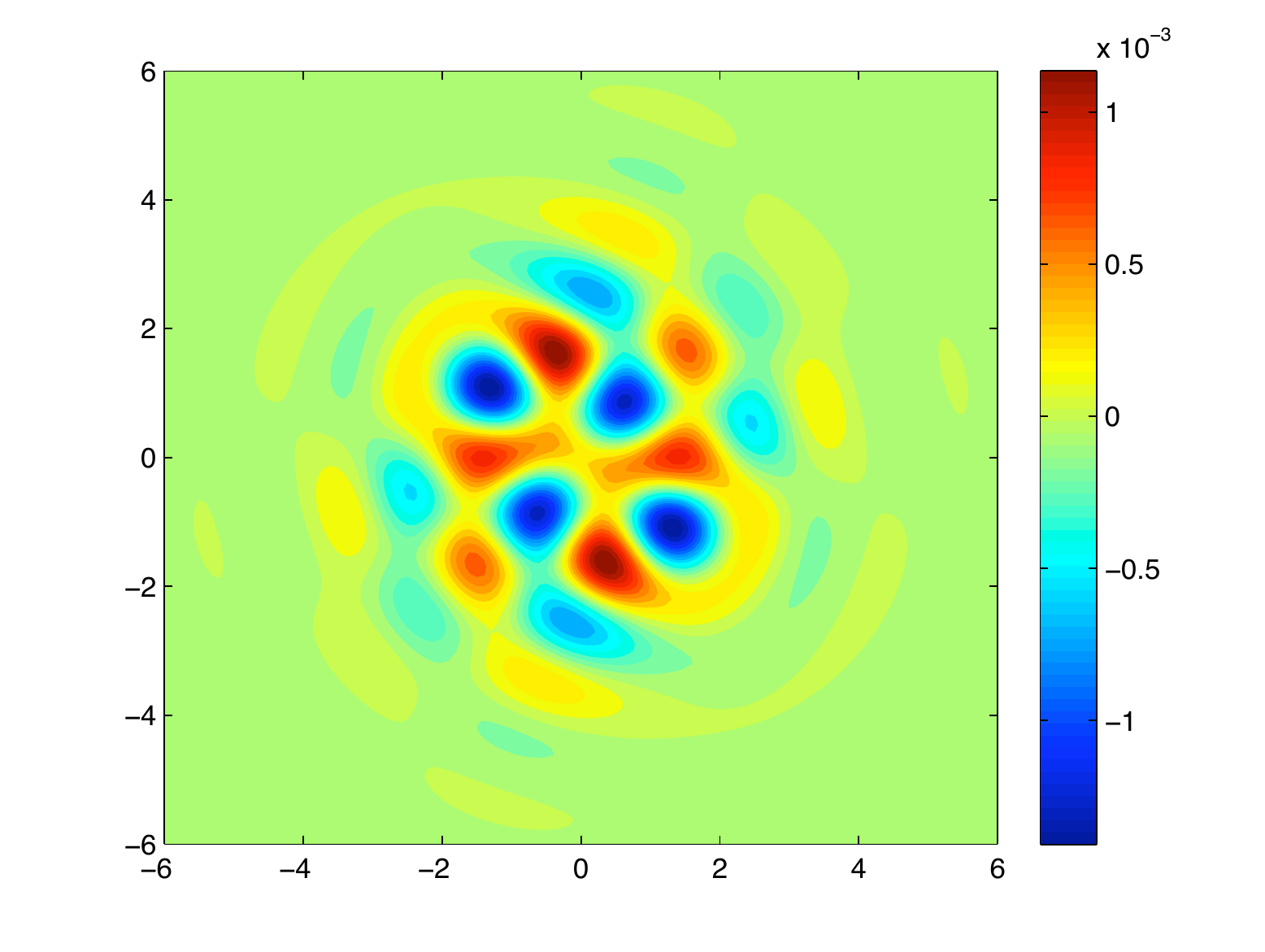}\\
\caption{The relative error at time $32$ between the solutions for (from left to right) $m=4$, $m=6$, $m=8$, $m=10$, $m=12$, $m=14$, $m=16$, $m=18$, $m=20$ and $m=22$ and the solution for $m=24$, with $\delta = 0.25$ and $Re=1000$.}\label{fig:errorPic}
\end{center}
\end{figure}

In addition to examining the size of the error as $m$ increases, it is interesting to consider the spatial distribution of the error. In Figure \ref{fig:errorPic}, we plot in the $xy$-plane the relative difference, { $$\frac{\omega_n(x,t) - \omega_{24}(x,t)}{||\omega_{24}||_\infty},$$ between the solution for $24$ moments and the solutions for a smaller number of moments at time $t=32$.}
 These numerical results suggest that in addition to increasing overall accuracy, the length scale of the error is declining as more moments are retained, indicating that resolution of the finer length scales improves as $m$ increases.


\subsection{Error analysis of the coarse grid simulation} \label{subsec:ErrorMMVM}
In section \ref{sec:CourseGrid}  two coarse grid approximations of early time tripole relaxation were computed using MMVM, the first a was classical vortex method (using Gaussian basis functions without higher Hermite moment corrections), i.e. $m=0$, and the second was a $m=2$ MMVM simulation.  When compared to both high precision tripole calculation, \cite{Barba:2004,Barba:2005} and a high order ($m=24$) single-particle MMVM calculation, the $m=2$ solution captured much more of the physical features associated to early time tripole relaxation as shown in figure \ref{fig:tripolet15}.  To quantify this improvement in error of $m=2$, as compared to $m=0$,  we computed the relative $L^2$,  and $L^\infty$ error, which we denote as $||e_m||_2$ and $||e_m||_\infty$ respectively for $m=0,2$. We use our single-particle MMVM $m=24$ calculation as the benchmark solution and present the result in Table \ref{tab:error}.

\begin{table}[!hbp]
\centering
\begin{tabular}{||c|cc|cc||}
    \hline
t & $||e_0||_2$ &$||e_2||_2$ & $||e_0||_\infty$ & $||e_2||_\infty$ \\
    \hline
 &  &  & &  \\
1 & 0.0125 & 0.0034 & 0.0113 &  0.0028 \\
2 & 0.0248 & 0.0067 & 0.0225 &  0.0058  \\
4 & 0.0488 & 0.0138 & 0.0443 & 0.0123  \\
8 & 0.0943 & 0.0309 & 0.0848 & 0.0309 \\
16& 0.1742 & 0.0821 & 0.1532 & 0.0956  \\
    \hline
    \hline
\end{tabular}
\caption{The relative error $||e_m||_2$ and $||e_m||_\infty$ for $m=0,2$ at times $t =1$, $2$, $4$, $8$, and $16$.}\label{tab:error}
\end{table}

It is immediately clear from the results in table \ref{tab:error}  that the $m=2$ MMVM simulation is a significant improvement in accuracy as compared to $m=0$. Specifically we can see that for early times the $m=2$ simulation is approximately $4$ times more accurate than $m=0$ in both $L^2$ and $L^\infty$ metrics.  This large improvement in accuracy eventually declines to around twice as accurate by $t=16$.   This decline in accuracy is due to the increased excitation of the second order moments which become too large and is analogous to the reduction in accuracy of Rossi's elliptical vortices when the aspect ratio  becomes to large, \cite{Rossi:2006A}. Thus, like Rossi's  elliptical  vortices, to maintain higher order accuracy over longer time scales one will need to stop the calculation and re-interpolate or re-initialize the vorticity, likely using the high order accuracy methods recently developed in \cite{Barba:2010}.

\subsection{A note on implementation and performance}

In our implementation of MMVM, we symbolically compute the equations \eqref{eq:FinalSingleI}-\eqref{eq:moment_evolutionFinish}, in the case of the single-particle MMVM, or equations (\ref{eq:Final_CrossCoef})-(\ref{eq:xjdot_Final}), in the case of the full MMVM, using Maple 10.5. The equations are then exported from Maple to Fortran code. We proceed to use this Fortran code to solve the equations in Matlab using ode45 with `RelTol' and `AbsTol' both set to $10^{-8}$. There is a one-time preprocessing cost associated with Maple's generation of the {Fortran} code and the subsequent writing to {.m} files. Once the .m files are created, the ode45 solver is relatively fast. For example, a typical run with initial conditions of the form \eqref{eq:IC_sd} and $Re=1000$ with $m=2$ moments took less than one second, for $m=18$ moments took $3.35$ minutes, and for $m=22$ moments took $10.65$ minutes to solve from $t=0$ to $t=2000$ on a $2.4$ GHz Intel Core 2 Duo processor with $2$ GB of RAM. Typical run times on the $6 \times 6$ course grid calculation with $m=2$ took about $6$ mins to run from $t=0$ to $t=50$.  These run times reflect the fact that there where several limitations of this simplistic implementation. Much improvement may be achieved  as no effort was made to optimizing the memory usage and the efficiency of storing and calling the system of equations (which can become quite large).  New, efficient and parallelizable  code is currently in development and is a necessary step for future large scale simulations using MMVM.


\section{Conclusion} \label{sec:conclusion}

In this paper, we have presented simplified equations for a new multi-moment vortex method for computing solutions to the 2D vorticity equation. The novel feature of MMVM as compared to classical vortex methods is that the higher Hermite moments allow the vortex particles to convect and deform without any of the usual computational difficulties associated to calculating the Biot-Savart kernel for anisotropic vortex elements.

We first considered a single-particle MMVM simulation and showed that the method captures the physical mechanism of shear diffusion (and the associated time scale $Re^{1/3}$) for low to medium Reynolds numbers.  We then presented two examples using the full MMVM which show how the inclusion of higher moments improves the calculation over non-deforming basis elements. The first is a simple model for vortex merger and we show that early convective merger behavior is only captured with the inclusion of higher Hermite moments. In the second example we perform a coarse grid calculation of tripole relaxation and show that by including just $m=2$ Hermite moments we show upwards of a {four fold improvement} in the reduction of error as compared to tradition vortex methods ($m=0$).  

One interesting feature about MMVM is that there are two parameters, the number of particles, $n$, and the order of the Hermite expansion, $m$, that can be used to tune the spatial accuracy of the method, whereas classical vortex methods has just the number of particles $n$. It was shown in Section \ref{sec:Numerical} that for a fixed $n=1$ the spatial convergence is exponential in $m$. One important avenue for future work is to better perform convergence studies and analysis of the spatial accuracy as a function of both $n$ and $m$. For a fixed $n$, increasing $m$ also introduces additional computational costs and in any practical implementation, one will need to find an optimal balance between computational efficiency and spatial accuracy by varying both $n$ and $m$.

The examples in this paper and the demonstrated improvement in the reduction of spatial error, establishes the promise of MMVM for larger scale calculations and thus the need for future work in this direction.  In addition to an efficient implementation of MMVM,  we will also need to incorporate  spatial adaptation of the vortex particles, such as the highly accurate methods recently developed in \cite{Barba:2010}, in order to push MMVM simulations to later times while maintaining accuracy.

\section{Acknowledgements}  The authors would like to thank Ricardo Cortez for many helpful suggestions in the writing of this paper.  The work of DU was supported in part by DMS-0902792 and a UC President's fellowship. CEW was supported in part by DMS-0908093. AB was supported in part by NSF grant DMS-0907931. Any findings, conclusions, opinions, or recommendations are those of the authors, and do not necessarily reflect the views of the NSF.

\appendix
\section{Combinatorial formulae}\label{sec:Combinatorial}

\subsection{The Single Vortex Case}\label{sec:single}
In this appendix we calculate explicitly the integral term in equation \eqref{eq:moment_evolution} found in Section \ref{sec:Review}, which has the form

\begin{equation}\label{eq:theintegral}
\int_{\real^2} H_{k_1,k_2}(\xx)
 ( D_{x_1}^{m_1} D_{x_2}^{m_2} \VV_{00}(\xx;\lambda) ) \cdot \mathbf{\nabla}_{\xx} ( D_{x_1}^{\ell_1} D_{x_2}^{\ell_2} \phi_{00}(\xx;\lambda) ) {\rm d}\xx ~.
\end{equation}

It is the goal of this appendix to further improve upon the simplification \eqref{eq:oneblobkernel} by deriving exact combinatorial expressions formulae for (\ref{eq:theintegral}) since such combinatorial formulae will allow for efficient numerical implementation of the Hermite spectral method.

We proceed by apply integrating by parts to (\ref{eq:theintegral}) to get:
\begin{eqnarray}
\int_{\real^2} H_{k_1,k_2}(\xx)
 ( D_{x_1}^{m_1} D_{x_2}^{m_2} \VV_{00}(\xx;\lambda) ) \cdot \mathbf{\nabla}_{\xx} ( D_{x_1}^{\ell_1} D_{x_2}^{\ell_2} \phi_{00}(\xx;\lambda) ) {\rm d}\xx = \\ -
\int_{\real^2} ( ( D_{x_1}^{m_1} D_{x_2}^{m_2} \VV_{00}(\xx;\lambda) ) \cdot \mathbf{\nabla}_{\xx}H_{k_1,k_2}(\xx) )
  D_{x_1}^{\ell_1} D_{x_2}^{\ell_2} \phi_{00}(\xx;\lambda) {\rm d}\xx = ~~-(\Gamma^{\rm I}+\Gamma^{II })
\end{eqnarray}
\begin{remark}
The term in the integration by parts formula in which the derivatives fall on $\VV_{00}$ do not appear because the fluid velocity is divergence free.
\end{remark}
where
\begin{eqnarray}
\Gamma^{\rm I} & = & \int_{\real^2} ( D_{x_1}^{m_1} D_{x_2}^{m_2} V_{00}^{1}(\xx;\lambda) )  D_{x_1}H_{k_1,k_2}(\xx)
  D_{x_1}^{\ell_1} D_{x_2}^{\ell_2} \phi_{00}(\xx;\lambda) {\rm d}\xx \\
\Gamma^{\rm II} & = & \int_{\real^2} ( D_{x_1}^{m_1} D_{x_2}^{m_2} V_{00}^{2}(\xx;\lambda) )  D_{x_2}H_{k_1,k_2}(\xx)
  D_{x_1}^{\ell_1} D_{x_2}^{\ell_2} \phi_{00}(\xx;\lambda) {\rm d}\xx
\end{eqnarray}
We first focus on equation $\Gamma^{\rm I}$. If we again integrate by parts $\ell_1$ times with respect to $x_1$ and $\ell_2$ times with respect to $x_2$ we arrive at:
\begin{eqnarray}
\Gamma^{\rm I} & = & \sum_{i=0}^{\ell_1} \sum_{j=0}^{\ell_2} {\ell_1 \choose i}{\ell_2 \choose j} (-1)^{\ell_1+\ell_2} \inttwo \phi_{00}D_{x_1}^{i+1}D_{x_2}^{j}H_{k_1,k_2}D_{x_1}^{m_1+\ell_1-i}D_{x_2}^{m_2+\ell_2-j}V_{00}^1{\rm d}\xx \\ & = & \sum_{i=0}^{min(\ell_1,k_1-1)} \sum_{j=0}^{min(\ell_2,k_2)} {\ell_1 \choose i}{\ell_2 \choose j}(-1)^{\ell_1+\ell_2} \left(\frac{2^{i+1}k_1!}{\lambda^{2(i+1)}(k_1-i-1)!}\right)\left(\frac{2^{j}k_2!}{\lambda^{2(j)}(k_2-j)!}\right) \times \nonumber\\ & & \times \inttwo \phi_{00}H_{k_1-i-1,k_2-j}D_{x_1}^{m_1+\ell_1-i}D_{x_2}^{m_2+\ell_2-j}V_{00}^1{\rm d}\xx \nonumber
\end{eqnarray}
Where in the second equality we used the fact that
\begin{eqnarray*}
\partial_{x_1}H_{n,m} &= &\left(\frac{2n}{\lambda^2}\right)H_{n-1,m}(x)\\
\partial_{x_2}H_{n,m} &= &\left(\frac{2m}{\lambda^2}\right)H_{n,m-1}(x).
\end{eqnarray*}
Now if we take a closer look at the last integral we can compute:
\begin{eqnarray*}
\inttwo \phi_{00}H_{k_1-i-1,k_2-j}D_{x_1}^{m_1+\ell_1-i}D_{x_2}^{m_2+\ell_2-j}V_{00}^1{\rm d}\xx & &
\end{eqnarray*}
\begin{eqnarray*}
\\ &=&(-1)^{k_1-i-1+k_2-j} \inttwo D_{x_1}^{k_1-i-1}D_{x_2}^{k_2-j}\phi_{00}D_{x_1}^{m_1+\ell_1-i}D_{x_2}^{m_2+\ell_2-j}V_{00}^1{\rm d}\xx \\ &=& (-1)^{k_1-i-1+k_2-j}(-1)^{k_1-i-1+k_2-j}\inttwo \phi_{00}D_{x_1}^{m_1+k_1-i-1+\ell_1-i}D_{x_2}^{m_2+k_2-j+\ell_2-j}V_{00}^1{\rm d}\xx \\ &=&
\inttwo \phi_{00}D_{x_1}^{m_1+k_1-i-1+\ell_1-i}D_{x_2}^{m_2+k_2-j+\ell_2-j}V_{00}^1{\rm d}\xx
\end{eqnarray*}
where the first equality comes from the fact that
\begin{eqnarray}
H_{n,m} = (-1)^{n+m}\phi_{00}^{-1}D_{x_1}^{n}D_{x_2}^{m}\phi_{00},
\end{eqnarray}
and the second inequality is a result of applying integration by parts. Thus in order to better understand expression (\ref{eq:theintegral}) we must simplify integrals of the form:
\begin{equation}\label{eq:lastint}
\inttwo \phi_{00}(x)D_{b_1}^{\alpha_1}D_{b_2}^{\alpha_2}V_{00}^1(\xx+\bb){\rm d}\xx|_{\bb=0}
\end{equation}

To do so we consider both components of the velocity field. By re-writing the velocity $V_{00}$ in terms of the vorticity $\phi_{00}$ we have:
\begin{equation}\label{eq:stream}
\VV_{00}(\xx+\bb;\lambda) = - \mathbf{\nabla}_{b}^* (\Delta_b)^{-1} \phi_{00}(\xx+\bb)\ ,
\end{equation}
where $\mathbf{\nabla}_{b}^* f = (\partial_{x_2} f,-\partial_{x_1} f)$. Thus we can study equation (\ref{eq:lastint}) as the first component of the equation (\ref{eq:integral_kernel}) below:
\begin{eqnarray}\label{eq:integral_kernel}
 && D_{b_1}^{\alpha_1}D_{b_2}^{\alpha_2} \int_{\real^2} \VV_{00}(\xx+\bb;\lambda)
\phi_{00}(\xx;\lambda) {\rm d}\xx \\ \nonumber && \qquad \qquad
= -D_{b_1}^{\alpha_1}D_{b_2}^{\alpha_2} \mathbf{\nabla}_b^* (\Delta_b)^{-1}  \int_{\real^2} \phi_{00}(\xx+\bb;\lambda)
\phi_{00}(\xx;\lambda) {\rm d}\xx.
\end{eqnarray}

As seen in \cite{SIADS} we can reduce (\ref{eq:integral_kernel}) to the form,
\begin{equation}\label{eq:Fullint}
\inttwo\phi_{00}D_{b_1}^{\alpha_1}D_{b_2}^{\alpha_2} V_{00}(x+\bb\lambda) = D_{b_1}^{\alpha_1}D_{b_2}^{\alpha_2}V_{00}(\bb;\sqrt{2}\lambda) |_{\bb=0}~.
\end{equation}

\begin{remark}
It is worth noting here that the differential equations governing the moments of the Hermite representation reduce to evaluating derivatives of the velocity field associated to the Lamb-Oseen vortex at zero. As we will see in the next section similar importance on the velocity field of the Lamb-Oseen vortex persist in the multi-vortex expansion.
\end{remark}

We consider the first component of (\ref{eq:Fullint}) which is relevant for equation (\ref{eq:lastint}). Expanding this component we get
\begin{eqnarray}
\frac{1}{2\pi}\frac{(-b_2)}{|\bb|^2}\left(1-e^{-\frac{|\bb|^2}{2\lambda^2}}\right) & = & \frac{-b_2}{2\pi} \sum_{n=0}^\infty \left(\frac{1}{2\lambda^2}\right)^{n+1}\frac{(-1)^n}{(n+1)!}|\bb|^{2n} \nonumber \\
& = & \frac{-1}{2\pi} \sum_{n=0}^\infty \left(\frac{1}{2\lambda^2}\right)^{n+1}\frac{(-1)^n}{(n+1)!}\sum_{k=0}^n {n\choose k} (b_1)^{2k}(b_2)^{2n-2k+1}~~. \nonumber \\
\end{eqnarray}
Thus we can evaluate:
\begin{equation}
D_{b_1}^{\alpha_1}D_{b_2}^{\alpha_2}\frac{1}{2\pi}\frac{(-b_2)}{|\bb|^2}\left(1-e^{-\frac{|\bb|^2}{2\lambda^2}}\right) |_{\bb=0}
\end{equation}
\begin{eqnarray}
 = \frac{-1}{2\pi} \sum_{n=0}^\infty \left(\frac{1}{2\lambda^2}\right)^{n+1}\frac{(-1)^n}{(n+1)!}\sum_{k=0}^n {n\choose k} \frac{(2k)!}{(2k-\alpha_1)!}(b_1)^{2k-\alpha_1} \frac{(2n-2k+1)!}{(2n-2k+1-\alpha_2)!}(b_2)^{2n-2k+1-\alpha_2} |_{\bb=0} \nonumber.
\end{eqnarray}
We can then see that the relevant values of $k^*$ and $n^*$ are:
\begin{eqnarray}
k^* &=& \frac{\alpha_1}{2} \nonumber \\
n^* &=& \frac{\alpha_2+\alpha_1-1}{2}. \nonumber
\end{eqnarray}
It is also clear that $\alpha_1$ must be even and $\alpha_2$ must be odd in order to have a non-zero contribution. This yields the final formula for the first component:
\begin{eqnarray}
D_{b_1}^{\alpha_1}D_{b_2}^{\alpha_2}\frac{1}{2\pi}\frac{(-b_2)}{|\bb|^2}\left(1-e^{-\frac{|\bb|^2}{2\lambda^2}}\right) |_{\bb=0} \nonumber &=&\\ \frac{-1}{2\pi} \left(\frac{1}{2\lambda^2}\right)^{n^*+1}\frac{(-1)^{n^*}}{(n^*+1)!} {n^*\choose k^*} \frac{(2k^*)!}{(2k^*-\alpha_1)!} \frac{(2n^*-2k^*+1)!}{(2n^*-2k^*+1-\alpha_2)!} &=&\\
 \frac{-1}{2\pi} \left(\frac{1}{2\lambda^2}\right)^{n^*+1}\frac{(-1)^{n^*}}{(n^*+1)!} {n^*\choose k^*} (\alpha_1)! (\alpha_2)!
\end{eqnarray}
or,
\begin{equation}
D_{b_1}^{\alpha_1}D_{b_2}^{\alpha_2}\frac{1}{2\pi}\frac{(-b_2)}{|\bb|^2}\left(1-e^{-\frac{|\bb|^2}{2\lambda^2}}\right) |_{\bb=0} = \mathcal{H}_1(\alpha_1,\alpha_2)
\end{equation}
where
\begin{equation*}
\mathcal{H}_1(\alpha_1,\alpha_2) = \left\{
\begin{array}{cl}
\frac{-1}{2\pi} \left(\frac{1}{2\lambda^2}\right)^{\frac{\alpha_2+\alpha_1+1}{2}}\frac{(-1)^{\frac{\alpha_2+\alpha_1-1}{2}}}{(\frac{\alpha_2+\alpha_1+1}{2})!} {\frac{\alpha_2+\alpha_1-1}{2}\choose \frac{\alpha_1}{2}} (\alpha_1)! (\alpha_2)! & \text{if $\alpha_1$ even and $\alpha_2$ odd} \\
0 & \text{otherwise. }
\end{array} \right.
\end{equation*}

If we compute the second component of (\ref{eq:integral_kernel}) in a similar fashion we see that

\begin{equation}
D_{b_1}^{\alpha_1}D_{b_2}^{\alpha_2}\frac{1}{2\pi}\frac{(b_1)}{|\bb|^2}\left(1-e^{-\frac{|\bb|^2}{2\lambda^2}}\right) |_{\bb=0} = \mathcal{H}_2(\alpha_1,\alpha_2)
\end{equation}
where
\begin{equation*}
\mathcal{H}_2(\alpha_1,\alpha_2) = \left\{
\begin{array}{cl}
\frac{1}{2\pi} \left(\frac{1}{2\lambda^2}\right)^{\frac{\alpha_2+\alpha_1+1}{2}}\frac{(-1)^{\frac{\alpha_2+\alpha_1-1}{2}}}{(\frac{\alpha_2+\alpha_1+1}{2})!} {\frac{\alpha_2+\alpha_1-1}{2}\choose \frac{\alpha_1-1}{2}} (\alpha_1)! (\alpha_2)! & \text{if $\alpha_1$ odd and $\alpha_2$ even} \\
0 & \text{otherwise.}
\end{array} \right.
\end{equation*}

We can plug these ``simple" formulas into { $\Gamma^{\rm I} \equiv \Gamma^{\rm I} [k_1,k_2,\ell_1,\ell_2,m_1,m_2;\lambda]$} and {$\Gamma^{\rm II} \equiv \Gamma^{\rm II} [k_1,k_2,\ell_1,\ell_2,m_1,m_2;\lambda]$} to get our finalized equation for the coefficients of the moments, though we must first note that if we follow the derivation of I we can find that $\Gamma^{\rm II}$ takes the form:
\begin{eqnarray}
{\Gamma^{\rm II}[k_1,k_2,\ell_1,\ell_2,m_1,m_2;\lambda]} & = & \sum_{i=0}^{\ell_1} \sum_{j=0}^{\ell_2} {\ell_1 \choose i}{\ell_2 \choose j} (-1)^{\ell_1+\ell_2} \inttwo \phi_{00}D_{x_1}^{i}D_{x_2}^{j+1}H_{k_1,k_2}D_{x_1}^{m_1+\ell_1-i}D_{x_2}^{m_2+\ell_2-j}V_{00}^2{\rm d}\xx \nonumber \\ & = & \sum_{i=0}^{min(\ell_1,k_1)} \sum_{j=0}^{min(\ell_2,k_2-1)} {\ell_1 \choose i}{\ell_2 \choose j}(-1)^{\ell_1+\ell_2} \left(\frac{2^{i}k_1!}{\lambda^{2(i)}(k_1-i)!}\right)\left(\frac{2^{j+1}k_2!}{\lambda^{2(j+1)}(k_2-j-1)!}\right) \times \nonumber\\ & & \times \inttwo \phi_{00}H_{k_1-i,k_2-j-1}D_{x_1}^{m_1+\ell_1-i}D_{x_2}^{m_2+\ell_2-j}V_{00}^2{\rm d}\xx
\end{eqnarray}

Thus we can use the equations above to write for both { $\Gamma^{\rm I}$ and $\Gamma^{\rm II}$}:
\begin{eqnarray}\label{eq:FinalSingleI}
{ \Gamma^{\rm I} [k_1,k_2,\ell_1,\ell_2,m_1,m_2;\lambda]} & = & \sum_{i=0}^{min(\ell_1,k_1-1)} \sum_{j=0}^{min(\ell_2,k_2)} {\ell_1 \choose i}{\ell_2 \choose j}(-1)^{\ell_1+\ell_2} \left(\frac{2^{i+1}k_1!}{\lambda^{2(i+1)}(k_1-i-1)!}\right)\left(\frac{2^{j}k_2!}{\lambda^{2(j)}(k_2-j)!}\right) \times \nonumber\\ & & \times \inttwo \phi_{00}D_{x_1}^{m_1+k_1-i-1+\ell_1-i}D_{x_2}^{m_2+k_2-j+\ell_2-j}V_{00}^1{\rm d}\xx \nonumber \\ & = & \sum_{i=0}^{min(\ell_1,k_1-1)} \sum_{j=0}^{min(\ell_2,k_2)} {\ell_1 \choose i}{\ell_2 \choose j}(-1)^{\ell_1+\ell_2} \left(\frac{2^{i+1}k_1!}{\lambda^{2(i+1)}(k_1-i-1)!}\right)\left(\frac{2^{j}k_2!}{\lambda^{2(j)}(k_2-j)!}\right) \times \nonumber\\& & \mathcal{H}_1(m_1+k_1-i-1+\ell_1-i,m_2+k_2-j+\ell_2-j)
\end{eqnarray}
\begin{eqnarray}\label{eq:FinalSingleII}
{ \Gamma^{\rm II} [k_1,k_2,\ell_1,\ell_2,m_1,m_2;\lambda]} & = & \sum_{i=0}^{min(\ell_1,k_1)} \sum_{j=0}^{min(\ell_2,k_2-1)} {\ell_1 \choose i}{\ell_2 \choose j}(-1)^{\ell_1+\ell_2} \left(\frac{2^{i}k_1!}{\lambda^{2(i)}(k_1-i)!}\right)\left(\frac{2^{j+1}k_2!}{\lambda^{2(j+1)}(k_2-j-1)!}\right) \times \nonumber\\ & & \times \inttwo \phi_{00}D_{x_1}^{m_1+k_1-i+\ell_1-i}D_{x_2}^{m_2+k_2-j-1+\ell_2-j}V_{00}^1{\rm d}\xx \nonumber\\ & = & \sum_{i=0}^{min(\ell_1,k_1)} \sum_{j=0}^{min(\ell_2,k_2-1)} {\ell_1 \choose i}{\ell_2 \choose j}(-1)^{\ell_1+\ell_2} \left(\frac{2^{i}k_1!}{\lambda^{2(i)}(k_1-i)!}\right)\left(\frac{2^{j+1}k_2!}{\lambda^{2(j+1)}(k_2-j-1)!}\right) \times \nonumber\\& & \mathcal{H}_2(m_1+k_1-i+\ell_1-i,m_2+k_2-j-1+\ell_2-j)
\end{eqnarray}

 Thus we have simplified the differential equations for $M[k_1,k_2](t)$ to:
\begin{equation} \label{eq:moment_evolutionFinish}
 \frac{d M}{dt}[k_1,k_2,t] = \frac{(-1)^{(k_1+k_2)} \lambda^{2 (k_1+ k_2)}}{2^{k_1+k_2} (k_1 !) (k_2 !) } \sum_{\ell_1,\ell_2}^\infty \sum_{m_1,m_2}^\infty
 M[\ell_1,\ell_2,t] M[m_1,m_2,t]{ \tilde{\Gamma} [k_1,k_2,\ell_1,\ell_2,m_1,m_2;\lambda],}
\end{equation}
{ where $\tilde{ \Gamma} [k_1,k_2,\ell_1,\ell_2,m_1,m_2;\lambda] =  \Gamma^{\rm I} [k_1,k_2,\ell_1,\ell_2,m_1,m_2;\lambda]+\Gamma^{\rm II} [k_1,k_2,\ell_1,\ell_2,m_1,m_2;\lambda]$.}


\subsection{The Multi-Vortex Expansion}\label{sec:multi}

We continue our work by analyzing the multi-vortex version of the model in a similar way. First we re-write equation \eqref{eq:omegaj} which are the governing equations for the multi-vortex model:
\begin{eqnarray}\label{eq:omegajA}\nonumber
\frac{\partial \omega^j}{\partial t}(\xx-\xx^j(t),t) &=& \nu \Delta \omega^j(\xx-\xx^j(t),t)
- \left(\sum_{\ell=1}^m
\uu^{\ell}(\xx-\xx^{\ell}(t),t) \right) \cdot \mathbf{\nabla} \omega^j(\xx-\xx^j(t),t) \\
&& \qquad + \dot{\xx}^j(t) \cdot \mathbf{\nabla} \omega^j(\xx-\xx^j(t),t)
\ ,\ j=1, \dots , M.
\end{eqnarray}
Each of the functions $\omega^j$ is then expanded in Hermite
moments as done in the previous section. Recall also that we must couple equations (\ref{eq:omegajA}) with the equations of motion for the centers defined above in equation \eqref{eq:xjdot} by:

\begin{equation}\label{eq:xjdotA}
\frac{{\rm d} \xx^j}{{\rm d}t}(t) = \frac{1}{M^j[0,0;t]} \sum_{ j^{\prime}=1}^m \inttwo \left(
\uu^{j^{\prime}} (\zz + \sss_{j,j^{\prime}},t) \omega^j(\zz,t) \right) {\rm d}\zz\ .
\end{equation}
where $\sss_{j,j^{\prime}} = -\xx^{j^{\prime}}(t) + \xx^j(t)$ and $\zz = \xx-\xx^j(t)$.

 To start we define
\begin{equation}\label{eq:moment_expansion_def_tw0}
\omega^j(\zz,t) = \sum_{k_1, k_2}^{\infty} M^j[k_1,k_2;t] \phi_{k_1,k_2}(\zz,t;\lambda)
\end{equation}
for $j=1..N$.
We make a similar expansion for the velocity field in terms of the functions
$\VV_{\ell_1,\ell_2}$, and insert the expansions into \eqref{eq:omegajA}.  Continuing to follow the work in section 4 of \cite{SIADS} we find that the differential equation associated for each coefficient is:
\begin{eqnarray}\label{eq:mom_multi}
&& \frac{{\rm d}M^j [k_1,k_2;t] }{{\rm d}t} = \\ \nonumber &&
\qquad - P_{k_1,k_2} \left[ \left( \sum_{j^{\prime}=1}^m
\sum_{\ell_1,\ell_2}^{\infty} M^{j^{\prime}} [\ell_1,\ell_2;t]
\VV_{\ell_1,\ell_2}(\zz+\sss_{j,j^{\prime}},t;\lambda) \right)
 \right. \\ \nonumber &&
\qquad \qquad \qquad \qquad \left. \cdot \mathbf{\nabla} \left(\sum_{m_1, m_2}^{\infty} M^j[m_1,m_2;t] \phi_{m_1,m_2}(\zz,t;\lambda) \right) \right] \nonumber\\
&& + P_{k_1,k_2} \left[\dot{\xx}^j(t)\cdot \mathbf{\nabla} \left(\sum_{m_1, m_2}^{\infty} M^j[m_1,m_2;t] \phi_{m_1,m_2}(\zz,t;\lambda) \right)  \right] \nonumber.
\end{eqnarray}

Equation (\ref{eq:mom_multi}) will be analyzed in several pieces. First note that when $j^{\prime}=j$ then $\sss_{j,j^{\prime}}=0$ and this case is reduced precisely to the work done in Section \ref{sec:one}. Thus there are only two new pieces that must be computed, the vortex interaction terms which include in the first term the case $j\ne j^{\prime}$ and the entire second term in (\ref{eq:mom_multi}).

For the case where $j\ne j^{\prime}$, a similar derivation found in the previous section will retain the same conclusions up until one reaches the step at equation (\ref{eq:Fullint}). We set $\bb = \sss_{j,j^{\prime}}$ instead of $\bb=0$ here. This is as far as one can simplify the equations for the case $j\ne j^{\prime}$. Thus we are left with the last term:
\begin{eqnarray} \label{eq:last_term}
P_{k_1,k_2} \left[\dot{\xx}^j(t)\cdot \mathbf{\nabla} \left(\sum_{m_1,m_2}^{\infty} M^j[m_1,m_2;t] \phi_{m_1,m_2}(\zz,t;\lambda) \right)  \right] = \quad \quad\quad\quad\\
 \frac{(-1)^{(k_1+k_2)} \lambda^{2 (k_1+ k_2)}}{2^{k_1+k_2} (k_1 !) (k_2 !) } \sum_{m1,m2} M^j[m_1,m_2;t] \inttwo H_{k_1,k_2}(\zz) \dot{\xx}^j(t)\cdot \nabla_\zz (D^{m_1}_{z_1}D^{m_2}_{z_2}\phi_{00}(\zz)) {\rm d}\zz \nonumber\\ ={\rm III + IV} \nonumber
\end{eqnarray}
where
\begin{eqnarray}
{\rm III} &=& \dot{x}_1^j(t) \inttwo H_{k_1,k_2}(\zz) D_{m_1+1,m_2}\phi_{00}(\zz){\rm d}\zz \\
{\rm IV} &=& \dot{x}_2^j(t) \inttwo H_{k_1,k_2}(\zz) D_{m_1+1,m_2}\phi_{00}(\zz){\rm d}\zz.
\end{eqnarray}
For now we consider just {\rm III}. We see that by applying integration by parts we have:
\begin{eqnarray}
{\rm III} & = & \dot{x}_1^j(t)(-1)^{m_1+m_2+1} \inttwo D^{m_1+1,m_2}H_{k_1,k_2}(\zz)\phi_{00}(\zz) {\rm d}\zz \nonumber\\
& = & (-1)^{m_1+m_2+1} \left(\frac{2^{m_1+1}k_1!}{\lambda^{2(m_1+1)}(k_1-m_1-1)!}\right)\left(\frac{2^{m_2}k_2!}{\lambda^{2(m_2)}(k_2-m_2)!}\right) \nonumber\\ & & \times \inttwo H_{k_1-m_1-1,k_2-m_2}(\zz) \phi_{00}(\zz) {\rm d}\zz.
\end{eqnarray}
But notice that
\begin{equation}
\inttwo H_{\alpha_1,\alpha_2}(\zz) \phi_{00}(\zz) {\rm d} \zz = \left\{ \begin{array}{cl}1 & \alpha_1,\alpha_2 = 0 \\ 0 & \text{otherwise}
\end{array}\right. ~.
\end{equation}
Similar work can be done for {\rm IV} and thus getting for both:
\begin{equation}\label{eq:III}
{\rm III} = \left\{ \begin{array}{cl} (-1)^{m_1+m_2+1} \left(\frac{2^{m_1+1}k_1!}{\lambda^{2(m_1+1)}}\right)\left(\frac{2^{m_2}k_2!}{\lambda^{2(m_2)}}\right) \dot{x}_1^j(t) & k_1 = m_1+1,~ k_2=m_2 \\ 0 & \text{otherwise}
\end{array}\right.
\end{equation}

\begin{equation}\label{eq:IV}
{\rm IV} = \left\{ \begin{array}{cl} (-1)^{m_1+m_2+1} \left(\frac{2^{m_1}k_1!}{\lambda^{2(m_1)}}\right)\left(\frac{2^{m_2+1}k_2!}{\lambda^{2(m_2+1)}}\right) \dot{x}_2^j(t) & k_1 = m_1,~ k_2=m_2+1 \\ 0 & \text{otherwise}
\end{array}\right.~.
\end{equation}
Combining the equation (\ref{eq:III}) and (\ref{eq:IV}) with (\ref{eq:last_term}) we { get the remarkably simple equation}:

\begin{eqnarray} \label{eq:last_termsimp}
P_{k_1,k_2} \left[\dot{\xx}^j(t)\cdot \mathbf{\nabla} \left(\sum_{m_1,m_2}^{\infty} M^j[m_1,m_2;t] \phi_{m_1,m_2}(\zz,t;\lambda) \right)  \right] = \quad \quad\quad\quad \nonumber \\ \dot{x}^j_1(t) M^j[k_1-1,k_2,t] +\dot{x}^j_2(t) M^j[k_1,k_2-1,t].
\end{eqnarray}

Lastly we must analyze equation (\ref{eq:xjdotA}). Again we sub in our expansion (\ref{eq:moment_expansion_def_tw0}) into (\ref{eq:xjdot}) and find componentwise:
\begin{eqnarray}
\frac{{\rm d} \xx^j}{{\rm d}t}(t)& = &\frac{1}{M^j[0,0;t]} \sum_{ j^{\prime}=1}^m \inttwo \left(
\uu^{j^{\prime}} (\zz + \sss_{j,j^{\prime}},t) \omega^j(\zz,t) \right) {\rm d}\zz \nonumber \\ & = &  \frac{1}{M^j[0,0;t]}\sum_{j^{\prime}=1}^m
\sum_{\ell_1,\ell_2}^{\infty}\sum_{m_1,m_2}^{\infty} M^{j^{\prime}} [\ell_1,\ell_2;t] M^{j} [m_1,m_2;t] \nonumber\\ & & \times
\inttwo \VV_{\ell_1,\ell_2}(\zz+\sss_{j,j^{\prime}},t;\lambda) \phi_{m_1,m_2}(\zz,t;\lambda) {\rm d}\zz~,
\end{eqnarray}
where we write out
\begin{eqnarray}
\inttwo \VV_{\ell_1,\ell_2}(\zz+\bb,t;\lambda) \phi_{m_1,m_2}(\zz,t;\lambda) {\rm d}\zz &=& \inttwo D^{\ell_1}_{b_1}D^{\ell_2}_{b_2} \VV_{00}(\zz+\bb)D^{m_1}_{z_1}D^{m_2}_{z_2}\phi_{00}(\zz) {\rm d}\zz \nonumber \\& & (-1)^{m_1+m_2} \inttwo D^{\ell_1+m_1}_{b_1}D^{\ell_2+m_2}_{b_2} \VV_{00}(\zz+\bb)\phi_{00}(\zz) {\rm d}\zz~. \nonumber
\end{eqnarray}
and now we are back precisely to equation (\ref{eq:Fullint}) with $\bb = \sss_{j,j^{\prime}}$. We can now conclude by writing down the full equations of motion for the coefficients for the two blob moments as:
\begin{equation}\label{eq:Final_CrossCoef}
\frac{{\rm d}M^j [k_1,k_2;t] }{{\rm d}t} = { A(k_1,k_2) + B(k_1,k_2) + C(k_1,k_2)},
\end{equation}
where
\begin{eqnarray}
{ A(k_1,k_2)} = \frac{(-1)^{(k_1+k_2)} \lambda^{2 (k_1+ k_2)}}{2^{k_1+k_2} (k_1 !) (k_2 !) } \sum_{\ell_1,\ell_2}^\infty \sum_{m_1,m_2}^\infty
 M^j[\ell_1,\ell_2,t] M^j[m_1,m_2,t] { \tilde{ \Gamma} [k_1,k_2,\ell_1,\ell_2,m_1,m_2;\lambda],}
\end{eqnarray}
where { $\tilde{ \Gamma} [k_1,k_2,\ell_1,\ell_2,m_1,m_2;\lambda]$ is defined in equations (\ref{eq:FinalSingleI}) and (\ref{eq:FinalSingleII}). $B(k_1,k_2)$} represents the $j \ne j^{\prime}$ terms in the first sum and can be written as
\begin{eqnarray}
{ B(k_1,k_2)} = \frac{(-1)^{(k_1+k_2)} \lambda^{2 (k_1+ k_2)}}{2^{k_1+k_2} (k_1 !) (k_2 !) }\sum_{j' \ne j}^m \sum_{\ell_1,\ell_2}^\infty \sum_{m_1,m_2}^\infty
 M^{j^{\prime}}[\ell_1,\ell_2,t] M^j[m_1,m_2,t]{ \Gamma_B [k_1,k_2,\ell_1,\ell_2,m_1,m_2;\lambda,\{ \sss_{j,j^{\prime}}\}]}
\end{eqnarray}
where ${ \Gamma_B [k_1,k_2;\ell_1,\ell_2,m_1,m_2;\lambda,\{ \sss_{j,j^{\prime}}\}]}= { \Gamma_B^{\rm I} [k_1,k_2;\ell_1,\ell_2,m_1,m_2;\lambda,\{ \sss_{j,j^{\prime}}\}]}+ { \Gamma_B^{\rm II} [k_1,k_2;\ell_1,\ell_2,m_1,m_2;\lambda,\{ \sss_{j,j^{\prime}}\}]}$ and
\begin{eqnarray}\label{eq:FinalCrossI}
 {\Gamma_B^{\rm I}} & = & \sum_{i=0}^{min(\ell_1,k_1-1)} \sum_{j=0}^{min(\ell_2,k_2)} {\ell_1 \choose i}{\ell_2 \choose j}(-1)^{\ell_1+\ell_2} \left(\frac{2^{i+1}k_1!}{\lambda^{2(i+1)}(k_1-i-1)!}\right)\left(\frac{2^{j}k_2!}{\lambda^{2(j)}(k_2-j)!}\right) \times \nonumber\\& & \mathcal{H}^B_1(m_1+k_1-i-1+\ell_1-i,m_2+k_2-j+\ell_2-j)
\end{eqnarray}
\begin{eqnarray}\label{eq:FinalCrossII}
 {\Gamma_B^{\rm II}} & = & \sum_{i=0}^{min(\ell_1,k_1)} \sum_{j=0}^{min(\ell_2,k_2-1)} {\ell_1 \choose i}{\ell_2 \choose j}(-1)^{\ell_1+\ell_2} \left(\frac{2^{i}k_1!}{\lambda^{2(i)}(k_1-i)!}\right)\left(\frac{2^{j+1}k_2!}{\lambda^{2(j+1)}(k_2-j-1)!}\right) \times \nonumber\\& & \mathcal{H}^B_2(m_1+k_1-i+\ell_1-i,m_2+k_2-j-1+\ell_2-j),
\end{eqnarray}
and
\begin{equation}
\mathcal{H}^B=D_{b_1}^{\alpha_1}D_{b_2}^{\alpha_2}\frac{1}{2\pi}\frac{(-b_2,b_1)}{|\bb|^2}\left(1-e^{-\frac{|\bb|^2}{2\lambda^2}}\right) |_{\bb=\sss_{j,j^{\prime}}}.
\end{equation}
The final term, $C$, in equation (\ref{eq:Final_CrossCoef}) is simply the contribution from (\ref{eq:last_termsimp}) i.e.
\begin{eqnarray}
C = \dot{x}^j_1(t) M^j[k_1-1,k_2,t] +\dot{x}^j_2(t) M^j[k_1,k_2-1,t].
\end{eqnarray}

The simplified equations of motion for the centers can also be written compactly as:
\begin{eqnarray}\label{eq:xjdot_Final}
\frac{{\rm d} \xx^j}{{\rm d}t}(t) & = &  \frac{1}{M^j[0,0;t]}\sum_{j^{\prime}=1}^m
\sum_{\ell_1,\ell_2}^{\infty}\sum_{m_1,m_2}^{\infty} M^{j^{\prime}} [\ell_1,\ell_2;t] M^{j} [m_1,m_2;t] \nonumber\\ & & \times (-1)^{m_1+m_2}
\mathcal{H}^B(m_1+\ell_1, m_2+\ell_2.)
\end{eqnarray}

Thus equations (\ref{eq:Final_CrossCoef})-(\ref{eq:xjdot_Final}) represent the reduced equations of motion for vorticity as given by the model.

\bibliographystyle{plain} \bibliography{multi_reference}

\begin{thebibliography}{10}

\bibitem{AgulloVerga:2001}
Olivier Agullo and Alberto Verga.
\newblock Effect of viscosity in the dynamics of two point vortices: Exact
  results.
\newblock {\em Physical Review E}, 63, 2001.

\bibitem{Anderson:1985}
Christopher Anderson and Claude Greengard.
\newblock On vortex methods.
\newblock {\em SIAM Journal on Numerical Analysis}, 22(3):413--440, 1985.

\bibitem{barba:2006}
L.~A. Barba and A.~Leonard.
\newblock Emergence and evolution of tripole vortices from net-circulation
  initial conditions.
\newblock {\em Physics of Fluids}, 19(1):017101, 2007.

\bibitem{Barba:2005}
L.~A. Barba, A.~Leonard, and C.~B. Allen.
\newblock Advances in viscous vortex methods - meshless spatial adaption based
  on radial basis function interpolation.
\newblock {\em International Journal for Numerical Methods in Fluids},
  47(5):387--421, 2005.

\bibitem{Barba:2010}
L.A. Barba and Louis~F. Rossi.
\newblock Global field interpolation for particle methods.
\newblock {\em Journal of Computational Physics}, 229(4):1292 -- 1310, 2010.

\bibitem{Barba:2004}
Lorena~A. Barba.
\newblock {\em Vortex Method of computing high-{R}eynolds number flows:
  Increased accuracy with a fully mesh-less formulation}.
\newblock Ph.D. Thesis, California Institute of Technology, 2004.

\bibitem{Beale:1982}
J.~Thomas Beale and Andrew Majda.
\newblock Vortex methods. {II}. {H}igher order accuracy in two and three
  dimensions.
\newblock {\em Math. Comp.}, 39(159):29--52, 1982.

\bibitem{Bernoff:1994}
Andrew~J. Bernoff and Joseph~F. Lingevitch.
\newblock Rapid relaxation of an axisymmetric vortex.
\newblock {\em Physical Fluids}, 6(11):3717--3723, 1995.

\bibitem{Canuto:2006}
Hussaini M.Y. Quarteroni A. Zang~Th.A. Canuto, C.G.
\newblock {\em {Spectral Methods: Fundamentals in Single Domains}}.
\newblock Springer, Berlin, 2006.

\bibitem{CerretelliWilliamson:2003}
C.~Cerretelli and C.~H.~K. Williamson.
\newblock The physical mechanism for vortex merging.
\newblock {\em Journal of Fluid Mechanics}, 475:41--77, 2003.

\bibitem{Chorin:1973}
A.J. Chorin and P.~Bernard.
\newblock Discretization of a vortex sheet, with an example of roll-up.
\newblock {\em Journal of Computation Physics}, 13:423--429, 1973.

\bibitem{Cottet:2000}
Georges-Henri Cottet and Petros~D. Koumoutsakos.
\newblock {\em Vortex methods}.
\newblock Cambridge University Press, Cambridge, 2000.
\newblock Theory and practice.

\bibitem{Eldredge:2002}
Jeff~D. Eldredge, Tim Colonius, and Anthony Leonard.
\newblock A vortex particle method for two-dimensional compressible flow.
\newblock {\em Journal of Computational Physics}, 179(2):371 -- 399, 2002.

\bibitem{GallayWayne:2002}
Thierry Gallay and C.~Eugene Wayne.
\newblock Invariant manifolds and the long-time asymptotics of the
  {N}avier-{S}tokes and vorticity equations on {$\bold R\sp 2$}.
\newblock {\em Archive for Rational Mechanics and Analysis}, 163(3):209--258,
  2002.

\bibitem{Greengard:1985}
Claude Greengard.
\newblock The core spreading vortex method approximates the wrong equation.
\newblock {\em Journal of Computational Physics}, 61(2):345 -- 348, 1985.

\bibitem{Huang:2009}
Mei-Jiau Huang, Huan-Xun Su, and Li-Chieh Chen.
\newblock A fast resurrected core-spreading vortex method with no-slip boundary
  conditions.
\newblock {\em Journal of Computational Physics}, 228(6):1916 -- 1931, 2009.

\bibitem{Josserand:2007}
Ch. Josserand and M.~Rossi.
\newblock The merging of two co-rotating vortices: a numerical study.
\newblock {\em European Journal of Mechanics - B/Fluids}, 26(6):779 -- 794,
  2007.

\bibitem{DizeVerga:2002}
St\'{e}phane Le~Diz\`{e}s and Alberto Verga.
\newblock Viscous interactions of two co-rotating vortices before merging.
\newblock {\em Journal of Fluid Mechanics}, 467:389--410, 2002.

\bibitem{Leonard:1980}
A.~Leonard.
\newblock Vortex methods for flow simulation.
\newblock {\em Journal of Computational Physics}, 37:289 -- 335, 1980.

\bibitem{Bernoff:1995}
Joseph~F. Lingevitch and Andrew~J. Bernoff.
\newblock Distortion and evolution of a localized vortex in an irrotational
  flow.
\newblock {\em Physical Fluids}, 7(5):1015--1026, 1995.

\bibitem{lundgren:1982}
T.~S. Lundgren.
\newblock Strained spiral vortex model for turbulent fine structure.
\newblock {\em Physics of Fluids}, 25(12):2193--2203, 1982.

\bibitem{majda:2002}
Andrew~J. Majda and Andrea~L. Bertozzi.
\newblock {\em Vorticity and incompressible flow}, volume~27 of {\em Cambridge
  Texts in Applied Mathematics}.
\newblock Cambridge University Press, Cambridge, 2002.

\bibitem{Melander:1984}
M.~V. Melander, A.~S. Styczek, and N.~J. Zabusky.
\newblock Elliptically desingularized vortex model for the two-dimensional
  {E}uler equations.
\newblock {\em Physical Review Letters}, 53:1222--1225, 1984.

\bibitem{Melander:1986}
M.~V. Melander, N.~J. Zabusky, and A.~S. Styczek.
\newblock A moment model for vortex interactions of the two-dimensional {E}uler
  equations. {P}art 1. {C}omputational validation of a {H}amiltonian elliptical
  representation.
\newblock {\em Journal of Fluid Mechanics}, 167:95--115, 1986.

\bibitem{meunier:2002}
P.~Meunier, U.~Ehrenstein, T.~Leweke, and M.~Rossi.
\newblock A merging criterion for two-dimensional co-rotating vortices.
\newblock {\em Physical Fluids}, 14(8):2757�2766, 2002.

\bibitem{Meunier:2005}
Patrice Meunier, St\'{e}phane~Le Diz\'{e}s, and Thomas Leweke.
\newblock Physics of vortex merging.
\newblock {\em Comptes Rendus Physique}, 6(4-5):431 -- 450, 2005.
\newblock Aircraft trailing vortices.

\bibitem{Moeleker:2001}
Piet Moeleker and Anthony Leonard.
\newblock Lagrangian methods for the tensor-diffusivity subgrid model.
\newblock {\em Journal of Computational Physics}, 167(1):1 -- 21, 2001.

\bibitem{nagem:2007}
Raymond~J. Nagem, Guido Sandri, and David Uminsky.
\newblock Vorticity dynamics and sound generation in two-dimensional fluid
  flow.
\newblock {\em The Journal of the Acoustical Society of America},
  122(1):128--134, 2007.

\bibitem{SIADS}
Raymond~J. Nagem, Guido Sandri, David Uminsky, and C.~Eugene Wayne.
\newblock Generalized {H}elmholtz-{K}irchhoff model for two dimensional
  distributed vortex motion.
\newblock {\em SIAM Journal on Applied Dynamical Systems}, 8(1):160--179, 2009.

\bibitem{Rhines:1982}
{P.B. Rhines} and {W.R. Young}.
\newblock How rapidly is a passive scalar mixed within closed streamlines?
\newblock {\em Journal of Fluid Mechanics}, 133:133--145, 1983.

\bibitem{Platte:2009}
Rodrigo~B. Platte, Louis~F. Rossi, and Travis~B. Mitchell.
\newblock Using global interpolation to evaluate the biot-savart integral for
  deformable elliptical gaussian vortex elements.
\newblock {\em SIAM Journal on Scientific Computing}, 31(3):2342--2360, 2009.

\bibitem{Rossi:1996}
Louis~F. Rossi.
\newblock {Resurrecting Core Spreading Vortex Methods: A New Scheme That Is
  Both Deterministic and Convergent}.
\newblock {\em SIAM Journal of Scientific Computing}, 17(2):370--397, 1996.

\bibitem{Rossi:2005}
Louis~F. Rossi.
\newblock Achieving high-order convergence rates with deforming basis
  functions.
\newblock {\em SIAM Journal on Scientific Computing}, 26(3):885--906, 2005.

\bibitem{Rossi:2006B}
Louis~F. Rossi.
\newblock A comparative study of lagrangian methods using axisymmetric and
  deforming blobs.
\newblock {\em SIAM Journal on Scientific Computing}, 27(4):1168--1180, 2006.

\bibitem{Rossi:2006A}
Louis~F. Rossi.
\newblock Evaluation of the biot--savart integral for deformable elliptical
  gaussian vortex elements.
\newblock {\em SIAM Journal on Scientific Computing}, 28(4):1509--1532, 2006.

\bibitem{rossi:1997}
Louis~F. Rossi, Joseph~F. Lingevitch, and Andrew~J. Bernoff.
\newblock Quasi-steady monopole and tripole attractors for relaxing vortices.
\newblock {\em Physics of Fluids}, 9(8):2329--2338, 1997.

\bibitem{Uminsky:2009}
David~T. Uminsky.
\newblock {\em The Viscous $N$ Vortex Problem: A Generalized
  Helmholtz/Kirchhoff Approach}.
\newblock Ph.D. Thesis, Boston University, 2009.

\end{thebibliography}

\end{document}